%% file: article.tex
\documentclass[12pt,french,english]{article}
\usepackage[T1]{fontenc}
\usepackage{amsmath}
\usepackage{color}
\usepackage{graphicx}
\usepackage{amssymb}

\makeatletter

\newcommand{\noun}[1]{\textsc{#1}}
\providecommand{\tabularnewline}{\\}
 \usepackage{amsthm, amsmath, amsfonts, amssymb}
 \theoremstyle{plain}    
 \newtheorem{thm}{Theorem}%[section] enlevé par moi pour eviter la numérotation par section

 \theoremstyle{remark}    
 \newtheorem{acknowledgement}[thm]{Acknowledgement} 
 \theoremstyle{plain}    
 \newtheorem{prop}[thm]{Proposition} %%Delete [thm] to re-start numbering

 \theoremstyle{definition}
 \newtheorem{defn}[thm]{Définition}
 \theoremstyle{plain}    
 \newtheorem{lem}{Lemma} %%Delete [thm] to re-start numbering
 \theoremstyle{plain}    
 \newtheorem{cor}[thm]{Corollary} %%Delete [thm] to re-start numbering

\usepackage{a4wide}
\def\defi{\stackrel{\rm def}{=}}
\usepackage{color}
\usepackage{graphics}

\usepackage{babel}
\addto\extrasfrench{\providecommand{\fg}{\ifdim\lastskip>\z@\unskip\fi~\frqq}}
\makeatother
\begin{document}

\title{Semi-classical formula beyond the Ehrenfest time in quantum chaos.
(I) Trace formula}

\author{Frédéric Faure\textit{}%
\thanks{Institut Fourier, 100 rue des Maths, BP74 38402 St Martin d'Heres,
France. e-mail:frederic.faure@ujf-grenoble.fr. http://www-fourier.ujf-grenoble.fr/\textasciitilde{}faure%
}}

\maketitle
Dedicated to Professor Yves Colin de Verdière on the occasion of his
60th birthday. 

\begin{abstract}
We consider a nonlinear area preserving Anosov map $M$ on the torus
phase space, which is the simplest example of a fully chaotic dynamics.
We are interested in the quantum dynamics for long time, generated
by the unitary quantum propagator $\hat{M}$. The usual semi-classical
Trace formula expresses $\mbox{Tr}\left(\hat{M}^{t}\right)$ for finite
time $t$, in the limit $\hbar\rightarrow0$, in terms of periodic
orbits of $M$ of period $t$. Recent work reach time $t\ll t_{E}/6$
where $t_{E}=\log\left(1/\hbar\right)/\lambda$ is the Ehrenfest time,
and $\lambda$ is the Lyapounov coefficient. Using a semi-classical
normal form description of the dynamics uniformly over phase space,
we show how to extend the trace formula for longer time of the form
$t\sim C.t_{E}$ where $C$ is any constant, with an error $\mathcal{O}\left(\hbar^{\infty}\right)$.
\end{abstract}

\tableofcontents{}

\section{Introduction}

Semi-classical analysis is a fruitful approach to understand wave
equations in the regime where the wave length is small in comparison
with the size of the domain, or with the size of the typical variation
of the potential, where the wave evolves. In that regime, one shows
that the evolution of a wave can be described in terms of Hamiltonian
classical dynamics in the same domain (or with the same potential).
For example, the Van-Vleck formula (1928) expresses the evolved wave
as a sum of the initial wave transported along several classical trajectories.
Because the wave formalism enters in many area of physics (acoustic
waves, seismic waves, electromagnetic waves, quantum waves ...), semi-classical
analysis is an important mathematical tool to understand physical
phenomena. Wave equations, and more generally Partial Differential
Equations is an important domain of mathematics, so semi-classical
analysis has also been extremely studied and developed in mathematics.

\subsection{Problematics of quantum chaos}

A common way to express the semi-classical limit or short wave-length
limit, is to introduce a dimensionless parameter $\hbar$ in the wave
equation, called the {}``Planck constant'', which corresponds to
$\hbar\simeq l/L$ where $l$ is the wave length and $L$ the typical
size of the domain. The semi-classical limit is then $\hbar\rightarrow0$.

One possible way to understand semi-classical correspondences is through
wave packets (or coherent states), which are waves localized as much
as possible in phase space, so that to mimic the motion of a classical
particle. Because of the uncertainty principle $\Delta q\Delta p\geq\hbar$,
in the phase space of position $q$ and momentum $p$, a wave packet
can not be localized better than a small domain of surface $\hbar$,
called the Planck cell. This dispersion of the wave packet in phase
space is responsible for its spreading during time evolution (in a
similar way as the spreading of a classical disk of surface $\hbar$).
After a finite time (compared to $\hbar\rightarrow0$), the spreading
is finite, the wave is still localized. For that reason, one can derive
the Van-Vleck formula giving the matrix elements of the propagator
or the semi-classical trace formula \cite{combescure-97,combescure-99}\cite{gutzwiller},
and obtain the Weyl formula giving the averaged density of states.

If the classical dynamics takes place in a compact energy surface,
and has exponential instabilities (e.g. hyperbolic dynamical systems),
the evolution of a wave packet is much harder to understand for longer
times. This is one of the goals of quantum chaos studies. If $\lambda$
denotes the classical Lyapounov exponent of instabilities, an heuristic
approach suggests that the wave packet which has initial length $\Delta q,\Delta p\geq\hbar$
should spreads into a long and thin distribution of length $L_{t}\geq\hbar e^{\lambda t}$,
which rolls up in the energy surface, as soon as $L_{t}\gg constant\Leftrightarrow t\gg\frac{1}{\lambda}\log\left(1/\hbar\right)$.
This introduces the Ehrenfest time $t_{E}=\frac{1}{\lambda}\log\left(1/\hbar\right)$
which is an important characteristic time scale in quantum chaos.
For longer times, $t\gg t_{E}$, the distribution may intersect all
the other elementary cells of surface $\hbar$ (Planck cells) in many
distinct components, whose number grows exponentially fast with $t$.
Each component is identified with a classical trajectory. As this
description suggests (and as been suggested by many works in the physical
literature \cite{semi3}), the matrix elements of the propagator operator
$\hat{M}_{t}$ between localized wave packets, for $t\gg t_{E}$,
could be expressed as a (exponentially large) sum over these distinct
trajectories. In particular the trace of the propagator could be expressed
as a sum over $\mathcal{N}_{t}\propto e^{\lambda t}$ periodic orbits.
This is the content of the Gutzwiller trace formula \cite{gutzwiller_71}\cite{gutzwiller}.
Although there are numerical calculations that suggest its validity,
such a semi-classical formula has not yet been proved to be valid
for time greater than $\frac{1}{6}t_{E}$. The difficulty relies in
the control of the errors of individual terms which add together.

A major goal of quantum chaos would be to describe the long time dynamics
of wave packets up to time $t_{H}\simeq1/\hbar$ (or more), called
the Heisenberg time%
\footnote{For a time independent dynamics with $d$ degrees of freedom, $t_{H}\simeq1/\hbar^{d}$.
For the model developed in this paper, $t_{H}\simeq1/\hbar$.%
}, which would allow semi-classical formula to resolve individual energy
levels. But at time $t\simeq t_{H}$ the number of classical trajectories
entering in a semi-classical formula is of order $\mathcal{N}_{t_{H}}\simeq\exp\left(\lambda/\hbar\right)$,
extremely large (!), compared to the dimension of the effective Hilbert
space which is $1/\hbar$. One famous observation and conjecture in
quantum chaos is that correlations between closed energy levels behave
as eigenvalues of random matrix ensembles \cite{bohigas_81}\cite{bohigas-89}.
Many works in physical literature assume that semi-classical expressions
such as the Gutzwiller formula are valid up to time of order $t_{H}\simeq1/\hbar$,
and deduce some heuristic explanations for the random matrix theory,
or other important quantum chaos phenomena \cite{eck95}\cite{haake_01}.
For recent reviews on the mathematical aspects of quantum chaos, see
\foreignlanguage{french}{}\cite{debievre-00b},\cite{debievre-05},\cite{zelditch-96}\cite{zelditch-05}. 

In this paper we consider a simple model of quantum chaotic dynamics,
namely a non linear uniform hyperbolic map on the torus, and we show
that semi-classical formula extends to time scales $t=C.t_{E}$, where
$C>0$ is \emph{any} constant. We do not reach the Heisenberg time.
The model is particular, but we believe that the methods presented
here could be extended to more general uniform hyperbolic dynamical
systems. In this paper, the objective is to discuss the Gutzwiller
trace formula, and in a second paper we will discuss a second application,
the {}``Van-Vleck formula'' which expresses matrix elements of the
propagator, also for times $t=Ct_{E}$.

\subsection{\label{sub:Characteristic-time-in}Characteristic time in quantum
evolution of wave packets discussed with a numerical example}

In this Section, in order to motivate the importance of the Ehrenfest
time, as a qualitative frontier in semi-classical evolution problems
in quantum chaos, we present and discuss some recent results concerning
the evolution of quantum states in a hyperbolic flow. One of the main
challenges in quantum chaos is to deal with both the long time limit
$t\rightarrow\infty$, and the semi-classical limit $\hbar\rightarrow0$.
Usual semi-classical results, such as the Ehrenfest or Egorov theorem,
concerns $\hbar\rightarrow0$ first, and $t\rightarrow\infty$ after.
The challenge is to try to reverse this order (in order to get information
on individual eigenfunctions and eigenvalues), or more modestly, make
$t$ depending on $\hbar$. 

\begin{figure}[tbph]
\begin{centering}\scalebox{0.9}[0.9]{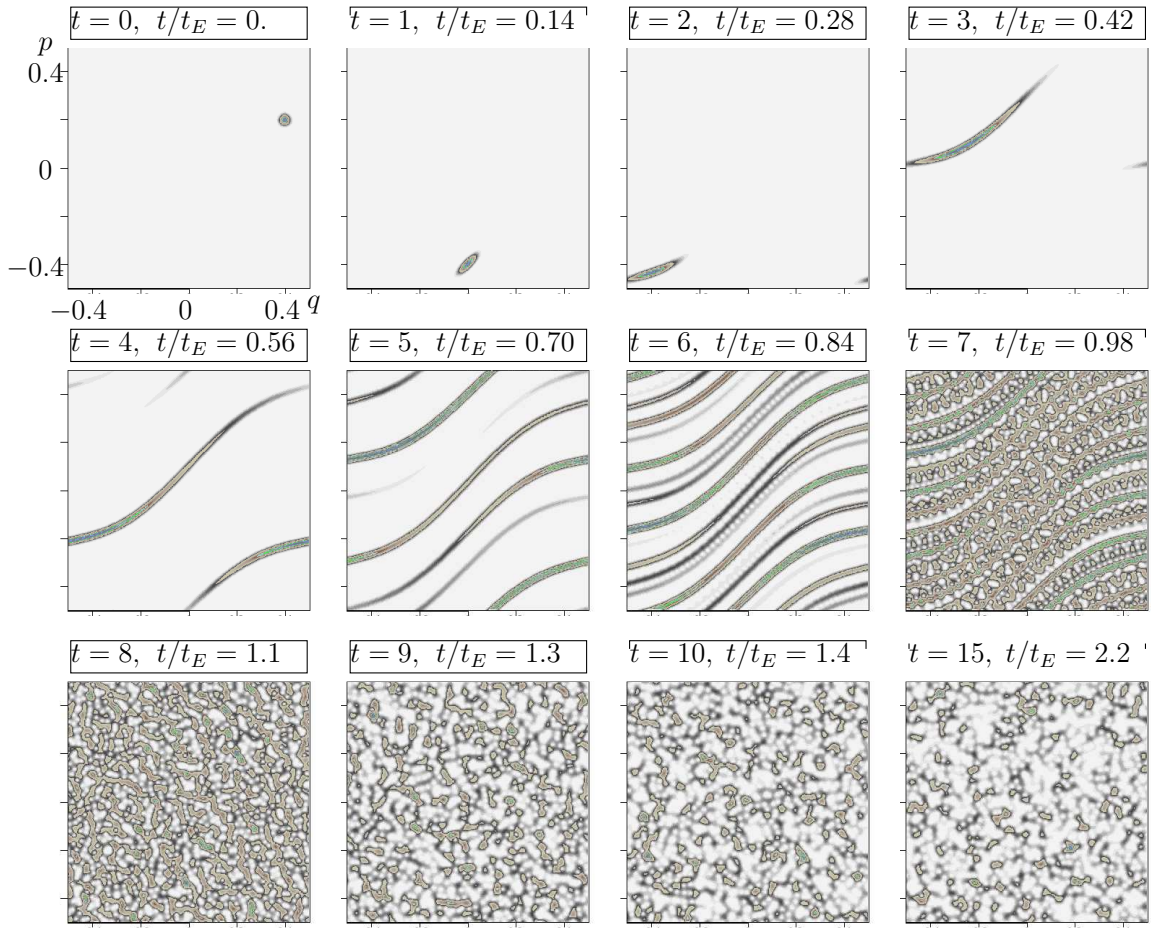}\par\end{centering}

\caption{\label{cap:Evolution-of-coherent-state}Phase space (Husimi) distribution
of the evolution of a (generic) coherent state at initial position
$x_{0}=\left(0.4,0.2\right)$, for different times $t=0,1,2,\ldots$
and $2\pi\hbar=10^{-3}$. }
\end{figure}

\begin{figure}[tbph]
\begin{centering}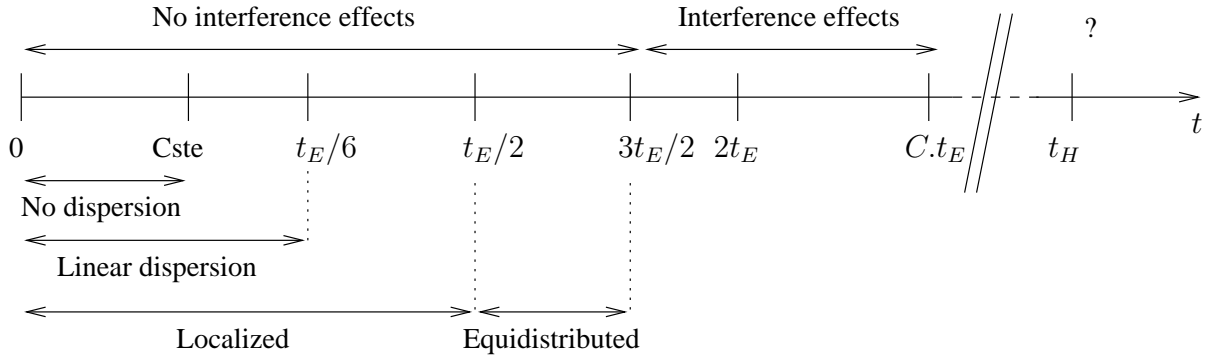\par\end{centering}

\caption{\label{cap:Characteristic-times}Characteristic times which appear
in the semi-classical limit $\hbar\rightarrow0$, for the evolution
of an initial coherent state. $t_{E}=\frac{1}{\lambda}\log\left(1/\hbar\right)$is
the Ehrenfest time, (very small {}``in practice'') compared to the
Heisenberg time $t_{H}=1/\hbar$.}
\end{figure}

The present discussion is made around the observation of the evolution
of a wave packet with a numerical example. We refer to Section \ref{sec:Quantum-map}
for a complete definition of the model. Let $M$ denotes a non linear
hyperbolic map on the torus phase space $\mathbb{T}^{2}$, and $\hat{M}$
the corresponding quantized map . A wave packet (a coherent state
$\psi_{x_{0}}$) is launched at time $t=0$, at a generic position
$x_{0}=\left(q_{0},p_{0}\right)\in\mathbb{T}^{2}$. Figure \ref{cap:Evolution-of-coherent-state}
shows the Husimi distribution (i.e. phase space representation) of
the evolved state $\psi\left(t\right)=\hat{M}^{t}\psi_{x_{0}}$ at
different time $t\in\mathbb{N}$. We remind that the Husimi distribution
of the initial state $\psi_{x_{0}}$ has typical width $\Delta_{0}\simeq\sqrt{\hbar}$,
(due to the uncertainty principle $\Delta q\Delta p\simeq\hbar$,
and the specific choice $\Delta q=\Delta p=\Delta_{0}\simeq\sqrt{\hbar}$).
During time evolution, the wave packet center is moving, and its distribution
spreads due to instabilities of the trajectories. Figure \ref{cap:Characteristic-times}
summarizes the main effects we discuss below.

\paragraph{Finite time regime with {}``no dispersion'':}

We first consider a fixed value of $t=Cste$, and $\hbar\rightarrow0$
(of course $t$ can be arbitrary large in principle). The evolved
state $\psi\left(t\right)$ is localized at the classical position
$x\left(t\right)=M^{t}x_{0}$. In more precise terms, the semi-classical
measure of $\psi\left(t\right)$ is a Dirac measure at $x\left(t\right)=M^{t}x_{0}$.
The evolved state $\psi\left(t\right)$ spreads but its width is $\Delta_{t}\simeq e^{\lambda t}\Delta_{0}\simeq\sqrt{\hbar}$,
still of order $\hbar^{1/2}$ \cite{squeez2}\cite{heller-75}\cite{litt1}.
Because $t$ can be chosen arbitrary large a priori, the ergodic nature
of the dynamics may have importance if $x\left(t\right)$ follows
a dense trajectory for example. Some well known semi-classical results
such as the semi-classical Egorov theorem \cite{zworski-03}, or the
Schnirelman quantum ergodicity theorem \cite{colin-85}\cite{zelditch-87}
use these finite range of time.

\paragraph{Linear dispersion regime:}

Some recent and very general results \cite{combescure-97}\cite{paul-99}\cite{joye-00}\cite{robert-02}
describe the evolved quantum state $\psi\left(t\right)$, in the linear
dispersion regime, which means that non linear effects on the dispersion
of the coherent state are supposed to be negligible with respect to
the linear effects. Because the first non linear effects correspond
to cubic terms in the Hamiltonian, this imposes that $\Delta_{t}^{3}\ll\hbar$,
equivalently $e^{\lambda t}\hbar^{1/2}\ll\hbar^{1/3}$, or $t\ll\frac{1}{6}t_{E}$.
In our numerical example $\frac{1}{6}t_{E}=1.2$.

\paragraph{Localized regime:}

After that time, the coherent state spreads more and more. But its
width is still of microscopic size if $\Delta_{t}\ll1$, i.e. $t\ll\frac{1}{2}t_{E}$.
In more precise terms, the semi-classical measure of $\psi\left(t\right)$
is still a Dirac measure at $x\left(t\right)$ in that range of time.
In our example $\frac{1}{2}t_{E}=3.6$. At a time around $t\simeq\frac{1}{2}t_{E}$,
the quantum state has size of order 1, and can be described as a {}``Lagrangian
W.K.B state'' \cite{nonnenmacher-04bis}.

\paragraph{Equidistribution regime:}

For time $t$ larger than $\frac{1}{2}t_{E}$, the wave packet spreads
and wraps around the torus phase space, along unstable manifolds,
like a classical probability measure. Thanks to classical mixing,
a smooth classical probability distribution is known to converge towards
the uniform Liouville measure for large time. The Husimi distribution
is expected to behave like a classical measure, and equidistributes,
if the different branches do not {}``interfere'' with each other
on phase space. After the time $\frac{1}{2}t_{E}$, we evaluate that
the distance between consecutive branches get smaller and smaller
like $d\sim e^{-\lambda(t-t_{E}/2)}$until the critical value $\hbar$
is obtained at time $t=\frac{1}{2}t_{E}+t_{E}=\frac{3}{2}t_{E}$ (This
is indeed the ultimate value, because if $d\gg\hbar$, one can still
insert a (squeezed) localized wave packet between two consecutive
branches, which means that the branches do not yet interfere). Correspondingly
to this description, in \cite{debievre-00}, the authors show that
for the \emph{linear} map, the semi-classical measure $\psi\left(t\right)$
converges towards the Liouville measure, in the range of time $\frac{1}{2}t_{E}\ll t\ll\frac{3}{2}t_{E}$.
J.M. Bouclet and S. De Bièvre in \cite{bouclet-04} obtain a similar
result for a non linear hyperbolic map, but for $t\ll\frac{2}{3}t_{E}$.
S. Nonnenmacher in \cite{nonnenmacher-04bis} reaches the time $t\ll\frac{3}{2}t_{E}$.
In \cite{schubert-04}, R. Schubert has described evolution of an
initial Lagrangian state under a hyperbolic flow. He has obtained
similar results, namely equidistribution up to time $t\ll t_{E}$.
This is indeed similar, because an initial coherent state becomes
a Lagrangian state at time $\frac{1}{2}t_{E}$. This range of time
is also considered and controlled in \cite{nalini-04,nalini-06}.

\paragraph{Longer time and interference effects:}

For longer time very little is known. Some arguments and numerical
observations in \cite{semi3} suggest that semi-classical formula
applies for longer time. This is the subject of the present work and
the second paper to come \cite{fred-VanVleck-06}. We show that in
the range of time $t\in\left[0,C.t_{E}\right]$, where $C$ is any
constant, the evolved state $\psi\left(t\right)$ described by its
Husimi distribution $Hus\left(x\right)=\left|\langle x|\psi\left(t\right)\rangle\right|^{2}$,
or by its Bargmann distribution $\langle x|\psi\left(t\right)\rangle$,
can be expressed in general as a (finite) sum over different classical
trajectories starting from the vicinity of the initial state $x_{0}$,
and ending at time $t$ in the vicinity of the point $x$ (similarly
to the semi-classical Van-Vleck formula). These trajectories give
unavoidable interferences effects for time $t\geq\frac{3}{2}t_{E}$,
because one can estimate that the sum involves more than one trajectory.
Similarly the trace formula expresses $\mbox{Tr}\left(\hat{M}^{t}\right)$
as a sum over periodic trajectories of period $t$. It is known that
in specific cases, revival may occur%
\footnote{For a linear map, these interferences effects are responsible for
exact revival of quantum states at time $t=2t_{E}$, and existence
of non ergodic invariant semi-classical measure (strong scars), as
shown in \cite{fred-steph-02,fred-steph-03}. Reaching the time $2t_{E}$
for the semi-classical description of a \emph{non linear} hyperbolic
map, was the main motivation of this work, although we have not yet
any precise idea if strong scarring effect may exist for general non
linear hyperbolic systems.%
} at time $t\simeq2t_{E}$ \cite{fred-steph-02}, however it is expected
that at least generically, these different contributions are somehow
uncorrelated, and as a result, the state $\psi\left(t\right)$ is
{}``generically'' equidistributed over phase space as can be observed
on figure \ref{cap:Evolution-of-coherent-state}.

An important characteristic time which is not considered here, because
far much larger than the actual semi-classical approach could reach,
is the \textbf{Heisenberg time} $t_{H}=1/h$ ($=1000$ in our example).
This time is related with the mean separation between eigenvalues
of $\hat{M}$. Some important effect of quantum chaos are numerically
observed at this range of time, and explained by a \textbf{Random
Matrix Theory} approach \cite{bohigas-89}. It allows to describe
statistical properties of individual eigenfunctions and eigenvalues.
Note that contrary to the mathematical works which are {}``stopped''
by the Ehrenfest time, the Heisenberg time is extremely discussed
in the physical literature, essentially with the random matrix theory.

\subsection{Results and organisation of the paper}

The main result of this paper is Theorem \ref{pro:formule_trace_semi_classique}
page \pageref{pro:formule_trace_semi_classique}, which shows that
Gutzwiller semi-classical trace formula is valid for long time $t\simeq C\log\left(1/\hbar\right)$,
with any $C>0$. This formula expresses $\mbox{Tr}\left(\hat{M}^{t}\right)$
in terms of semi-classical invariants associated to periodic orbits
of period $t$, up to an arbitrary small error: for any $K$, there
exists $D_{C,k}>0$, such that\[
\left|\mbox{Tr}\left(\hat{M}^{t}\right)-\mbox{T}_{semi,t,J}\right|\leq D_{C,K}\hbar^{K}\]
with \[
\mbox{T}_{semi,t,J}=\sum_{x=M^{t}x}\left|\mbox{Det}\left(D_{x}M^{t}-1\right)\right|^{-1/2}\exp\left(i\mathcal{A}_{x,t}/\hbar\right)\left(1+\hbar E_{1,x,t}+\hbar^{2}E_{2,x,t}+\ldots\hbar^{J}E_{J,x,t}\right)\]
with $J>2\left(K+C\right)$, and where $\mathcal{A}_{x,t}\equiv\oint pdq-Hdt$
is the action of the periodic orbit, and $E_{j,x,t}$ depends on semi-classical
Normal forms coefficients of the periodic orbit.

This formula has already been derived in a similar form (i.e. with
Normal form invariants) many time \cite{guillemin-96}\cite{zelditch-98}\cite{iantchenko-02b}\cite{sjostrand-02},
but for finite time $t$ (with respect to $\hbar$) giving: \[
\left|\mbox{Tr}\left(\hat{M}^{t}\right)-\mbox{T}_{semi,t,K}\right|\leq C_{t,K}\hbar^{K}.\]
The main difficulty to extend this last formula for larger time was
twice: how to control that neither the error term $C_{t,K}$, nor
the normal form coefficients $E_{j,x,t}$ do diverge, when $t\simeq C\log\left(1/\hbar\right)$
is so large. Such divergences could a priori be expected, due to exponential
increase of the complexity of the dynamics with $t$. For a linear
hyperbolic map, there are no semiclassical corrections, the trace
formula is exact, and has been derived by J.P. Keating \cite{keating-91b}.

In Section \ref{sec:Quantum-map}, we define the dynamics and recall
important properties of uniform hyperbolicity. In Section \ref{sec:The-map-as-PO},
we describe the periodic points and prepare the Trace Formula. In
Section \ref{sec:Semi-classical-description-of}, we show how to control
the semi-classical dynamics over large time, which will give us a
control of the above error term $C_{t,K}$. In Section \ref{sec:Semi-classical-De-Latte-non-stationnary},
we give a global semi-classical Normal form description of the hyperbolic
dynamics, derived from a work of David DeLatte \cite{delatte_92}
in the classical case, and as a result, this gives us a control of
the above coefficients $E_{j,x,t}$ uniformly over $x$ and $t$.
Considering the results together, we deduce the semi-classical Trace
Formula in Section \ref{sec:Trace-of-M}. The appendices, that can
be skipped in a first reading, give proofs of the Propositions used
in this paper. The results are illustrated with numerical examples
in Section \ref{sub:Numerical-results-and}. It is observed there
that our bounds on the errors are not sharp at all, and that the validity
of semi-classical Gutzwiller formula seems to be much better, as already
suggested in \cite{semi3}. We discuss again this fact in the conclusion,
where we suggest an alternative approach with prequantum dynamics,
and some perspectives.

\begin{acknowledgement}
The author gratefully acknowledges Yves Colin de Verdière, Patrick
Gérard, San VuNgoc for numerous discussions, and in particular Stéphane
Nonnenmacher for helpful remarks and suggestions. We acknowledge a
support by {}``Agence Nationale de la Recherche'' under the grant
ANR-05-JCJC-0107-01.
\end{acknowledgement}

\section{\label{sec:Quantum-map}Quantum non linear Anosov map on the torus}

In this section we describe the dynamical system considered in this
paper, a non linear area preserving map $M$ on the torus $\mathbb{T}^{2}$.
This map is constructed as a perturbation of a linear hyperbolic map
$M_{0}$, and is supposed to be uniformly hyperbolic (which is true
for small enough perturbations because of structural stability theorem).
In order to quantize the map $M$ in a natural way, we construct it
from a Hamiltonian flow. This section recalls some well known results
about this construction.

\subsection{Classical Dynamics}

\subsubsection{The linear map $M_{0}$}

Consider a quadratic Hamiltonian on phase space $x=\left(q,p\right)\in\mathbb{R}^{2}$
with symplectic two form $\omega=dq\wedge dp$:\begin{equation}
H_{0}\left(q,p\right)=\frac{1}{2}\alpha q^{2}+\frac{1}{2}\beta p^{2}+\gamma qp,\label{eq:H0}\end{equation}

with coefficients $\alpha,\beta,\gamma\in\mathbb{R}$. The Hamilton
equation of motion for the trajectory $x\left(t\right)$ are $dq(t)/dt=\partial H_{0}/\partial p=\gamma q+\beta p$
and $dp(t)/dt=-\partial H_{0}/\partial q=-\alpha q-\gamma p$. We
deduce that the flow after time 1 is $x(1)=M_{0}x(0)$ with the matrix\begin{equation}
M_{0}\defi\left(\begin{array}{cc}
A & B\\
C & D\end{array}\right)=\exp\left(\begin{array}{cc}
\gamma & \beta\\
-\alpha & -\gamma\end{array}\right)\in\mbox{SL}\left(2,\mathbb{R}\right),\label{eq:M0}\end{equation}

\begin{figure}[tbph]
\begin{centering}\scalebox{0.9}[0.9]{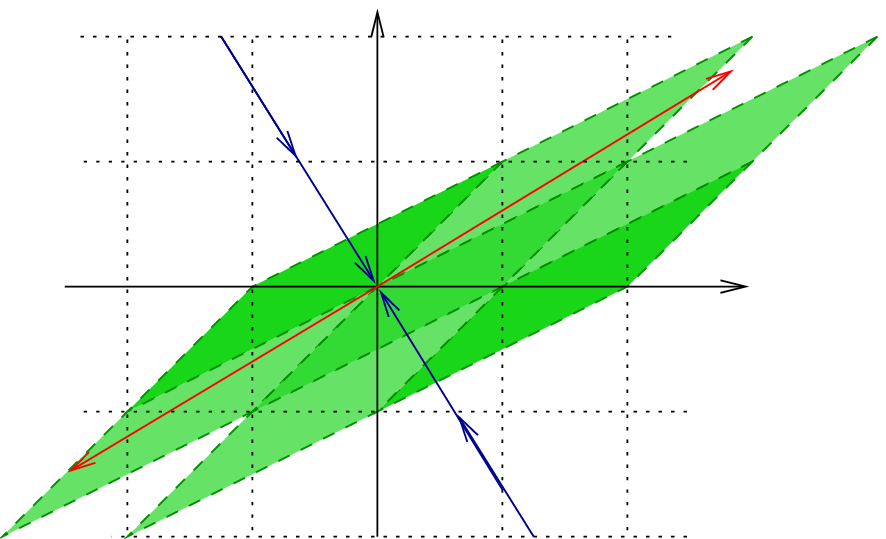}\scalebox{0.9}[0.9]{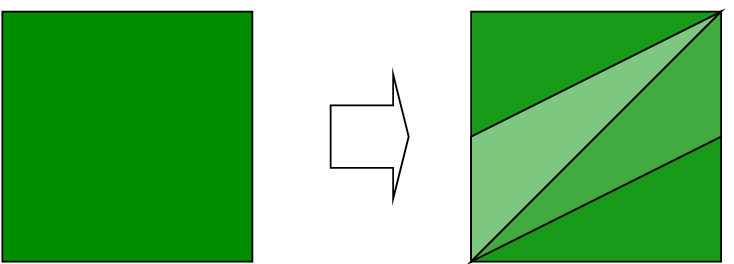}\par\end{centering}

\caption{\label{cap:Dynamics-of-M0}Dynamics of $M_{0}=\left(\protect\begin{array}{cc}
2 & 1\protect\\
1 & 1\protect\end{array}\right)\in SL\left(2,\mathbb{Z}\right)$ on $\mathbb{R}^{2}$ and on $\mathbb{T}^{2}$.}
\end{figure}

\textit{i.e.} $\det\left(M_{0}\right)=AD-BC=1$. We assume that $\gamma^{2}>\alpha\beta$
(equivalently $\mbox{Tr}\left(M_{0}\right)=A+D>2$), so that $M_{0}$
is a \textbf{hyperbolic map} with two real eigenvalues $e^{\pm\lambda_{0}}$
where $\lambda_{0}=\sqrt{\gamma^{2}-\alpha\beta}>0$ is called the
\textbf{Lyapounov exponent}. The two associated real eigenvectors
denoted by $u_{0},s_{0}\in\mathbb{R}^{2}$, correspond to an unstable
and a stable direction for the dynamics. Suppose moreover that $A,B,C,D\in\mathbb{Z}$,
i.e. $M_{0}\in SL\left(2,\mathbb{Z}\right)$. Then for any $x\in\mathbb{R}^{2},n\in\mathbb{Z}^{2}$,
\[
M_{0}\left(x+n\right)=M_{0}\left(x\right)+M_{0}\left(n\right)\equiv M_{0}\left(x\right)\,\textrm{mod}1\]
so $M_{0}$ induces a map on the torus phase space $\mathbb{T}^{2}=\mathbb{R}^{2}/\mathbb{Z}^{2}$,
see figure \ref{cap:Dynamics-of-M0}. This map is Anosov (uniformly
hyperbolic), with strong chaotic properties, such as ergodicity and
mixing, see \cite{katok_hasselblatt} p. 154.

\subsubsection{Hamiltonian perturbation}

We introduce a non linear perturbation of the previous map. Consider
a $C^{\infty}$ function on the torus $H_{1}:\mathbb{T}^{2}\rightarrow\mathbb{R}$
(i.e. $H_{1}$ is a periodic function on $\mathbb{R}^{2}$). Let $M_{1}$
be the map on $\mathbb{R}^{2}$ given by the Hamiltonian flow generated
from $H_{1}$ after time 1. We compose the two maps and define\begin{equation}
\boxed{M\defi M_{1}.M_{0}}\label{eq:map_M}\end{equation}

Which induces a map on $\mathbb{T}^{2}$ also denoted by $M$.

\paragraph{Remarks:}

\begin{enumerate}
\item The final dynamics $M^{t}:\mathbb{R}^{2}\rightarrow\mathbb{R}^{2}$,
$t\in\mathbb{Z}$, might be seen as generated by a time dependant
Hamiltonian $H\left(x,t\right)$ (periodic in time $t\in\mathbb{R}$
but discontinuous). 
\item There is a useful relation:\begin{equation}
M\left(x+n\right)=M\left(x\right)+M_{0}\left(n\right),\qquad\forall x\in\mathbb{R}^{2},n\in\mathbb{Z}^{2}.\label{eq:M(x+n)}\end{equation}

\end{enumerate}
\begin{proof}
From $M_{0}\left(x+n\right)=M_{0}\left(x\right)+M_{0}\left(n\right)$,
and $M_{1}\left(x+n\right)=M_{1}\left(x\right)+n$, one deduces $M\left(x+n\right)=M_{1}M_{0}\left(x+n\right)=M_{1}\left(M_{0}\left(x\right)+M_{0}\left(n\right)\right)=M\left(x\right)+M_{0}\left(n\right)$,
because $M_{0}\left(n\right)\in\mathbb{Z}^{2}$.
\end{proof}

\subsubsection{Hyperbolic dynamics and structural stability \label{sub:Unstable-and-stable} }

\paragraph{Structural stability theorem:}

The structural stability theorem states that if the perturbation $H_{1}$
is small enough with respect to the $C^{2}$ norm, then $M$ is an
hyperbolic map on $\mathbb{T}^{2}$ conjugated to $M_{0}$ by a Hölder
continuous map $\mathbf{H}$ (\cite{arnold-ed2},\cite{katok_hasselblatt}
p. 89):

\selectlanguage{french}
\vspace{0.cm}\begin{center}\fbox{\parbox{16cm}{

\selectlanguage{english}
\begin{thm}
\label{thm:structural_stability}There exists $\varepsilon>0$, such
that if $\left\Vert H_{1}\right\Vert _{C^{2}}<\varepsilon$, then
\begin{equation}
M=\mathbf{H}.M_{0}.\mathbf{H}^{-1}\label{eq:Q_homeom}\end{equation}
with $\mathbf{H}:\mathbb{T}^{2}\rightarrow\mathbb{T}^{2}$ Hölder
continuous, and $M$ is uniformly hyperbolic in the following sense.

Let $DM_{x}:T_{x}\mathbb{T}^{2}\rightarrow T_{M\left(x\right)}\mathbb{T}^{2}$
be the tangential map at $x\in\mathbb{T}^{2}$. For any $x\in\mathbb{T}^{2}$,
there exists a frame of tangent vectors $\left(u_{x},s_{x}\right)\in\left(T_{x}\mathbb{T}^{2}\right)$,
called \textbf{respectively unstable} and \textbf{stable directions},
(chosen such that\begin{equation}
u_{x}\wedge s_{x}=1\label{eq:ux_sx}\end{equation}
i.e. they form a symplectic basis. We can also impose that $\left\Vert u_{x}\right\Vert =\left\Vert s_{x}\right\Vert $,
in order to fix the choice) and an \textbf{expansion rate} $\lambda_{x}$
such that:\begin{equation}
DM_{x}\left(u_{x}\right)=e^{\lambda_{x}}u_{M\left(x\right)},\quad DM_{x}\left(s_{x}\right)=e^{-\lambda_{x}}s_{M\left(x\right)},\label{eq:DM}\end{equation}
and more important: $u_{x},s_{x},\lambda_{x}$ are Hölder continuous
functions of $x\in\mathbb{T}^{2}$. For any $x\in\mathbb{T}^{2}$,
\begin{equation}
0<\lambda_{min}\leq\lambda_{x}\leq\lambda_{max},\quad\mbox{with\,\,}\lambda_{min}\defi\min_{x\in\mathbb{T}^{2}}\lambda_{x},\quad\lambda_{max}\defi\max_{x\in\mathbb{T}^{2}}\lambda_{x}.\label{eq:lmin_lmax}\end{equation}

\end{thm}
\selectlanguage{french}
}}\end{center}\vspace{0.cm}

\selectlanguage{english}
In all the paper, we will suppose the perturbation to be weak enough
so that the previous theorem holds. See Figure \ref{cap:variete}.

\paragraph{Remarks}

\begin{itemize}
\item Of course, for a null perturbation $H_{1}=0$, $M=M_{0}$, then $\lambda_{x}=\lambda_{0}$
for every $x$.
\item The conjugation Eq.(\ref{eq:Q_homeom}) between $M$ and $M_{0}$
is very useful for topological purpose. For example, it implies that
the periodic trajectories of $M$ (and more generally all the symbolic
dynamics) are in one-to-one correspondence with those of $M_{0}$.
As a consequence, $M$ has also strong chaotic properties such as
topological mixing. However some more refined characteristic quantities
of the dynamical systems $M_{0}$ and $M$ are different. For example
the Lyapounov coefficients of corresponding trajectories of $M$ and
$M_{0}$ are different, and this is related to the fact that the transformation
map $\mathbf{H}$ is not $C^{1}$, and does not conserve area. To
understand the quantum (or semi-classical) dynamics of $M$, the equivalence
Eq.(\ref{eq:Q_homeom}) is not strong enough, because in semiclassical
analysis, we need symplectic conjugations (here, $\mathbf{H}$ does
not conserve area and can not be quantized).
\item In \cite{hurder-90}, it is shown that $u_{x},s_{x},\lambda_{x}$
are $C^{2-\delta}$ functions of $x$, but we will not use it. 
\item For practical purpose, we introduce $Q_{x}\in SL\left(2,\mathbb{R}\right)$,
$x\in\mathbb{R}^{2}$, the symplectic matrix which transforms the
canonical basis $\left(e_{q},e_{p}\right)$ of $\mathbb{R}^{2}$ to
the basis $\left(u_{x},s_{x}\right)$:\begin{equation}
Q_{x}=\left(\begin{array}{cc}
\left(u_{x}\right)_{q} & \left(s_{x}\right)_{q}\\
\left(u_{x}\right)_{p} & \left(s_{x}\right)_{p}\end{array}\right).\label{eq:Qx}\end{equation}
$Q_{x}$ depends continuously on $x\in\mathbb{T}^{2}$. We can write:\begin{equation}
DM_{x}=Q_{Mx}\left(\begin{array}{cc}
\exp\left(\lambda_{x}\right) & 0\\
0 & \exp\left(-\lambda_{x}\right)\end{array}\right)Q_{x}^{-1}\label{eq:DMx_with_Qx}\end{equation}

\item \textbf{Explicit expression of $u_{x}$ and $s_{x}$:} Let $u_{0}\in\mathbb{R}^{2}$
be any given vector, and denote $\left[u_{0}\right]\in P\left(\mathbb{R}^{2}\right)$
its direction in the projective space and $\left[DM_{x}\right]$ the
action of $DM_{x}$ in the projective space. Then the direction of
$u_{x}$ is given by\[
\left[u_{x}\right]=\lim_{t\rightarrow-\infty}\left[DM_{M^{-t}\left(x\right)}^{t}\right]\left[u_{0}\right]\]
which converges for almost every $u_{0}$. Similarly the stable direction
is given by\[
\left[s_{x}\right]=\lim_{t\rightarrow+\infty}\left[DM_{M^{t}\left(x\right)}^{-t}\right]\left[s_{0}\right]\]
where $s_{0}$ is a given vector.
\end{itemize}

\subsubsection{Example for numerical illustrations\label{sub:Example-for-numerical}}

In this paper we will illustrate the results by numerical calculations
in a specific example. The linear non perturbed map is: \begin{equation}
M_{0}=\left(\begin{array}{cc}
2 & 1\\
1 & 1\end{array}\right)\label{eq:example_M0}\end{equation}

with Lyapounov coefficient $\lambda_{0}$, given by $e^{\lambda_{0}}=\frac{3+\sqrt{5}}{2}=2,62\ldots$.
The perturbation $M_{1}$ is generated by the Hamiltonian\[
H_{1}\left(q,p\right)=a\cos\left(2\pi q\right),\qquad a=0.01\]
giving a hyperbolic map on $\mathbb{T}^{2}$ \begin{equation}
M\defi M_{1}M_{0}\label{eq:example_M}\end{equation}

\begin{figure}[tbph]
\begin{centering}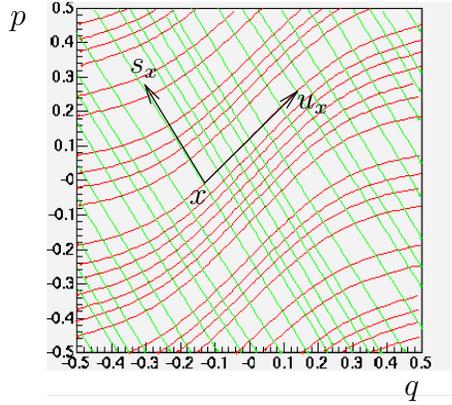\par\end{centering}

\caption{\label{cap:variete} Part of Stable and unstable manifolds of the
Anosov map $M$, Eq.(\ref{eq:example_M}), passing through the origin
$\left(0,0\right)$. At every point $x\in\mathbb{T}^{2}$, $u_{x},s_{x}$
are vectors tangent to the foliation.}
\end{figure}

Figure \ref{cap:variete} shows the torus phase space foliated by
stable and unstable manifolds.

\subsection{\label{sub:Quantum-mechanics}Quantum mechanics}

\subsubsection{Quantum mechanics on the plane}

The Hilbert space associated to the plane phase space $\mathbb{R}^{2}$
is $\mathcal{H}_{plane}\defi L^{2}\left(\mathbb{R}\right)$. The Hamiltonian
operator $\hat{H}_{0}$ is obtained by Weyl quantization of $H_{0}$,
Eq.(\ref{eq:H0}):\[
\hat{H}_{0}=Op_{Weyl}\left(H_{0}\right)=\frac{\alpha}{2}\hat{q}^{2}+\frac{\beta}{2}\hat{p}^{2}+\frac{\gamma}{2}\left(\hat{q}\hat{p}+\hat{p}\hat{q}\right),\]

Where the operators are defined by $\left(\hat{q}\varphi\right)\left(q\right)\defi q\varphi\left(q\right)$
and $\left(\hat{p}\varphi\right)\left(q\right)\defi-i\hbar\left(\frac{d\varphi}{dq}\right)\left(q\right)$,
with $\varphi\in L^{2}\left(\mathbb{R}\right)$ in suitable domains,
and where the {}``Planck constant'' $\hbar>0$ has been introduced.
The Schrödinger equation defines the evolution of $\varphi\left(t\right)\in\mathcal{H}_{plane}=L^{2}\left(\mathbb{R}\right)$:

\[
\frac{d\varphi\left(t\right)}{dt}=-\frac{i}{\hbar}\hat{H}_{0}\varphi\left(t\right),\]
and generates a unitary evolution operator between time $t=0\rightarrow1$,
written $\hat{M}_{0}$:\begin{equation}
\varphi(1)=\hat{M}_{0}\varphi\left(0\right),\quad\hat{M}_{0}=\exp\left(-\frac{i}{\hbar}\hat{H}_{0}\right).\label{eq:quantum_map_M0}\end{equation}

Similarly, we define $\hat{H}_{1}=Op_{Weyl}\left(H_{1}\right)$ to
be the Weyl quantization%
\footnote{Explicitely, if $H_{1}$ is expanded in Fourier series:\[
H_{1}\left(q,p\right)=\sum_{n=\left(n_{1},n_{2}\right)\in\mathbb{Z}^{2}}c_{n}e^{i2\pi\left(n_{1}q+n_{2}p\right)}\]
(with $c_{-n}=\overline{c_{n}}$, so that $H_{1}$ is real valued),
then \begin{equation}
\hat{H}_{1}=Op_{Weyl}\left(H_{1}\right)=\sum_{n=\left(n_{1},n_{2}\right)\in\mathbb{Z}^{2}}c_{n}e^{i2\pi\left(n_{1}\hat{q}+n_{2}\hat{p}\right)}\label{eq:H1_quantum}\end{equation}
} of $H_{1}$, and $\hat{M}_{1}=\exp\left(-\frac{i}{\hbar}\hat{H}_{1}\right)$
the evolution operator after time $1$. Finally the quantum map is\begin{equation}
\hat{M}=\hat{M}_{1}.\hat{M}_{0}:\qquad\mathcal{H}_{plane}\rightarrow\mathcal{H}_{plane}\label{eq:map_M_hat}\end{equation}
also written $\hat{M}_{plane}$ in the following, and not to be confused
with the quantum map $\hat{M}_{torus}$ for the torus phase space
introduced below.

\paragraph{Unitary translation operators:}

For $v=\left(v_{1},v_{2}\right)\in\mathbb{R}^{2}$, let \begin{equation}
T_{v}:\begin{cases}
\mathbb{R}^{2} & \rightarrow\mathbb{R}^{2}\\
x & \mapsto T_{v}\left(x\right)=x+v\end{cases}\label{eq:T_v_classique}\end{equation}
be the translation on classical phase space. This translation can
be expressed as the flow of the Hamiltonian function $f\left(q,p\right)=\left(v_{1}p-v_{2}q\right)$
after time 1. The corresponding quantum translation operator is defined
by: \begin{equation}
\hat{T}_{v}\defi\exp\left(-\frac{i}{\hbar}\left(v_{1}\hat{p}-v_{2}\hat{q}\right)\right).\label{e:quantum translation}\end{equation}
 These quantum translations satisfy the algebraic relation of the
Weyl-Heisenberg group\cite{folland-88}\cite{perelomov1}: \begin{equation}
\hat{T}_{v}\,\hat{T}_{v'}=e^{iS/\hbar}\,\hat{T}_{v+v'},\label{e:Heisenberg}\end{equation}
 with $S=\frac{1}{2}\left(v_{2}v_{1}'-v_{1}v_{2}'\right)=-\frac{1}{2}v\wedge v'$.
For any matrix $M_{0}\in{\textrm{SL}}(2,\mathbb{R})$, one trivially
has $M_{0}\left(x+v\right)=M_{0}\left(x\right)+M_{0}\left(v\right)$
which rewrites $M_{0}T_{v}=T_{M_{0}v}M_{0}$. This intertwining persists
at the quantum level: \begin{equation}
\hat{M}_{0}\hat{T}_{v}=\hat{T}_{M_{0}v}\hat{M}_{0}.\label{e: M and T}\end{equation}

We deduce a relation for the non linear quantum map $\hat{M}$ analogous
to the classical relation Eq.(\ref{eq:M(x+n)}) :\begin{equation}
\hat{M}\hat{T}_{n}=\hat{T}_{M_{0}\left(n\right)}\hat{M},\qquad\forall n\in\mathbb{Z}^{2}\label{eq:M_Tn}\end{equation}

\begin{proof}
From Eq.(\ref{eq:H1_quantum}) and Eq.(\ref{eq:T_v_classique}), we
can write $\hat{H}_{1}=\sum_{n=\left(n_{1},n_{2}\right)\in\mathbb{Z}^{2}}c_{n}\hat{T}_{-2\pi\hbar n}$.
From Eq.(\ref{e:Heisenberg}), one has $\left[\hat{T}_{-2\pi\hbar n},\hat{T}_{n'}\right]=0$.
We deduce that $\left[\hat{T}_{n'},\hat{H}_{1}\right]=0$, $\left[\hat{T}_{n'},\hat{M}_{1}\right]=0$,
for any $n'\in\mathbb{Z}^{2}$, which reflects the periodicity of
the function $H_{1}$ at the quantum level. Then using Eq.(\ref{e: M and T}),
$\hat{M}\hat{T}_{n}=\hat{M}_{1}\hat{M}_{0}\hat{T}_{n}=\hat{M}_{1}\hat{T}_{M_{0}\left(n\right)}\hat{M}_{0}=\hat{T}_{M_{0}\left(n\right)}\hat{M}$.
\end{proof}

\subsubsection{Quantum mechanics on the torus}

At the classical level, the torus phase space was obtained by introducing
periodicity on $\mathbb{R}^{2}$ with respect to the $\mathbb{Z}^{2}$
lattice, generated by translations $T_{\left(1,0\right)}$ and $T_{\left(0,1\right)}$.
The same construction can be done in quantum mechanics. The difference
is that we have to check that the quantum translation operators $\hat{T}_{\left(1,0\right)}$
and $\hat{T}_{\left(0,1\right)}$ commute before we consider their
common eigenspaces. From Eq.(\ref{e:Heisenberg}), we have $\hat{T}_{\left(1,0\right)}\hat{T}_{\left(0,1\right)}=e^{-i/\hbar}\hat{T}_{\left(0,1\right)}\hat{T}_{\left(1,0\right)}$,
so $\left[\hat{T}_{\left(1,0\right)},\hat{T}_{\left(0,1\right)}\right]=0$
if and only if $\hbar$ is such that \begin{equation}
\boxed{N\defi\frac{1}{2\pi\hbar}\in\mathbb{N}^{*}}\label{eq:N_h}\end{equation}

\textbf{\emph{We will suppose condition Eq.(\ref{eq:N_h}) from now
on}}\emph{.} We define

\[
\boxed{\mathcal{H}_{torus}\defi\left\{ \varphi\in\mathcal{S}^{'}\left(\mathbb{R}\right)\,/\,\,\hat{T}{}_{\left(1,0\right)}\varphi=\varphi,\quad\hat{T}{}_{\left(0,1\right)}\varphi=\varphi\right\} }\]

In order to have a concrete expression of a wave function $\varphi\in\mathcal{H}_{torus}$%
\footnote{$\mathcal{H}_{torus}$ is the well known space for Finite Fourier
Transform.%
}, remark that using $\hbar$-Fourier-Transform defined by $\tilde{\varphi}\left(p\right)\defi\frac{1}{\sqrt{2\pi\hbar}}\int dq\,\varphi\left(q\right)e^{-ipq/\hbar}$,
then $\varphi\in\mathcal{H}_{torus}$ is characterized by the periodicity
conditions $\tilde{\varphi}\left(p+1\right)=\tilde{\varphi}\left(p\right)$
and $\varphi\left(q+1\right)=\varphi\left(q\right)$. The periodicity
of $\tilde{\varphi}$ implies that $\varphi\left(q\right)=\sum_{n\in\mathbb{Z}}a_{n}\delta\left(q-\frac{n}{N}\right)$,
with $a_{n}\in\mathbb{C}$. The periodicity of $\varphi\left(q\right)$
implies that $a_{n+N}=a_{n}$ for any $n$. So $\varphi\in\mathcal{H}_{torus}$
is specified by $\left(a_{n}\right)_{n=1\rightarrow N}\in\mathbb{C}^{N}$.
We deduce that:

\[
\boxed{\textrm{dim}\mathcal{H}_{torus}=N=\frac{1}{2\pi\hbar}.}\]

For simplicity we also assume, that $N$ \textbf{is even}, so that,
using Eq.(\ref{e:Heisenberg}), $\hat{T}_{n=\left(n_{1},n_{2}\right)}=\hat{T}_{\left(1,0\right)}^{n_{1}}\hat{T}_{\left(0,1\right)}^{n_{2}}$
for any $n\in\mathbb{Z}^{2}$. We define a onto projector $\hat{\mathcal{P}}:\mathcal{H}_{plane}\rightarrow\mathcal{H}_{torus}$
(with a dense domain which contains $\mathcal{S}\left(\mathbb{R}\right)$),
which makes a quantum state periodic with respect to the lattice $\mathbb{Z}^{2}$
in phase space: \begin{equation}
\hat{\mathcal{P}}\defi\sum_{(n_{1},n_{2})\in\mathbb{Z}^{2}}\:\hat{T}_{\left(1,0\right)}^{n_{1}}\:\hat{T}_{\left(0,1\right)}^{n_{2}}=\sum_{n\in\mathbb{Z}^{2}}\:\hat{T}_{n}.\label{e:Projector}\end{equation}
 From Eq.(\ref{e: M and T}), we deduce:

\begin{equation}
\hat{M}\hat{\mathcal{P}}=\hat{\mathcal{P}}\hat{M}\label{e:dual map}\end{equation}

In other words we have a commutative diagram:\begin{eqnarray*}
\mathcal{H}_{plane} & \underrightarrow{\,\,\hat{M}\,\,\,} & \mathcal{H}_{plane}\\
\downarrow\hat{\mathcal{P}} &  & \downarrow\hat{\mathcal{P}}\\
\mathcal{H}_{torus} & \underrightarrow{\,\,\hat{M}\,\,\,} & \mathcal{H}_{torus}\end{eqnarray*}

Which means that $\hat{M}$ induces an endomorphism:\begin{equation}
\boxed{\hat{M}_{torus}=\hat{M}_{1}.\hat{M}_{0}:\,\,\mathcal{H}_{torus}\rightarrow\mathcal{H}_{torus}}\label{eq:M_torus}\end{equation}

When no confusing is possible, we will write $\hat{M}$ for this operator
$\hat{M}_{torus}$.

\subsubsection{Standard coherent states\label{sub:Standard-coherent-states}}

We will use in many instances some particular quantum states, the
standard coherent states \cite{perelomov1}\cite{ec1}, which are
semi-classically localized on phase space. For $x=\left(q,p\right)\in\mathbb{R}^{2}$,
the coherent state $\varphi_{x}\in\mathcal{H}_{plane}$ is $\varphi_{x}\left(q'\right)\defi\frac{1}{\left(\pi\hbar\right)^{1/4}}\exp\left(-i\frac{pq}{2\hbar}\right)\exp\left(i\frac{pq'}{\hbar}\right)\exp\left(-\frac{\left(q'-q\right)^{2}}{2\left(\sqrt{\hbar}\right)^{2}}\right)$.
Semi-classical localization of $\varphi_{x}$ at $x\in\mathbb{R}^{2}$
in phase space comes from the fact that its modulus is a Gaussian
{}``localized around'' $q$ with width $\Delta q\simeq\sqrt{\hbar}$
(which vanishes for $\hbar\rightarrow0$). Its $\hbar$-Fourier Transform
is $\tilde{\varphi}_{x}\left(p'\right)=\frac{1}{\left(\pi\hbar\right)^{1/4}}\exp\left(-i\frac{pq}{2\hbar}\right)\exp\left(-i\frac{qp'}{\hbar}\right)\exp\left(-\frac{\left(p'-p\right)^{2}}{2\left(\sqrt{\hbar}\right)^{2}}\right)$,
similarly localized around $p$. More algebraically, $\varphi_{0}\in\mathcal{H}_{plane}$
(with $x=0$), is the ground state of the Harmonic Oscillator and
is defined by $a\varphi_{0}=0$, with $a=\left(\hat{q}+i\hat{p}\right)/\sqrt{2\hbar}$.
The coherent state $\varphi_{x}$ is then obtained by translation
\begin{equation}
\varphi_{x}\defi\hat{T}_{x}\varphi_{0},\quad x=(q,p)\in\mathbb{R}^{2}.\label{e:|z>}\end{equation}
For short, we will also write $|x\rangle\defi\varphi_{x}$ for a coherent
state. The \textbf{Husimi distribution} of a quantum state $\psi\in\mathcal{H}_{plane}$
is the positive measure on phase space:\[
\mbox{Hus}_{\psi}\left(x\right)\defi\frac{1}{2\pi\hbar}\left|\langle x|\psi\rangle\right|^{2}\]

The closure relation is (\cite{perelomov1} p. 15)\begin{equation}
\hat{\mbox{Id}}_{/\mathcal{H}_{plane}}=\int_{\mathbb{R}^{2}}\frac{dx}{2\pi\hbar}|x\rangle\langle x|\label{eq:Closure_plane}\end{equation}

A \textbf{coherent state on the torus} is defined by periodicity,
using Eq. (\ref{e:Projector}):\[
|x\rangle_{torus}\defi\hat{\mathcal{P}}|x\rangle\in\mathcal{H}_{torus}\]
They provide the closure relation:\begin{equation}
\hat{\mbox{Id}}_{/\mathcal{H}_{torus}}=\int_{\left[0,1\right]^{2}}\frac{dx}{2\pi\hbar}|x\rangle_{torus}\langle x|\label{eq:Closure_torus}\end{equation}

\section{The map as a sum over periodic points\label{sec:The-map-as-PO}}

In the first part of this section we characterize and count the number
of periodic points of the map $M$ with a given period $t\in\mathbb{Z}$.
These points are shown to play an important role in the second part,
where we will express $\hat{M}_{torus}^{t}$, defined in Eq.(\ref{eq:M_torus}),
and its trace in terms of them. Some parts of this section can be
found in \cite{keating-91a} or \cite{pollicott-98}.

\subsection{Periodic points of $M$}

Consider a discrete time $t\in\mathbb{Z}\mathcal{n}\left\{ 0\right\} $.
A point $x\in\mathbb{R}^{2}$ gives a periodic point $\left[x\right]\in\mathbb{T}^{2}$
of $M$ with period $t$ if \begin{eqnarray*}
M^{t}\left[x\right] & = & \left[x\right]\\
\Leftrightarrow &  & \exists n\in\mathbb{Z}^{2}\,/\, M^{t}x=x+n\\
\Leftrightarrow &  & \exists n\in\mathbb{Z}^{2}\,/\,\mathcal{T}_{t}\left(x\right)=n\end{eqnarray*}
with the map\[
\mathcal{T}_{t}\defi\left\{ \begin{array}{ccc}
\mathbb{R}^{2} & \rightarrow & \mathbb{R}^{2}\\
x & \rightarrow & \mathcal{T}_{t}\left(x\right)=M^{t}\left(x\right)-x\end{array}\right.\]

\paragraph{Periodicity of $\mathcal{T}_{t}$:}

From Eq.(\ref{eq:M(x+n)}), we get\[
\mathcal{T}_{t}\left(x+m\right)=\mathcal{T}_{t}\left(x\right)+\mathcal{T}_{0,(t)}\left(m\right),\qquad\forall x\in\mathbb{R}^{2},\,\,\forall m\in\mathbb{Z}^{2}\]
Where $\mathcal{T}_{0,(t)}\left(x\right)\defi M_{0}^{t}\left(x\right)-x$.
In particular $\mathcal{T}_{0,(t)}\left(m\right)\in\mathbb{Z}^{2}$
for any $m\in\mathbb{Z}^{2}$. We introduce the sub-lattice of $\mathbb{Z}^{2}$\noun{:}
\begin{equation}
\Lambda_{t}\defi\mathcal{T}_{0,\left(t\right)}\left(\mathbb{Z}^{2}\right)\subset\mathbb{Z}^{2}\label{eq:lattice_Gamma}\end{equation}
generated by $e_{1}=\mathcal{T}_{0,(t)}\left(1,0\right)\in\mathbb{Z}^{2}$
and $e_{2}=\mathcal{T}_{0,(t)}\left(0,1\right)\in\mathbb{Z}^{2}$.
We denote by $\mathcal{C}_{t}\subset\mathbb{Z}^{2}$ the elements
which belong to the origin cell of $\Lambda_{t}$:\[
\mathcal{C}_{t}\defi\left\{ n\in\mathbb{Z}^{2}\quad/\quad\mathcal{T}_{0,(t)}^{-1}\left(n\right)\in[0,1[^{2}\right\} \]
 Of course $\mathcal{C}_{t}\equiv\mathbb{Z}^{2}/\Lambda_{t}$.

\vspace{0.cm}\begin{center}\fbox{\parbox{16cm}{

\begin{prop}
\label{pro:The-periodic-points}For a small enough perturbation, $\mathcal{T}_{t}$
is a diffeomorphism, for every $t$. Periodic points with period $t$
can be labeled by $n\in\mathbb{Z}^{2}$:\begin{equation}
x_{n,t}\defi\mathcal{T}_{t}^{-1}\left(n\right),\qquad n\in\mathbb{Z}^{2}\label{eq:periodic_fixed_point_xn}\end{equation}
Different values $n,n'$ may give the same point $\left[x_{n,t}\right]=\left[x_{n',t}\right]$.
The periodic points $\left[x_{n,t}\right]\in\mathbb{T}^{2}$ on the
torus are in one to one correspondence with $n\in\mathcal{C}_{t}$.
The number of periodic points with period $t$ is\begin{eqnarray}
\mathcal{N}_{t} & \defi\sharp\left(\mathcal{C}_{t}\right) & =\left|\det\left(\mathcal{T}_{0,(t)}\right)\right|=\left|\det\left(M_{0}^{t}-I\right)\right|\label{eq:Nt_periodic_points}\\
 & = & \mbox{e}^{\lambda_{0}t}-2+\mbox{e}^{-\lambda_{0}t}\qquad\left(\simeq e^{\lambda_{0}t}\,\textrm{for}\,\,\, t\gg1\right)\nonumber \end{eqnarray}

\end{prop}
}}\end{center}\vspace{0.cm}

\begin{proof}
Using Eq.(\ref{eq:DMx_with_Qx}), the matrix of $\left(D\mathcal{T}_{t}\right)_{x}$is
\[
\left(D\mathcal{T}_{t}\right)_{x}=\left(DM^{t}\right)_{x}-Id=Q_{M^{t}x}\left(\begin{array}{cc}
\exp\left(\Lambda_{x,t}\right) & 0\\
0 & \exp\left(-\Lambda_{x,t}\right)\end{array}\right)Q_{x}^{-1}-Id\]
with $\Lambda_{x,t}=\sum_{t'=0}^{t-1}\lambda_{M^{t'}x}$ , so $\lambda_{min}t\leq\Lambda_{x,t}\leq\lambda_{max}t$.
We have supposed that $\lambda_{min}>0$. Then\[
\det\left(\left(D\mathcal{T}_{t}\right)_{x}\right)=\det\left(\left(\begin{array}{cc}
\exp\left(\Lambda_{x,t}\right) & 0\\
0 & \exp\left(-\Lambda_{x,t}\right)\end{array}\right)-Q_{M^{t}x}^{-1}Q_{x}\right)\]
 The matrix $Q_{x'}^{-1}Q_{x}$ goes to identity (uniformly in $x,x'\in\mathbb{T}^{2}$)
when the perturbation vanishes. Therefore, we can write $Q_{x'}^{-1}Q_{x}=Id+\varepsilon A_{x,x'}$,
where $A_{x,x'}$ has matrix elements bounded by $1$ in absolute
value, and $\varepsilon$ goes to zero when the perturbation vanishes.
One computes \[
\det\left(\left(D\mathcal{T}_{t}\right)_{x}\right)=2\left(1-\cosh\Lambda_{x,t}\right)+\varepsilon A_{11}\left(e^{-\Lambda_{x,t}}-1\right)+\varepsilon A_{22}\left(e^{\Lambda_{x,t}}-1\right)+\varepsilon^{2}\textrm{det}\left(A\right)\]
and deduces \[
\left|\det\left(\left(D\mathcal{T}_{t}\right)_{x}\right)\right|>2\left(1-\varepsilon\right)\left(\cosh\left(\lambda_{min}t\right)-1\right)-3\varepsilon>0\]
for small enough $\varepsilon$. We deduce that $\mathcal{T}_{t}$
is a diffeomorphism on $\mathbb{R}^{2}$.
\end{proof}

\paragraph{Remarks:}

\begin{figure}[tbph]
\begin{centering}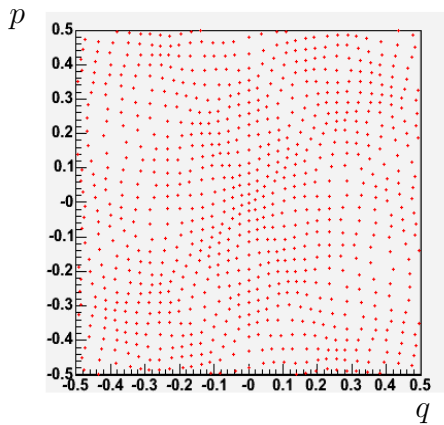\par\end{centering}

\caption{\label{cap:Periodic-points-of-M}Periodic points of map $M$, Eq.
(\ref{eq:example_M}), with period $t=7$. There are $\mathcal{N}_{t}=841$
periodic points.}
\end{figure}

\begin{itemize}
\item Note that two periodic points $x_{n',t},\, x_{n,t}$ may belong to
the same periodic orbit of $M_{torus}$. Explicitely:\[
x_{n',t}=M\left(x_{n,t}\right)\Leftrightarrow n'=M_{0}\left(n\right)\]
so periodic orbits of $M$ on $\mathbb{T}^{2}$ with period $t$,
are in one to one correspondence with periodic orbits of $M_{0}$
acting on the finite set $\mathcal{C}_{t}\equiv\mathbb{Z}^{2}/\Lambda_{t}$.
\item Here is the number of periodic points with period $t$, for the example
Eq. (\ref{eq:example_M0}), see figure \ref{cap:Periodic-points-of-M}:\\
\begin{tabular}{|c|c|c|c|c|c|c|c|c|c|c|c|c|}
\hline 
$t$&
$1$&
$2$&
$3$&
$4$&
$5$&
$6$&
$7$&
$8$&
$9$&
$10$&
$11$&
$\ldots$\tabularnewline
\hline
$\mathcal{N}_{t}$&
$1$&
$5$&
$16$&
$45$&
$121$&
$320$&
$841$&
$2205$&
$5776$&
$15125$&
$39601$&
$\ldots$\tabularnewline
\hline
\end{tabular}\\

\end{itemize}

\subsection{Expression of the quantum map $\hat{M}_{torus}^{t}$, using periodic
points\label{sub:Expression-of-M-using-periodic-points}}

For a given $t\in\mathbb{Z}$, we consider \[
\left(\hat{\mathcal{P}}\hat{M}_{plane}^{t}\right):\mathcal{H}_{plane}\rightarrow\mathcal{H}_{torus}\]
defined on a suitable dense domain (which contains the Schwartz space
$\mathcal{S}\left(\mathbb{R}\right)$), where $\hat{\mathcal{P}}$
is defined in Eq.(\ref{e:Projector}), and $\hat{M}_{plane}$ is defined
in Eq.(\ref{eq:map_M_hat}) . This operator $\left(\hat{\mathcal{P}}\hat{M}_{plane}^{t}\right)$
is important to look at, because if one wants any matrix element of
the quantum map $\langle\tilde{\psi}_{2}|\hat{M}_{torus}^{t}|\tilde{\psi}_{1}\rangle$,
with $|\tilde{\psi}_{1}\rangle,|\tilde{\psi}_{2}\rangle\in\mathcal{H}_{torus}$,
then one just has to consider {}``lifted states on the plane'' $|\psi_{1}\rangle$,
$|\psi_{2}\rangle\in\mathcal{H}_{plane}$, such that $|\tilde{\psi}_{i}\rangle=\hat{\mathcal{P}}|\psi_{i}\rangle$,
$i=1,2$, and then \begin{equation}
\langle\tilde{\psi}_{2}|\hat{M}_{torus}^{t}|\tilde{\psi}_{1}\rangle=\langle\psi_{2}|\hat{\mathcal{P}}\hat{M}_{plane}^{t}|\psi_{1}\rangle\label{eq:Mtorus_Lifted}\end{equation}

We will now write $\hat{M}$ instead of $\hat{M}_{plane}$. One writes:\begin{equation}
\boxed{\left(\hat{\mathcal{P}}\hat{M}^{t}\right)=\sum_{n\in\mathbb{Z}^{2}}\hat{T}_{-n}\hat{M}^{t}=\sum_{n\in\mathbb{Z}^{2}}\hat{M}_{n,t}}\label{eq:MP}\end{equation}

with\begin{equation}
\boxed{\hat{M}_{n,t}\defi\hat{T}_{-n}\hat{M}^{t}\qquad:\mathcal{H}_{plane}\rightarrow\mathcal{H}_{plane}}\label{eq:Mnt}\end{equation}

where $n\in\mathbb{Z}^{2}$, and $t\in\mathbb{Z}$. The corresponding
classical map is of course\begin{equation}
M_{n,t}:\left\{ \begin{array}{ccc}
\mathbb{R}^{2} & \rightarrow & \mathbb{R}^{2}\\
x & \rightarrow & M_{n,t}\left(x\right)=M^{t}\left(x\right)-n\end{array}\right.\label{eq:classical_M_nt}\end{equation}
The map $M_{n,t}$ will be used many times in this paper. It is an
hyperbolic map (translation of $M^{t}$), whose unique fixed point
is the periodic point $x_{n,t}$, given by Eq.(\ref{eq:periodic_fixed_point_xn}),
because: $M_{n,t}\left(x_{n,t}\right)=M^{t}\left(x_{n,t}\right)-n=x_{n,t}$.

\subsubsection{Periodicity of the decomposition}

We decompose now the sum over $n\in\mathbb{Z}^{2}$ in Eq. (\ref{eq:MP}),
with respect to the lattice $\Lambda_{t}$ defined by Eq. (\ref{eq:lattice_Gamma}).
First, any $n'\in\mathbb{Z}^{2}$ can be decomposed in the unique
way:\[
n'=n+\mathcal{T}_{0}\left(m\right),\qquad n\in\mathcal{C}_{t},\quad m\in\mathbb{Z}^{2}\]

Observe that for $n,m\in\mathbb{Z}^{2}$ (we use Eq.(\ref{eq:M_Tn})),\begin{eqnarray*}
\hat{M}_{n+\mathcal{T}_{0}\left(m\right),t} & = & \hat{T}_{-n-\mathcal{T}_{0}\left(m\right)}\hat{M}^{t}=\hat{T}_{-n-M_{0}\left(m\right)+m}\hat{M}^{t}\\
 & = & e^{i\mathcal{F}\left(n,m\right)/\hbar}\hat{T}_{m}\hat{T}_{-n}\hat{T}_{-M_{0}\left(m\right)}\hat{M}^{t}=e^{i\mathcal{F}\left(n,m\right)/\hbar}\hat{T}_{m}\hat{T}_{-n}\hat{M}^{t}\,\hat{T}_{-m}\\
 & = & e^{i\mathcal{F}\left(n,m\right)/\hbar}\hat{T}_{m}\hat{M}_{n,t}\hat{T}_{-m}\end{eqnarray*}
 The phase $\mathcal{F}$ comes from Eq.(\ref{e:Heisenberg}) and
is given by:\[
\mathcal{F}\left(n,m\right)=\frac{1}{2}\left(n\wedge M_{0}\left(m\right)-m\wedge\left(n+M_{0}\left(m\right)\right)\right)\]
But $2\mathcal{F}$ is an integer so $e^{i\mathcal{F}\left(n,m\right)/\hbar}=e^{i2\pi N\mathcal{F}\left(n,m\right)}=+1$
because we have supposed $N$ even. Therefore\begin{equation}
\boxed{\left(\hat{\mathcal{P}}\hat{M}^{t}\right)=\sum_{n\in\mathbb{Z}^{2}}\hat{M}_{n,t}=\sum_{m\in\mathbb{Z}^{2}}\hat{T}_{m}\hat{M}_{\mathcal{C}_{t}}\hat{T}_{-m}}\label{eq:Mt_tore}\end{equation}

with\[
\boxed{\hat{M}_{\mathcal{C}_{t}}\defi\sum_{n\in\mathcal{C}_{t}}\hat{M}_{n,t}}\]
From Proposition \ref{pro:The-periodic-points}, this last expression
of $\hat{M}_{\mathcal{C}_{t}}$ is a finite sum over periodic points,
with $\mathcal{N}_{t}$ terms.

\subsubsection{Trace of $\hat{M}_{torus}^{t}$}

Remark that $\textrm{Tr}\left(\hat{M}_{torus}^{t}\right)$ is well
defined because $\hat{M}_{torus}^{t}$ acts in $\mathcal{H}_{torus}$
which is a finite dimensional space. On the opposite $\hat{M}_{n,t}=\hat{T}_{-n}\hat{M}^{t}$
is a unitary operator in $\mathcal{H}_{plane}=L^{2}\left(\mathbb{R}\right)$
and is not of Trace class. Nevertheless we will define a linear functional
which can be thought as a trace.

\vspace{0.cm}\begin{center}\fbox{\parbox{16cm}{

\begin{prop}
\label{pro:Trace_Mt}For any $t\in\mathbb{Z}$, and any $n\in\mathbb{Z}^{2}$,
the following integral is absolutely convergent:\begin{equation}
\mbox{T}\left(\hat{M}_{n,t}\right)\defi\int_{\mathbb{R}^{2}}\frac{dx}{2\pi\hbar}\langle x|\hat{M}_{n,t}|x\rangle\label{eq:Integrale}\end{equation}
where $|x\rangle$ is a standard coherent state at $x\in\mathbb{R}^{2}$
defined in Eq.(\ref{e:|z>}). We have the relation\begin{equation}
\mbox{Tr}\left(\hat{M}_{torus}^{t}\right)=\sum_{n\in\mathcal{C}_{t}}\mbox{T}\left(\hat{M}_{n,t}\right)\label{eq:Trace_Mt}\end{equation}

\end{prop}
}}\end{center}\vspace{0.cm}

Eq.(\ref{eq:Trace_Mt}) is an exact formula (not semi-classical).
It is a sum over \emph{$\mathcal{N}_{t}$} terms. This formula does
not use the assumption that $M$ is hyperbolic.

\begin{proof}
A coherent state $|x\rangle$ belongs to $\mathcal{S}\left(\mathbb{R}\right)$.
Using closure relations Eq.(\ref{eq:Closure_plane}) and Eq.(\ref{eq:Closure_torus}),
together with Eq.(\ref{eq:Mtorus_Lifted}) and Eq.(\ref{eq:Mt_tore}),
we compute:

\begin{align*}
\textrm{Tr}\left(\hat{M}_{torus}^{t}\right) & =\int_{\left[0,1\right]^{2}}\frac{dx}{2\pi\hbar}\langle x|\hat{M}_{torus}^{t}|x\rangle_{torus}=\int_{\left[0,1\right]^{2}}\frac{dx}{2\pi\hbar}\langle x|\hat{\mathcal{P}}\hat{M}_{plane}^{t}|x\rangle\\
 & =\int_{\left[0,1\right]^{2}}\frac{dx}{2\pi\hbar}\sum_{n\in\mathcal{C}_{t}}\sum_{m\in\mathbb{Z}^{2}}\langle x|\hat{T}_{m}\hat{M}_{n,t}\hat{T}_{-m}|x\rangle\\
 & =\sum_{n\in\mathcal{C}_{t}}\int_{\left[0,1\right]^{2}}\sum_{m\in\mathbb{Z}^{2}}\frac{dx}{2\pi\hbar}\langle x-m|\hat{M}_{n,t}|x-m\rangle\end{align*}

The sums are absolutely convergent. In particular: \[
\int_{\left[0,1\right]^{2}}\sum_{m\in\mathbb{Z}^{2}}\frac{dx}{2\pi\hbar}\langle x-m|\hat{M}_{n,t}|x-m\rangle=\int_{\mathbb{R}^{2}}\frac{dx}{2\pi\hbar}\langle x|\hat{M}_{n,t}|x\rangle=\mbox{T}\left(\hat{M}_{n,t}\right),\]
and we obtain $\textrm{Tr}\left(\hat{M}_{torus}^{t}\right)=\sum_{n\in\mathcal{C}_{t}}\mbox{T}\left(\hat{M}_{n,t}\right)$.
\end{proof}

\section{\label{sec:Semi-classical-description-of}Semi-classical description
of the dynamics}

We do not have yet considered the semi-classical limit $\hbar\rightarrow0$.
In this section, we give the essential Theorem (Theorem \ref{pro:voisinage_trajectoire}
below) which shows that in the semi-classical limit, long time quantum
dynamics can be expressed in terms of individual classical trajectories.

\subsection{Neighborhood of a point in phase space and localized states}

We first introduce the notion of a semi-classical neighborhood of
a point in phase space. Let $x\in\mathbb{R}^{2}$ be a point in phase
space, and $0<\alpha<1/2$. Let \begin{equation}
D_{x,\alpha}\defi\left\{ y\in\mathbb{R}^{2}\,\,\slash\left|y-x\right|<\hbar^{1/2-\alpha}\right\} \label{eq:set_D_alpha}\end{equation}
 be the disk of center $x$ and radius $\hbar^{1/2-\alpha}$, which
shrinks to zero as $\hbar\rightarrow0$. We define the \textbf{truncation
operator}: \begin{equation}
\boxed{\hat{P}_{x,\alpha}\defi\int_{y\in D_{x,\alpha}}\frac{dy}{2\pi\hbar}|y\rangle\langle y|=Op_{AW}\left(\chi_{D_{x,\alpha}}\right)}\label{eq:def_PD}\end{equation}
 being the Anti-Wick quantization of the characteristic function of
the disk $D_{x,\alpha}$ \cite{perelomov1}. We will often drop the
index $\alpha$, and write $\hat{P}_{x}\defi\hat{P}_{x,\alpha}$.
In the case $x=0$, we will drop the index $x$, and write: $\hat{P}_{\alpha}\defi\hat{P}_{x=0,\alpha}$.
Notice that in comparison with Eq.(\ref{eq:Closure_plane}), the integral
is restricted to the disk $D_{x,\alpha}$. The meaning of the operator
$\hat{P}_{x}$ is that it truncates a quantum states in the vicinity
of the point $x$. A quantum state is said to be localized at point
$x$, if this truncation has no effect on it in the semi-classical
limit. More precisely:

\begin{defn}
A sequence of normalized quantum state $\psi_{\hbar}\in\mathcal{H}_{plane}$
(sequence which depends on $\hbar\rightarrow0$) is said to be \textbf{$\alpha-$localized
at point} $x\in\mathbb{R}^{2}$, if\begin{equation}
\left\Vert \psi-\hat{P}_{x,\alpha}\psi\right\Vert _{L^{2}}=\mathcal{O}\left(\hbar^{\infty}\right)\label{eq:Def_etat_localise}\end{equation}

\end{defn}
Examples: a coherent state $|x\rangle$ is $\alpha-$localized at
$x$, for any $0<\alpha<1/2$. For any \emph{fixed} $n\in\mathbb{N}$,
let $|n\rangle$ be the eigenstate of the harmonic oscillator: $\frac{1}{2}\left(\hat{p}^{2}+\hat{q}^{2}\right)|n\rangle=\hbar\left(n+\frac{1}{2}\right)|n\rangle$,
and let $|x,n\rangle=\hat{T}_{x}|n\rangle$. Then $|x,n\rangle$ is
$\alpha-$localized at $x$, for any $0<\alpha<1/2$.

\subsection{Semi-classical evolution in a neighborhood of a classical trajectory}

Let $x\in\mathbb{R}^{2}$, and consider the classical trajectory $x\left(t\right)=M^{t}x\in\mathbb{R}^{2}$,
for $t\in\mathbb{N}$. The following Theorem will be useful in order
to compute matrix elements of the quantum evolution operator, of the
form $\langle\psi_{2}|\hat{M}^{t}|\psi_{1}\rangle$, where $\psi_{1}$
is localized at $x\left(0\right)$, and $\psi_{2}$ is localized at
$x\left(t\right)$. From Eq.(\ref{eq:Def_etat_localise}), $\langle\psi_{2}|\hat{M}^{t}|\psi_{1}\rangle=\langle\psi_{2}|\hat{P}_{x\left(t\right)}\hat{M}^{t}\hat{P}_{x\left(0\right)}|\psi_{1}\rangle+\mathcal{O}\left(\hbar^{\infty}\right)$,
so the computation involves the operator $\hat{P}_{x\left(t\right)}\hat{M}^{t}\hat{P}_{x\left(0\right)}$.

\vspace{0.cm}\begin{center}\fbox{\parbox{16cm}{

\begin{thm}
\label{pro:voisinage_trajectoire}For any $C>0$, any $t$, such that
$1\leq t<C\log\left(1/\hbar\right)$, any $0<\alpha<1/2$, any $K>0$,
there exists $C_{K}>0$, such that for any $x=x\left(0\right)\in\mathbb{R}^{2}$,
then in $L^{2}$ operator norm:\begin{equation}
\left\Vert \hat{P}_{x\left(t\right)}\hat{M}^{t}\hat{P}_{x\left(0\right)}-\hat{P}_{x\left(t\right)}\hat{M}\hat{P}_{x\left(t-1\right)}\hat{M}\hat{P}_{x\left(t-2\right)}\ldots\hat{M}\hat{P}_{x\left(0\right)}\right\Vert _{L^{2}}\leq C_{K}\hbar^{K}\label{eq:voisinage_trajectoire}\end{equation}

\end{thm}
}}\end{center}\vspace{0.cm}

The proof is given in appendix \ref{sec:Proof-of-Theorem_1} page
\pageref{sec:Proof-of-Theorem_1}, but we give the main idea below.

\paragraph{Remarks:}

\begin{itemize}
\item To shorten we write that the error is $\mathcal{O}\left(\hbar^{\infty}\right)$
uniformly with respect to $x\in\mathbb{R}^{2}$.
\item This relation means that in order to compute $\hat{P}_{x\left(t\right)}\hat{M}^{t}\hat{P}_{x\left(0\right)}$,
we can introduce truncation operators all along the trajectory, without
changing the result significantly in the semi-classical limit. In
other words the operator $\hat{P}_{x\left(t\right)}\hat{M}^{t}\hat{P}_{x\left(0\right)}$
depends only on the dynamics in the vicinity of the trajectory $x\left(0\right)\rightarrow x\left(t\right)$.
This result will allow us to use normal forms in the next Section,
which give a nice description of the dynamics in the vicinity of any
trajectory. 
\item \textbf{Idea of the proof:} Let us explain here the main idea of the
proof but at the level of classical dynamics. The proof follows the
same idea. From definition Eq.(\ref{eq:def_PD}), the operator $\hat{P}_{x\left(t\right)}=\int_{x\in D_{t}}\frac{dx}{2\pi\hbar}|x\rangle\langle x|$
truncates quantum states on the disk $D_{t}$ of small radius $\hbar^{1/2-\alpha}$,
and center $x\left(t\right)$. The transcription of Eq.(\ref{eq:voisinage_trajectoire})
in classical dynamics is \begin{equation}
D_{t}\cap M^{t}\left(D_{0}\right)=D_{t}\cap M\left(D_{t-1}\ldots M\left(D_{1}\cap M\left(D_{0}\right)\right)\right)\label{eq:Inter2}\end{equation}
To show this, let $G_{1}=M\left(D_{0}\right)\setminus D_{1}$. Let
$F_{1}$ such that $\mathbb{R}^{2}=D_{1}\cup F_{1}\cup G_{1}$ is
a disjoint union. So $F_{1}\cap M\left(D_{0}\right)=\emptyset$. Then
\begin{align}
D_{t}\cap M^{t}\left(D_{0}\right) & =D_{t}\cap M^{t-1}\left(\mathbb{R}^{2}\cap M\left(D_{0}\right)\right)\label{eq:Inter1}\\
 & =D_{t}\cap M^{t-1}\left(\left(D_{1}\cup F_{1}\cup G_{1}\right)\cap M\left(D_{0}\right)\right)\nonumber \\
 & =\left(D_{t}\cap M^{t-1}\left(D_{1}\cap M\left(D_{0}\right)\right)\right)\cup\left(D_{t}\cap M^{t-1}\left(G_{1}\right)\right)\nonumber \end{align}
Now observe that $G_{1}$ is the set of points coming from $D_{0}$
but who leave the set $D_{1}$ in the unstable direction. Due to uniform
hyperbolicity and the fact that we consider the dynamics on the cover
$\mathbb{R}^{2}$, the set $G_{1}$ goes away from the trajectory
$x\left(t\right)$ in the unstable direction. This gives: $D_{t}\cap M^{t-1}\left(G_{1}\right)=\emptyset$.
Then Eq.(\ref{eq:Inter1}) simplifies to $D_{t}\cap M^{t}\left(D_{0}\right)=D_{t}\cap M^{t-1}\left(D_{1}\cap M\left(D_{0}\right)\right)$.
Repeating this argument, we get Eq. (\ref{eq:Inter2}). (For illustration,
see Figure \ref{cap:schema_D_G_F} page \pageref{cap:schema_D_G_F}).
\item From the idea of the proof given above, it is clear that Eq.(\ref{eq:voisinage_trajectoire})
holds because it concerns the phase space cover $\mathbb{R}^{2}$.
Points who leave the trajectory in the unstable direction never come
back. At the semi-classical level, there is no interference effects.
The interference effects come when passing to the torus which is compact,
and are due to the sum Eq.(\ref{eq:MP}).
\end{itemize}

\paragraph{Consequence for the trace}

There is a consequence of Theorem \ref{pro:voisinage_trajectoire},
which shows that the integral Eq.(\ref{eq:Integrale}) up to a negligible
error, can be restricted to a neighborhood of the fixed point $x_{n,t}$
of the map $M_{n,t}$ , Eq.(\ref{eq:classical_M_nt}). This Lemma
will be useful in order to obtain the semiclassical Trace formula.

\begin{lem}
\label{lem:integrale_localisee_trace}For any $C>0$, any $t$, such
that $1\leq t<C\log\left(1/\hbar\right)$, any $0<\alpha<1/2$:\begin{align}
T\left(\hat{M}_{n,t}\right) & =\int_{x\in D_{x_{n,t}}}\frac{dx}{2\pi\hbar}\langle x|\hat{M}_{n,t}|x\rangle+\mathcal{O}\left(\hbar^{\infty}\right)\label{eq:lem_1}\\
 & =\mbox{Tr}\left(\hat{M}_{n,t}\hat{P}_{x_{n,t}}\right)+\mathcal{O}\left(\hbar^{\infty}\right)=\mbox{Tr}\left(\hat{P}_{x_{n,t}}\hat{M}_{n,t}\hat{P}_{x_{n,t}}\right)+\mathcal{O}\left(\hbar^{\infty}\right)\nonumber \end{align}
The error is uniform with respect to the point $x_{n,t}$ (i.e. with
respect to $t$ and $n\in\mathbb{Z}^{2}$).
\end{lem}
Note that $\hat{P}_{x}$ is a trace class operator, $\hat{M}_{n,t}$
is bounded, so $\left(\hat{M}_{n,t}\hat{P}_{x_{n,t}}\right)$ is also
trace class. Although intuitive, our proof of Lemma \ref{lem:integrale_localisee_trace}
is quite long and is given in Appendix \ref{sec:Proof-of-Lemma_trace}.

\section{\label{sec:Semi-classical-De-Latte-non-stationnary}Semi-classical
non-stationary Normal Form}

In the previous Section, we have explained that long time quantum
dynamics can be expressed with the operator $\hat{P}_{x\left(t\right)}\hat{M}^{t}\hat{P}_{x\left(0\right)}$,
where $x\left(t\right)=M^{t}x\left(0\right)$ is a given classical
trajectory. Using truncation operators all along the trajectory, we
have shown in Theorem \ref{pro:voisinage_trajectoire}, that the operator
$\hat{P}_{x\left(t\right)}\hat{M}^{t}\hat{P}_{x\left(0\right)}$ depends
in fact only of the vicinity of the trajectory $x\left(t'\right)$,
$t'\in[0,t]$. There appeared operators like $\hat{P}_{M\left(x\right)}\hat{M}\hat{P}_{x}$,
with $x\in\mathbb{R}^{2}$. This suggests to use a local description
of the dynamics along the trajectory, in terms of a Taylor expansion
up to a given order $J$. Using convenient canonical coordinates,
we can make this description in its simplest form, called a normal
form expansion. This is Theorem \ref{thm:compose_normal_form_quantique}
below. A normal form expression of the dynamics is particularly interesting
for large time, because along a given trajectory we will obtain a
product of normal forms operators which is quite easy to calculate
(because they commute together). In particular, for a periodic orbit
the computation of the trace will be explicit.

For classical hyperbolic map on the torus, David DeLatte in \cite{delatte_92},
has shown that there exists a global and uniform normal form expression,
called non-stationary normal form, which is unique up to co-boundary
terms. In this section we develop the semi-classical version of his
result, and use it to control long time evolution. Semi-classical
normal form for \emph{individual} unstable trajectories is already
a well known and very useful tool in semiclassical analysis, see \cite{CPI,CPII,CPIII},
\cite{zelditch-97,zelditch-98},\cite{sjostrand_hyp_02},\cite{ianchenko-02c,ianchenko-98}.
Our approach of semi-classical non stationary normal forms follows
closely the exposition of J. Sjöstrand in \cite{sjostrand_hyp_02},
but will be adapted to the normal form approach of David DeLatte \cite{delatte_92},
which is \emph{uniform} over phase space and not individual for periodic
orbits. Therefore it will give a satisfactory control of the normal
form expressions for long (periodic) orbits.

In this Section we give the main result useful for our purpose, and
in appendix \ref{sec:Proof-of-theorem}, we give the proofs and more
details, in particular we present there the semi-classical non stationary
normal form theory in terms of Hamiltonian flows. This could be useful
in order to extend the present results to more general hyperbolic
Hamiltonian flows.

\subsection{Semi-classical non-stationary normal form}

We give here the semi-classical version of a Theorem of David DeLatte
\cite{delatte_92} about non stationary normal forms. 

\vspace{0.cm}\begin{center}\fbox{\parbox{16cm}{

\begin{thm}
\label{pro:semi-classical-normal-form} Let $J\geq2$ even, be the
order of the normal form calculation. Let $0<\alpha<1/2$. The evolution
operator $\hat{M}$ in the neighborhood of any point $x\in\mathbb{R}^{2}$
is well approximated by a normal form operator $\hat{N}_{x}$:\begin{equation}
\left\Vert \hat{P}_{M\left(x\right)}\hat{M}\hat{P}_{x}-\hat{P}_{M\left(x\right)}\left(\hat{\mathcal{T}}_{M\left(x\right)}\hat{N}_{x}\hat{\mathcal{T}}_{x}^{-1}\right)\hat{P}_{x}\right\Vert _{L^{2}}\leq C\hbar^{A}\label{eq:quantum_non_stat_normal_form}\end{equation}

with $A=\left(\frac{1}{2}-\alpha'\right)\left(J+1\right)-1$, any
$\alpha<\alpha'<1/2$, and $C$ independent of $x$. The operator
\begin{equation}
\hat{N}_{x}=\exp\left(-\frac{i}{\hbar}\hat{K}_{x}\right)\label{eq:Def_Nx}\end{equation}
 is generated by a Hamiltonian with a total Weyl symbol $K_{x}\left(q,p\right)$
which is a normal form up to order $J$:\begin{align}
K_{x}\left(q,p\right) & =\sum_{0\leq l+j\leq J/2}\lambda_{l,j,\left(x\right)}\hbar^{l}\left(qp\right)^{j}\label{eq:H_Jx}\end{align}
$\lambda_{0,1,\left(x\right)}=\lambda_{x}$ is the local expansion
rate already introduced in Eq.(\ref{eq:DM}), and the meaning of the
other $\lambda_{l,j,\left(x\right)}$ is discussed below. $\hat{\mathcal{T}}_{x}$
is a product of unitary operators:\begin{equation}
\hat{\mathcal{T}}_{x}=\hat{T}_{x}\hat{Q}_{x}\hat{U}_{G_{1,\mathbf{x}}}\hat{U}_{G_{2,\mathbf{x}}}\ldots\hat{U}_{G_{n_{max},\mathbf{x}}}\label{eq:def_Tau_x}\end{equation}
where $\hat{T}_{x}$ is the translation operator Eq.(\ref{e:quantum translation}),
$\hat{Q}_{x}$ is the Weyl quantization of the linear symplectic map
$Q_{x}$, Eq.(\ref{eq:Qx}). The next operators give non linear corrections:
the operator $\hat{U}_{G_{n,\mathbf{x}}}=\exp\left(-\frac{i}{\hbar}\hat{G}_{n,x}\right)$
is generated by $\hat{G}_{n,x}$ whose Weyl symbol is equal to \[
G_{n,x}\left(q,p\right)=g_{l,a,b,(x)}\hbar^{l}q^{a}p^{b},\qquad\mbox{with}\,\,\,3\leq2l+a+b\leq J\]
in a neighborhood of the origin (the index $n$ enumerates all the
indices $\left(l,a,b\right)$ in the range $3\leq2l+a+b\leq J$, and
with increasing order of $\left(2l+a+b\right)$). The functions $g_{l,a,b,\left(x\right)}$
and $\lambda_{l,j,\left(x\right)}$ for $\left(l,j\right)\neq\left(0,0\right)$,
are continuous with respect to $x\in\mathbb{T}^{2}$. The function
$\lambda_{0,0,\left(x\right)}$ is continuous with respect to $x\in\mathbb{R}^{2}$
(but not periodic).
\end{thm}
}}\end{center}\vspace{0.cm}

The proof is given in appendix \ref{sec:Proof-of-theorem}.

\paragraph{Remarks:}

\begin{figure}[tbph]
\begin{centering}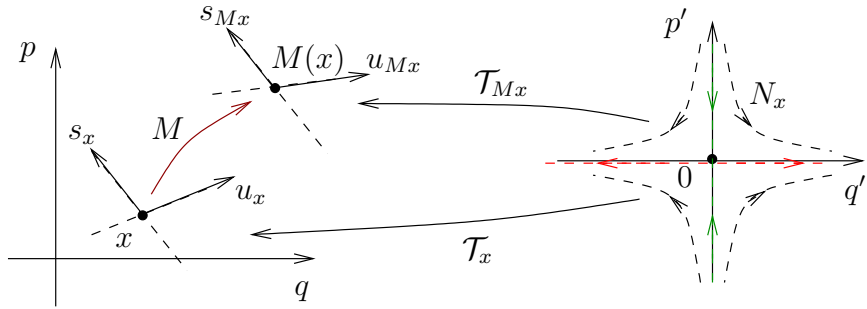\par\end{centering}

\caption{\label{cap:normalization}This picture traduces the conjugation relation
Eq.(\ref{eq:quantum_non_stat_normal_form}) of the non stationary
normal form. Here, in a neighborhood of a point $x$, the classical
map $M\equiv\mathcal{T}_{M\left(x\right)}N_{x}\mathcal{T}_{x}^{-1}$
is conjugated to a map $N_{x}$ which is a normal form up to order
$J$, and has $0$ as hyperbolic fixed point.}
\end{figure}

\begin{itemize}
\item Eq.(\ref{eq:quantum_non_stat_normal_form}) is interesting if the
error vanishes in the semi-classical limit $\hbar\rightarrow0$, so
if $\left(J+1\right)\left(\frac{1}{2}-\alpha\right)-1>0\Leftrightarrow J\geq\frac{1+2\alpha}{1-2\alpha}$,
for example if $J=2$ and $0<\alpha<1/6$.
\item If $J=2$, the normal form description is just at the level of linear
approximation. It is the quantum version of Eq.(\ref{eq:DMx_with_Qx})
and relies on uniform hyperbolicity. In the proof, it will be clear
that the next order terms are obtained iteratively from this linear
approximation, as usual in normal form calculations.
\item Eq.(\ref{eq:quantum_non_stat_normal_form}) has a classical version,
with maps $N_{x}$, $\mathcal{T}_{x}$ instead of operators $\hat{N}_{x}$,
$\hat{\mathcal{T}}_{x}$ respectively, and with similar Taylor expansions,
but without the semiclassical terms $\hbar^{l},\, l\geq1$. The classical
normal form is expressed in Figure \ref{cap:normalization}, and corresponds
to the theorem 1.1, p. 238 given in \cite{delatte_92} by David DeLatte
(the formal version). In \cite{delatte_92}\cite{delatte_95}, David
DeLatte proved a more stronger result: if $M$ is analytic, then normal
form series of $K_{x}$ and $\mathcal{T}_{x}$ are convergent in a
neighborhood of $\left(q,p\right)=\left(0,0\right)$, for $J\rightarrow\infty$.
Thanks to truncation operators, we will not need this result.
\end{itemize}

\begin{itemize}
\item The main non trivial result in Theorem \ref{pro:semi-classical-normal-form}
is the continuity and periodicity of $N_{x}$ and $\mathcal{T}_{x}$
with respect to $x$. This is a global constraint over the torus,
and relies on the uniform hyperbolicity of the dynamics. This continuity
is already given in Theorem \ref{thm:structural_stability} for the
expansion rate $\lambda_{x}$. In the proof, this appears in Lemma
\ref{lem:(David-Delatte)}.
\item $\lambda_{0,0,\left(x\right)}$ is called the Action of the path $x\rightarrow M\left(x\right)$.
It is given by the integral $\lambda_{0,0,\left(x\right)}=-\int\frac{1}{2}\left(pdq-qdp\right)-Hdt$
along the trajectory, as explained in Eq.(\ref{eq:action_l00}) page
\pageref{eq:action_l00}. $\lambda_{0,0,\left(x\right)}$ is a constant
term in the function Eq.(\ref{eq:H_Jx}) and does not appear in the
classical version of the Theorem, but is fundamental to explain the
interference effects in quantum mechanics. For a geometric interpretation
of $\lambda_{0,0}\left(x\right)$ in terms of parallel transport on
a Complex line bundle over the phase space (called the prequantum
bundle), see \cite{fred-PreQ-06} and references therein. 
\item We will use later on, the fact that for $\left(j,l\right)\neq\left(0,0\right)$,
$\lambda_{j,l,\left(x\right)}$ is a continuous function of $x\in\mathbb{T}^{2}$,
and is therefore bounded:\begin{equation}
\lambda_{j,l,min}\defi\mbox{min}_{x\in\mathbb{T}^{2}}\left(\lambda_{j,l,\left(x\right)}\right)\leq\lambda_{j,l,\left(x\right)}\leq\lambda_{j,l,max}\defi\mbox{max}_{x\in\mathbb{T}^{2}}\left(\lambda_{j,l,\left(x\right)}\right)\label{eq:lambda_bounded}\end{equation}

\item The function $\lambda_{1,0,\left(x\right)}=\lambda_{x}$ is equal
to the expansion rate introduced in Eq.(\ref{eq:DM}), and is called
the \textbf{Lyapounov cocycle.} The function $\lambda_{2,0,\left(x\right)}$
of $x$, is called the \textbf{Anosov cocycle} and is an obstruction
to $C^{2}$ regularity of the unstable/stable foliation, see \cite{hurder-90},
or \cite{hasselblatt_02} p. 289. These two cocycles are the first
terms of the series $\lambda_{j,l}$ of (semi-classical) cocycles.
\end{itemize}

\subsection{\label{sub:Semi-classical-Normal-form_long_orbit}Semi-classical
normal form for a long orbit}

We will now combine Eq.(\ref{eq:voisinage_trajectoire}) and Eq.(\ref{eq:quantum_non_stat_normal_form})
along a given trajectory $x\left(t\right)=M^{t}x$ starting from $x\in\mathbb{R}^{2}$.
Let us first define

\[
\hat{N}_{x,t}\defi\hat{N}_{M^{t-1}\left(x\right)}\hat{N}_{M^{t-2}\left(x\right)}\ldots\hat{N}_{x}\]
to be the product of Normal forms $\hat{N}_{x}=\exp\left(-\frac{i}{\hbar}\hat{K}_{x}\right)$
along the trajectory. Quantum normal forms Hamiltonian $\hat{K}_{x}$
at different point $x$ commute together (this is proved in Eq.(\ref{eq:F,G-commutte})
page \pageref{eq:F,G-commutte}). So the product $\hat{N}_{x,t}$
can be written

\begin{equation}
\boxed{\hat{N}_{x,t}=\exp\left(-\frac{i}{\hbar}t\hat{K}_{x,t}\right)}\label{eq:Nxt_expK}\end{equation}

where $\hat{K}_{x,t}=Op_{Weyl}\left(K_{x,t}\right)$ has total Weyl
symbol:\[
K_{x,t}\left(q,p\right)\defi\frac{1}{t}\sum_{s=0}^{t-1}K_{M^{s}x}\left(q,p\right)\]
which is just a time average along the trajectory. Using Eq.(\ref{eq:H_Jx})
we can write \begin{equation}
\boxed{K_{x,t}\left(q,p\right)=\sum_{0\leq l+j\leq J/2}\mu_{l,j,\left(x\right),t}\hbar^{l}\left(qp\right)^{j}}\label{eq:H_moy_mu}\end{equation}
with coefficients\begin{equation}
\boxed{\mu_{j,l,\left(x\right),t}\defi\frac{1}{t}\sum_{s=0}^{t-1}\lambda_{j,l,\left(M^{s}x\right)}}\label{eq:def_mu_ij}\end{equation}
From Eq.(\ref{eq:lambda_bounded}), we deduce that $\mu_{j,l,\left(x\right),t}$
is bounded uniformly with respect to $x\in\mathbb{T}^{2}$ and $t\in\mathbb{Z}$
(i.e. for any orbit):

\begin{equation}
\lambda_{j,l,min}\leq\mu_{j,l,\left(x\right),t}\leq\lambda_{j,l,max}\label{eq:l_bounded}\end{equation}
We can now state:

\vspace{0.cm}\begin{center}\fbox{\parbox{16cm}{

\begin{thm}
\label{thm:compose_normal_form_quantique}For any $J\geq2$, any $C>0$,
any $0<\alpha<1/2$, there exists $C_{1}>0$, such that for any $0<t\leq C\log\left(1/\hbar\right)$,
and any initial point $x\in\mathbb{R}^{2}$, any $1/2>\alpha'>\alpha$,\begin{equation}
\left\Vert \hat{P}_{M^{t}\left(x\right)}\hat{M}^{t}\hat{P}_{x}-\hat{P}_{M^{t}\left(x\right)}\left(\hat{\mathcal{T}}_{M^{t}\left(x\right)}\hat{N}_{x,t}\hat{\mathcal{T}}_{x}^{-1}\right)\hat{P}_{x}\right\Vert _{L^{2}}\leq t\, C_{1}\,\hbar^{\left(J+1\right)\left(\frac{1}{2}-\alpha'\right)-1}\label{eq:compose_normal_form_quantique}\end{equation}
As a corollary, for a periodic orbit, i.e. any fixed point $x_{n,t}$
of $M_{torus}^{t}$ given by Eq.(\ref{eq:periodic_fixed_point_xn}):

\begin{equation}
\left|\mbox{T}\left(\hat{M}_{n,t}\right)-e^{iA_{n}/\hbar}\textrm{Tr}\left(\hat{P}_{\alpha}\hat{N}_{\left(x_{n,t}\right),t}\hat{P}_{\alpha}\right)\right|\leq tC_{1}\,\hbar^{\left(J+1\right)\left(\frac{1}{2}-\alpha'\right)-1-\alpha'}\label{eq:Trace_Mt_semi_1}\end{equation}
where $\hat{M}_{n,t}$ is defined in Eq.(\ref{eq:Mnt}), $\mbox{T}\left(\hat{M}_{n,t}\right)$
is defined in Eq.(\ref{eq:Integrale}), and \begin{equation}
A_{n,t}\defi\frac{1}{2}n\wedge x_{n,t}\label{eq:An_orbite_periodique}\end{equation}

\end{thm}
}}\end{center}\vspace{0.cm}

\paragraph{Remarks:}

\begin{itemize}
\item The constant $A_{n,t}$ and $\mu_{0,0,\left(x_{n,t}\right),t}=-\frac{1}{t}\int\frac{1}{2}\left(pdq-qdp\right)-Hdt$
defined in Eq.(\ref{eq:def_mu_ij}), contribute to the total \textbf{action
of the periodic orbit} defined by:\begin{equation}
\mathcal{A}_{n,t}\defi A_{n,t}-t\mu_{0,0,\left(x_{n,t}\right),t}\label{eq:total_action_An}\end{equation}

\end{itemize}
\begin{proof}
The proof of Eq.(\ref{eq:compose_normal_form_quantique}) combines
Theorem \ref{pro:voisinage_trajectoire} and Theorem \ref{pro:semi-classical-normal-form}
in three steps. First from Eq.(\ref{eq:quantum_non_stat_normal_form}),
we deduce that\begin{align*}
\left\Vert \hat{P}_{x\left(t\right)}\hat{M}\hat{P}_{x\left(t-1\right)}\hat{M}\hat{P}_{x\left(t-2\right)}\ldots\hat{M}\hat{P}_{x\left(0\right)}\right.\\
\left.-\hat{P}_{x\left(t\right)}\left(\hat{\mathcal{T}}_{x\left(t\right)}\hat{N}_{x\left(t-1\right)}\hat{\mathcal{T}}_{x\left(t-1\right)}^{-1}\right)\hat{P}_{x\left(t-1\right)}\left(\hat{\mathcal{T}}_{x\left(t-1\right)}\hat{N}_{x\left(t-2\right)}\hat{\mathcal{T}}_{x\left(t-2\right)}^{-1}\right)\hat{P}_{x\left(t-2\right)}\ldots\hat{P}_{x\left(0\right)}\right\Vert _{L^{2}} & =Ct\hbar^{A}\end{align*}
We can now apply Theorem \ref{pro:voisinage_trajectoire} to the sequence
of hyperbolic maps $\left(\mathcal{T}_{M\left(x\right)}N_{x}\mathcal{T}_{x}^{-1}\right)$
in order to take off the intermediate operators $\hat{P}_{x\left(s\right)}$.
We obtain:\begin{align*}
\left\Vert \hat{P}_{x\left(t\right)}\left(\hat{\mathcal{T}}_{M^{t}\left(x\right)}\hat{N}_{x,t}\hat{\mathcal{T}}_{x}^{-1}\right)\hat{P}_{x\left(0\right)}\right.\\
\left.-\hat{P}_{x\left(t\right)}\left(\hat{\mathcal{T}}_{x\left(t\right)}\hat{N}_{x\left(t-1\right)}\hat{\mathcal{T}}_{x\left(t-1\right)}^{-1}\right)\hat{P}_{x\left(t-1\right)}\left(\hat{\mathcal{T}}_{x\left(t-1\right)}\hat{N}_{x\left(t-2\right)}\hat{\mathcal{T}}_{x\left(t-2\right)}^{-1}\right)\hat{P}_{x\left(t-2\right)}\ldots\hat{P}_{x\left(0\right)}\right\Vert  & _{L^{2}}=\mathcal{O}\left(\hbar^{\infty}\right)\end{align*}
Combining the last two equations with Eq.(\ref{eq:voisinage_trajectoire}),
we obtain finally Eq.(\ref{eq:compose_normal_form_quantique}). Now
we will prove Eq.(\ref{eq:Trace_Mt_semi_1}). First Eq.(\ref{eq:compose_normal_form_quantique})
for $x=x_{n,t}$ gives \[
\left\Vert \hat{P}_{M^{t}\left(x_{n,t}\right)}\hat{M}^{t}\hat{P}_{x_{n,t}}-\hat{P}_{M^{t}\left(x_{n,t}\right)}\left(\hat{\mathcal{T}}_{M^{t}\left(x_{n,t}\right)}\hat{N}_{x_{n,t},t}\hat{\mathcal{T}}_{x_{n,t}}^{-1}\right)\hat{P}_{x_{n,t}}\right\Vert _{L^{2}}\leq t\, C_{1}\,\hbar^{\left(J+1\right)\left(\frac{1}{2}-\alpha'\right)-1}\]
But $M^{t}x_{n,t}=x_{n,t}+n$, so $\hat{P}_{M^{t}\left(x_{n,t}\right)}=\hat{T}_{n}\hat{P}_{x_{n,t}}\hat{T}_{-n}$.
From Eq.(\ref{eq:def_Tau_x}), and $\hat{Q}_{M^{t}x_{n,t}}=\hat{Q}_{x_{nt}}$
(due to periodicity), we have $\hat{\mathcal{T}}_{M^{t}\left(x_{n,t}\right)}=\hat{T}_{M\left(x_{n,t}\right)}\hat{T}_{-x_{n,t}}\hat{\mathcal{T}}_{x_{n,t}}=e^{iA_{n,t}/\hbar}\hat{T}_{n}\hat{\mathcal{T}}_{x_{n,t}}$,
with $A_{n,t}\defi\frac{1}{2}n\wedge x_{n,t}$. The last equality
comes from Eq.(\ref{e:Heisenberg}). We obtain:\[
\left\Vert \hat{P}_{x_{n,t}}\hat{M}_{n,t}\hat{P}_{x_{n,t}}-e^{iA_{n,t}/\hbar}\hat{P}_{x_{n,t}}\left(\hat{\mathcal{T}}_{x_{n,t}}\hat{N}_{x_{n,t},t}\hat{\mathcal{T}}_{x_{n,t}}^{-1}\right)\hat{P}_{x_{n,t}}\right\Vert _{L^{2}}\leq t\, C_{1}\,\hbar^{\left(J+1\right)\left(\frac{1}{2}-\alpha'\right)-1}\]
Lemma \ref{lem:norme_1_et_norme_2} page \pageref{lem:norme_1_et_norme_2}
allows us to pass from operator norm to Trace norm estimates. It gives:\begin{equation}
\left|\mbox{Tr}\left(\hat{P}_{x_{n,t}}\hat{M}_{n,t}\hat{P}_{x_{n,t}}\right)-e^{iA_{n,t}/\hbar}\mbox{Tr}\left(\hat{P}_{x_{n,t}}\left(\hat{\mathcal{T}}_{x_{n,t}}\hat{N}_{x_{n,t},t}\hat{\mathcal{T}}_{x_{n,t}}^{-1}\right)\hat{P}_{x_{n,t}}\right)\right|\leq t\, C_{1}\,\hbar^{\left(J+1\right)\left(\frac{1}{2}-\alpha'\right)-1-\alpha'}\label{eq:etape_2}\end{equation}
We want now to take off the operator $\hat{\mathcal{T}}_{x_{n,t}}$.
Remind that $\mathcal{T}_{x}$ is a smooth canonical transformation
which send point $0$ to $x$. If $0<\beta<\alpha$, we have in operator
norm $\hat{P}_{x,\beta}\hat{\mathcal{T}}_{x}\hat{P}_{0,\alpha}=\hat{P}_{x,\beta}\hat{\mathcal{T}}_{x}+\mathcal{O}\left(\hbar^{\infty}\right)$,
$\hat{P}_{0,\alpha}\hat{\mathcal{T}}_{x}\hat{P}_{x,\beta}=\hat{\mathcal{T}}_{x}\hat{P}_{0,\beta}+\mathcal{O}\left(\hbar^{\infty}\right)$,
and $\hat{P}_{x,\beta}\hat{P}_{0,\alpha}=\hat{P}_{x,\beta}+\mathcal{O}\left(\hbar^{\infty}\right)$.
So, with $\beta>\gamma>\alpha$, \begin{align}
\mbox{Tr}\left(\hat{P}_{x_{n,t},\alpha}\left(\hat{\mathcal{T}}_{x_{n,t}}\hat{N}_{x_{n,t},t}\hat{\mathcal{T}}_{x_{n,t}}^{-1}\right)\hat{P}_{x_{n,t},\alpha}\right) & =\mbox{Tr}\left(\hat{P}_{x_{n,t},\alpha}\left(\hat{\mathcal{T}}_{x_{n,t}}\hat{P}_{0,\gamma}\hat{N}_{x_{n,t},t}\hat{P}_{0,\gamma}\hat{\mathcal{T}}_{x_{n,t}}^{-1}\right)\hat{P}_{x_{n,t},\alpha}\right)+\mathcal{O}\left(\hbar^{\infty}\right)\label{eq:etape_3}\\
 & =\mbox{Tr}\left(\hat{P}_{x_{n,t},\beta}\left(\hat{\mathcal{T}}_{x_{n,t}}\hat{P}_{0,\gamma}\hat{N}_{x_{n,t},t}\hat{P}_{0,\gamma}\hat{\mathcal{T}}_{x_{n,t}}^{-1}\right)\hat{P}_{x_{n,t},\beta}\right)+\mathcal{O}\left(\hbar^{\infty}\right)\nonumber \\
 & =\mbox{Tr}\left(\hat{\mathcal{T}}_{x_{n,t}}\hat{P}_{0,\gamma}\hat{N}_{x_{n,t},t}\hat{P}_{0,\gamma}\hat{\mathcal{T}}_{x_{n,t}}^{-1}\right)+\mathcal{O}\left(\hbar^{\infty}\right)\nonumber \\
 & =\mbox{Tr}\left(\hat{P}_{0,\gamma}\hat{N}_{x_{n,t},t}\hat{P}_{0,\gamma}\right)+\mathcal{O}\left(\hbar^{\infty}\right)\nonumber \end{align}
For the second line, we have used the property that the trace does
not depend on the choice of $\beta$, up to $\mathcal{O}\left(\hbar^{\infty}\right)$.
Finally, Eq.(\ref{eq:etape_2},\ref{eq:etape_3},\ref{eq:lem_1})
together give Eq.(\ref{eq:Trace_Mt_semi_1}).
\end{proof}

\subsection{Post-Normalization \label{sub:Post-Normalization}}

\subsubsection{Post Normalization corrections}

The Weyl symbol $K_{x,t}\left(q,p\right)$ given in Eq.(\ref{eq:H_moy_mu})
is a function of the product $\left(qp\right)$ only. Let us write
it $K_{x,t}\left(qp\right)$. The quantized operator we have to consider
in Eq.(\ref{eq:Nxt_expK}) is $\hat{K}_{x,t}=Op_{Weyl}\left(K_{x,t}\left(qp\right)\right)$.
First remark that, in order to compute the trace of the propagator,
or its matrix elements, it is easier to deal with a function of the
single operator $Op_{Weyl}\left(qp\right)$, and second, $Op_{Weyl}\left(K_{x,t}\left(qp\right)\right)\neq K_{x,t}\left(Op_{Weyl}\left(qp\right)\right)$
in general (see e.g. Eq.(\ref{eq:recurrence_relation}) below). So
we have to find a function $\tilde{K}_{x,t}$ of a single variable,
such that \begin{equation}
\boxed{\hat{K}_{x,t}=Op_{Weyl}\left(K_{x,t}\left(qp\right)\right)=\tilde{K}_{x,t}\left(Op_{Weyl}\left(qp\right)\right)}\label{eq:K_post_n}\end{equation}

\selectlanguage{french}
\vspace{0.cm}\begin{center}\fbox{\parbox{16cm}{

\selectlanguage{english}
\begin{prop}
If $K_{x,t}\left(qp\right)=\sum_{0\leq l+j\leq J/2}\mu_{l,j,x,t}\hbar^{l}\left(qp\right)^{j}$
is expressed as a power series in $\left(qp\right)$ and $\hbar$,
(as we get in Eq.(\ref{eq:H_moy_mu})), then $\tilde{K}_{x,t}$ is
given by:\begin{equation}
\tilde{K}_{x,t}\left(qp\right)=\sum_{l,j\geq0}\tilde{\mu}_{l,j,x,t}\hbar^{l}\left(qp\right)^{j}\label{eq:Ktilde_series_post_n}\end{equation}
where the coefficients $\tilde{\mu}_{l,j,x,t}$ are given explicitely
from $\mu_{l,j,x,t}$ by:\begin{equation}
\tilde{\mu}_{l,j,x,t}\defi\sum_{0\leq m\leq\left[l/2\right]}E_{j',m}\mu_{l',j',x,t},\qquad l'=l-2m,\, j'=j+2m\label{eq:mu_tilde_i_j}\end{equation}
where $E_{j',m}$ is a numerical factor determined by recursive formula:\begin{eqnarray}
E_{j+1,m} & = & E_{j,m}-\frac{j^{2}}{4}E_{\left(j-1\right),\left(m-1\right)},\textrm{ if }\,\, j,m\geq1\nonumber \\
E_{j,0} & =1, & E_{0,m}=0,\, E_{1,m}=0,\,\textrm{for\,\,\,}m\geq1\label{eq:def_E_jm}\end{eqnarray}

\end{prop}
\selectlanguage{french}
}}\end{center}\vspace{0.cm}

\selectlanguage{english}

\paragraph{Remarks:}

\begin{itemize}
\item In particular, at the principal and sub-principal level (i.e. order
$l=0,1$ in $\hbar^{l}$), there is no change: i.e. $\tilde{\mu}_{l,j,x,t}=\mu_{l,j,x,t}$,
if $l=0$ or $l=1$. 
\item In \cite{garcia-saz-04}, paragraph 6, Alfonso Garcia-Saz gives such
expressions as a special case of a more general problem. See also
appendix I in \cite{cargo_05}.
\end{itemize}
\begin{proof}
We use the Moyal start product $\sharp$ defined in Eq.(\ref{eq:start_product}),
which is equivalent to the product of operators. In other words, we
want to express $\left(qp\right)^{j}$ in terms of product of monomial
terms $\left(qp\right)^{\sharp j}\defi\left(qp\right)\sharp\left(qp\right)\sharp\ldots\sharp\left(qp\right)$.
We just apply Moyal product formula Eq.(\ref{eq:Moyal_product}),
with operators $A=qp$, $B=\left(qp\right)^{j}$, and observe that
$\left(A\mathcal{D}^{n}B\right)=0$, for $n=1$ and $n\geq3$, and
$\left(A\mathcal{D}^{2}B\right)=-2j^{2}\left(qp\right)^{j-1}$. This
gives \begin{equation}
\left(qp\right)\sharp\left(qp\right)^{j}=\left(qp\right)^{j+1}+\hbar^{2}\frac{j^{2}}{4}\left(qp\right)^{j-1}\label{eq:formule_sharp}\end{equation}
Or equivalently\begin{equation}
Op_{Weyl}\left(qp\right)^{j+1}=\left(Op_{Weyl}\left(qp\right)\right)\left(Op_{Weyl}\left(qp\right)^{j}\right)-\hbar^{2}\frac{j^{2}}{4}\left(Op_{Weyl}\left(qp\right)^{j-1}\right)\label{eq:recurrence_relation}\end{equation}
We deduce that \begin{equation}
\left(qp\right)^{j}=\sum_{m=0}^{\left[j/2\right]}E_{j,m}\hbar^{2m}\left(qp\right)^{\sharp\left(j-2m\right)}\label{eq:recursive_post_normalization}\end{equation}
This last formula says that each resonant term of $Op_{Weyl}\left(K_{x,t}\right)$
can be written \[
\mu_{l,j,\left(x\right),t}\hbar^{l}Op_{Weyl}\left(\left(qp\right)^{j}\right)=\mu_{l,j,\left(x\right),t}\sum_{m=0}^{\left[j/2\right]}E_{j,m}\hbar^{l+2m}\left(Op_{Weyl}\left(qp\right)\right)^{j-2m}\]
 as desired.
\end{proof}
\begin{prop}
$\left(qp\right)^{n}$ and $\left(qp\right)^{m}$ commute with respect
to the star product:\[
\left[\left(qp\right)^{n},\left(qp\right)^{m}\right]_{\sharp}=\left(qp\right)^{n}\sharp\left(qp\right)^{m}-\left(qp\right)^{m}\sharp\left(qp\right)^{n}=0\]
therefore if $F,G\in C^{\infty}\left(\mathbb{R}\right)$, \begin{equation}
\left[Op_{Weyl}\left(F\left(qp\right)\right),Op_{Weyl}\left(G\left(qp\right)\right)\right]=0\label{eq:F,G-commutte}\end{equation}

\end{prop}
\begin{proof}
By induction on the power: suppose that $\left[\left(qp\right)^{n},\left(qp\right)^{m}\right]_{\sharp}=0$,
for $n,m\leq N$. Using Eq.(\ref{eq:formule_sharp}), we compute\begin{eqnarray*}
\left[\left(qp\right)^{N+1},\left(qp\right)^{n}\right]_{\sharp} & = & \left(qp\right)^{N+1}\sharp\left(qp\right)^{n}-\left(qp\right)^{n}\sharp\left(qp\right)^{N+1}\\
 & = & \left(qp\right)\sharp\left(qp\right)^{N}\sharp\left(qp\right)^{n}-\hbar^{2}\frac{N^{2}}{4}\left(qp\right)^{N-1}\sharp\left(qp\right)^{n}\\
 &  & -\left(qp\right)^{n}\sharp\left(qp\right)\sharp\left(qp\right)^{N}+\hbar^{2}\frac{N^{2}}{4}\left(qp\right)^{n}\sharp\left(qp\right)^{N-1}\\
 & = & 0\end{eqnarray*}
from associativity of the star product. There is a more direct proof,
using the post-normalization transformation $F\rightarrow\tilde{F}$
described above: \[
\left[Op_{Weyl}\left(F\left(qp\right)\right),Op_{Weyl}\left(G\left(qp\right)\right)\right]=\left[\tilde{F}\left(Op_{Weyl}\left(qp\right)\right),\tilde{G}\left(Op_{Weyl}\left(qp\right)\right)\right]=0.\]

\end{proof}

\subsubsection{Semi-classical Trace formula for a periodic orbit}

Semi-classical expansions for the trace of hyperbolic normal forms
is explicitly done in \cite{iantchenko-02b}. It can be used to give
$\textrm{Tr}\left(\hat{P}_{\alpha}\hat{N}_{x_{n},t}\hat{P}_{\alpha}\right)$
in terms of the semi-classical post-normalized cocycles $\tilde{\mu}_{l,j,x_{n},t}$
defined in Eq.(\ref{eq:mu_tilde_i_j}). We do this in appendix \ref{sub:Appendix:-Calculus_Normal_Forms},
Proposition \ref{pro:appendix_trace_normale_form}, page \pageref{pro:appendix_trace_normale_form}.
We can deduce the following proposition:

\begin{prop}
\label{pro:Tn_semi_xnt}Let $J\geq2$. For any $C>0$, any $0<\alpha<1/2$,
there exists $C_{1}>0$ such that, for any time $0<t\leq C\log\left(1/\hbar\right)$,
any $n\in\mathcal{C}_{t}$, one has a semi-classical expression:\begin{equation}
\left|\textrm{Tr}\left(\hat{P}_{\alpha}\hat{N}_{x_{n,t},t}\hat{P}_{\alpha}\right)-T_{semi,J}\left(x_{n,t}\right)\right|\leq C_{1}\hbar^{\left[J/2\right]}\label{eq:trace_plane_N_normal}\end{equation}
where $x_{n,t}$ is the fixed point of the map $M_{n,t}$, Eq.(\ref{eq:classical_M_nt}),
and\[
T_{semi,J}\left(x_{n,t}\right)\defi\exp\left(-it\frac{\mu_{0,0,x_{n},t}}{\hbar}\right)\exp\left(-it\mu_{1,0,x_{n},t}\right)\frac{1}{2\sinh\left(\frac{\mu_{0,1,x_{n},t}t}{2}\right)}\mathcal{E}_{x_{n,t}}\]
with a semi-classical expansion\[
\mathcal{E}_{x_{n,t}}=1+\hbar E_{1}+\hbar^{2}E_{2}+\ldots+\hbar^{\left[J/2\right]-1}E_{\left(\left[J/2\right]-1\right)}\]
where $E_{s}$ depends on $t$ and $\tilde{\mu}_{l,j,x_{n},t}$ ,
with $\left(l+j\right)\leq s+1$. We have the bound \[
\left|E_{s}\right|\leq t^{s}E_{max,s}\]
where $E_{max,s}$ does not depend on $t$ nor on $n$. With Eq.(\ref{eq:Trace_Mt_semi_1})
we deduce that\begin{equation}
\boxed{\left|\mbox{T}\left(\hat{M}_{n,t}\right)-e^{iA_{n}/\hbar}T_{semi,J}\left(x_{n,t}\right)\right|\leq C_{1}\hbar^{\left[J/2\right]}}\label{eq:TMnt-TsemiJ}\end{equation}

\end{prop}

\paragraph{Remark:}

Explicit expressions of $E_{s}$, $s\geq1$ are quite complicated
for large $s$, and are given in Eq.(\ref{eq:E1_E2}) page \pageref{eq:E1_E2},
and Eq.(\ref{eq:I1_I2}) page \pageref{eq:I1_I2}. It appears in appendix
\ref{sub:Appendix:-Calculus_Normal_Forms} that with Weyl quantization
then $\mu_{1,0,x_{n},t}=0$ (and $\mu_{l,j}=0$ for $l$ odd). We
will use this simplification below.

\begin{proof}
Proposition \ref{pro:appendix_trace_normale_form}, page \pageref{pro:appendix_trace_normale_form},
gives $\left|\textrm{Tr}\left(\hat{P}_{\alpha}\hat{N}_{x_{n,t},t}\hat{P}_{\alpha}\right)-T_{semi,J}\left(x_{n,t}\right)\right|\leq C_{1}B$,
where $C_{1}$ does not depend on $t$ and $n$, and with $B=\max_{n,t}\left(\left(2\sinh\left(\frac{\tilde{\mu}_{0,1}t}{2}\right)\right)^{-1}E_{\left[J/2\right]}\hbar^{\left[J/2\right]}\right)$
given by the maximum over periodic orbits of the next order term in
the series of $T_{semi,J}\left(x_{n,t}\right)$. The uniform control
of the semi-classical cocycles, over the periodic orbits, Eq.(\pageref{eq:l_bounded})
gives $\tilde{\mu}_{0,1}\geq\lambda_{0,1,min}=\lambda_{min}$, where
$\lambda_{min}>0$ is defined in Eq.(\ref{eq:lmin_lmax}). Thus $\left(2\sinh\left(\frac{\tilde{\mu}_{0,1}t}{2}\right)\right)^{-1}\leq C_{2}e^{-\lambda_{min}t/2}\leq C_{2}\hbar^{\left(\frac{t\lambda_{min}}{2t_{E}}\right)}$,
because we can write $e^{-\lambda_{min}t/2}=e^{-\lambda_{0}t_{E}\left(\frac{t\lambda_{min}}{2t_{E}}\right)}=\hbar^{\left(\frac{t\lambda_{min}}{2t_{E}}\right)}$.
The uniform control of the semi-classical cocycles together with the
bound Eq.(\ref{eq:bound_Es}) also gives that $\left|E_{\left[J/2\right]}\right|\leq t^{s}E_{max,s}$
where $E_{max,s}$ does not depend on $n,t$. Now if $\left|t\right|\leq\mathcal{O}\left(\log\left(1/\hbar\right)\right)$
then $\left|t^{s}\right|<\hbar^{-\varepsilon}$ for any $\varepsilon>0$,
so at final $C_{1}B\leq C_{3}\hbar^{\left(\frac{t\lambda_{min}}{2t_{E}}\right)}\hbar^{-\varepsilon}\hbar^{\left[J/2\right]}\leq C_{4}\hbar^{\left[J/2\right]}$,
if we take $\varepsilon=\frac{1}{2}\left(\frac{\lambda_{min}}{2t_{E}}\right)$.
This gives Eq.(\ref{eq:trace_plane_N_normal}).

From Eq.(\ref{eq:trace_plane_N_normal}) and Eq.(\ref{eq:Trace_Mt_semi_1}),
we deduce that $\left|\mbox{T}\left(\hat{M}_{n,t}\right)-e^{iA_{n}/\hbar}T_{semi,J}\left(x_{n,t}\right)\right|\leq C_{1}\hbar^{\left[J/2\right]}+C'_{1}\,\hbar^{\left(J+1\right)\left(\frac{1}{2}-\alpha'\right)-1-\alpha'}$.
We observe that for large $J$, this gives a bound $\mathcal{O}\left(\hbar^{\infty}\right)$.
So we deduce that the actual bound is given by the next order term
in the series of $T_{semi,J}\left(x_{n,t}\right)$, namely $B=\mathcal{O}\left(\hbar^{\left[J/2\right]}\right)$
already computed above: in more precise terms, this is because $\left|\mbox{T}\left(\hat{M}_{n,t}\right)-e^{iA_{n}/\hbar}T_{semi,J\infty}\left(x_{n,t}\right)\right|=\mathcal{O}\left(\hbar^{\infty}\right)$,
and $\left|T_{semi,J\infty}\left(x_{n,t}\right)-T_{semi,J}\left(x_{n,t}\right)\right|=\mathcal{O}\left(\hbar^{\left[J/2\right]}\right)$.
\end{proof}

\section{\label{sec:Trace-of-M}Trace of the quantum map}

\subsection{Semi-classical trace formula}

The following Theorem is one of the main result of this paper. It
gives a semi-classical expression for the trace $\mbox{Tr}\left(\hat{M}_{torus}^{t}\right)$,
as a finite sum over the fixed points $x_{n,t}$ of the classical
map $M_{torus}^{t}$.

\selectlanguage{french}
\vspace{0.cm}\begin{center}\fbox{\parbox{16cm}{

\selectlanguage{english}
\begin{thm}
\label{pro:formule_trace_semi_classique}For any $K>0$ , any $C>0$,
any $J\geq2\left(K+C\right)$, there exists $C_{1}>0$, such that
for any time $\left|t\right|<Ct_{E}$, with $t_{E}=\frac{1}{\lambda_{0}}\log\frac{1}{\hbar}$,
any admissible value of $\hbar$ (see Eq.(\ref{eq:N_h})), the trace
of $\hat{M}_{torus}^{t}$ is given by the \textbf{semi-classical formula}:\begin{equation}
\left|\mbox{Tr}\left(\hat{M}_{torus}^{t}\right)-T_{semi,t,J}\right|\leq C_{1}\hbar^{K}\label{eq:Trace_Mt_semi}\end{equation}
\begin{equation}
T_{semi,t,J}=\sum_{n\in\mathcal{C}_{t}}\exp\left(i\frac{\mathcal{A}_{n,t}}{\hbar}\right)\frac{1}{2\sinh\left(\frac{\mu_{0,1,x_{n},t}t}{2}\right)}\,\mathcal{E}_{x_{n,t}}\label{eq:T_semi_t_J}\end{equation}
where each term of the finite sum is associated with a fixed point
$x_{n,t}$ of $M_{torus}^{t}$ given in Eq.(\ref{eq:periodic_fixed_point_xn}).
$\mathcal{A}_{n,t}=A_{n}-\mu_{0,0,x_{n},t}t$ is the action of the
periodic orbit, defined in Eq.(\ref{eq:total_action_An}). $\mu_{l,j,x_{n},t}$
and $\tilde{\mu}_{l,j,x_{n},t}$ are the semi-classical normal form
coefficients (cocycles) of the periodic orbit computed up to a given
order $J\geq2$ (i.e. with $2\left(j+l\right)\leq J$). $\mathcal{E}_{x_{n,t}}$
is a semi-classical series:

\[
\mathcal{E}_{x_{n,t}}=1+\sum_{1\leq s\leq\left[J/2\right]-1}\hbar^{s}E_{s}=1+\hbar E_{1}+\hbar^{2}E_{2}+\ldots+\hbar^{\left[J/2\right]-1}E_{\left[J/2\right]-1}\]

where $E_{s}$ depends on $t$ and $\tilde{\mu}_{l,j,x_{n},t}$ (with
$\left(l+j\right)\leq s+1$), and in particular is bounded by\[
\left|E_{s}\right|\leq t^{s}E_{max,s}\]
where $E_{max,s}$ does not depend on $t$ nor on $x_{n,t}$.
\end{thm}
\selectlanguage{french}
}}\end{center}\vspace{0.cm}

\selectlanguage{english}
\begin{proof}
We use Eq.(\ref{eq:Trace_Mt}), which is a sum with $\mathcal{N}_{t}=e^{\lambda_{0}t}-2+e^{-\lambda_{0}t}$
terms (from Eq.(\ref{eq:Nt_periodic_points})). We write $\mathcal{N}_{t}\leq e^{\lambda_{0}t}=e^{-\lambda_{0}t_{E}\left(-t/t_{E}\right)}=\hbar^{\left(-t/t_{E}\right)}\leq\hbar^{-C}$.
Each term $\mbox{T}\left(\hat{M}_{n,t}\right)$ is approximated by
a semiclassical expression given in Eq.(\ref{eq:TMnt-TsemiJ}), with
an uniform error $C_{1}\hbar^{\left[J/2\right]}$. By a simple triangular
inequality (i.e. we sum the magnitudes of the error bounds) we deduce
that

\[
\left|\mbox{Tr}\left(\hat{M}_{torus}^{t}\right)-T_{semi,t,J}\right|\leq\mathcal{N}_{t}C_{1}\hbar^{\left[J/2\right]}\leq C_{1}\hbar^{\left[J/2\right]-C}\]
This is $\mathcal{O}\left(\hbar^{K}\right)$ if $J>2\left(K+C\right)$
(for $J$ even). 
\end{proof}

\paragraph{Remarks:}

\begin{itemize}
\item If we look at the whole proof, the limitation to these logarithmic
time (any multiple of the Ehrenfest time $t_{E}=\frac{1}{\lambda_{0}}\log\frac{1}{\hbar}$)
is present many times. But the main limitation is due to exponential
proliferation of periodic orbits in the hyperbolic system, $\mathcal{N}_{t}\simeq e^{\lambda_{0}t}$.
\item The bound on the errors could be improved at many places in the proof,
so the condition $J>2\left(K+C\right)$ is not sharp. But the main
limitation in our proof is that we simply estimate the total error
as the sum of the absolute values of the error from each periodic
orbit. However its seems that all these errors compensate each other,
as observed in the next Section. However, a rigorous analysis of these
compensations is out of reach for the moment.
\item Let us comment again on the estimation used in the proof. We have
$\left(2\sinh\left(\frac{\mu_{0,1,x_{n},t}t}{2}\right)\right)^{-1}\simeq e^{-\lambda_{0}t/2}$
for large $t$, so, if we write Eq.(\ref{eq:T_semi_t_J}) as $T_{semi,t,J}=\sum_{n\in\mathcal{C}_{t}}T_{n,t}$,
each term associated to an individual periodic orbit $\left|T_{n,t}\right|\leq e^{-\lambda_{0}t/2}$
decreases, but $\mathcal{N}_{t}=\sharp\left(\mathcal{C}_{t}\right)\simeq e^{\lambda_{0}t}$
increases faster. We get the bound $\left|T_{semi,t,J}\right|\leq e^{\lambda_{0}t/2}$
which is greater than $\mbox{dim}\left(\mathcal{H}\right)=N=1/\left(2\pi\hbar\right)$
for $t\geq\frac{2}{\lambda_{0}}\log\left(1/\hbar\right)=2t_{E}$.
This shows that for $t\geq2t_{E}$, there are necessarily some cancellations
among the complex amplitudes involved in Eq.(\ref{eq:T_semi_t_J}),
so that $\left|T_{semi,t,J}\right|\leq\mbox{dim}\left(\mathcal{H}\right)=N$
is always satisfied. These cancellations are due to the leading complex
terms $\exp\left(-i\frac{\mathcal{A}_{n,t}}{\hbar}\right)$.
\item In some specific examples, namely the linear map $M_{0}$, Eq.(\ref{eq:M0}),
with particular values of $2\pi\hbar$, we observed in \cite{fred-steph-02}
that at $t\simeq2t_{E}$, all the actions $\exp\left(-i\frac{\mathcal{A}_{n,t}}{\hbar}\right)=1$
are equal and add together. This implies that the upper bound $\mbox{Tr}\left(\hat{M}_{torus}^{t=2t_{E}}\right)=\mbox{dim}\left(\mathcal{H}\right)=N$
is reached, and therefore that $\hat{M}_{torus}^{t\simeq2t_{E}}\propto\hat{Id}$
which implies revivals of quantum states and existence of scarred
eigenstates, i.e. non equidistributed invariant semiclassical measures.
\item Let us give now a {}``philosophical'' remark which shows again that
$t=2t_{E}$ seems to be a critical value of time. If we define the
{}``complexity'' of the trace formula to be equal to the number
of periodic orbits, it grows like $e^{\lambda_{0}t}$. This complexity
is larger than the {}``complexity'' of the linear quantum problem
(equal to the size of the matrix $\hat{M}$: $N^{2}\simeq e^{2\lambda_{0}t_{E}}$),
for $t\geq2t_{E}$. Again this shows that for $t\geq2t_{E}$, periodic
orbits manifest themselves in the semiclassical trace formula through
collective averaging effects (see Berry in \cite{houches-qc}, or
\cite{fred-PreQ-06} where this open problem is also discussed).
\end{itemize}

\subsection{Numerical results and observation of self-averaging effects\label{sub:Numerical-results-and}}

We illustrate the semi-classical formula for the trace Eq.(\ref{eq:Trace_Mt_semi}),
with the example defined by Eq.(\ref{eq:example_M}). We have computed
$\textrm{Tr}\left(\hat{M}^{t}\right)$ numerically for $t=0\rightarrow11$
(this is easy because the Hilbert space is finite dimensional, and
we expect the numerical result to have a good accuracy). We compare
it with the semi-classical approximation $\mbox{T}_{semi,t,J}$, Eq.(\ref{eq:T_semi_t_J}),
where every normal forms has been computed numerically%
\footnote{The algorithm to compute the normal forms is explained in Section
\ref{sub:Appendix:-Calculus_Normal_Forms}.%
} up to a given order $J\geq2$. We choose successively $N=1/\left(2\pi\hbar\right)=10$,
$N=1/\left(2\pi\hbar\right)=100$ with $J=2$ and $J=4$. The observed
error is denoted by $\mbox{Error}_{J}\left(t\right)\defi\left|\textrm{Tr}\left(\hat{M}^{t}\right)-T_{semi,t,J}\right|$.
The practical limitation in this numerical calculation is the increasing
number $\mathcal{N}_{t}\simeq e^{\lambda_{0}t}$ of periodic orbits;
already $\mathcal{N}_{t=11}=39601$. In Theorem \ref{pro:formule_trace_semi_classique},
we have estimated that

\[
\mbox{Error}_{J}\left(t\right)<C_{1}\varepsilon_{J}\left(t\right),\quad\textrm{with}\,\quad\varepsilon_{J}\left(t\right)=\hbar^{K}=\hbar^{\frac{J}{2}-C}=\hbar^{J/2-t/t_{E}}\]
with $t_{E}=\frac{1}{\lambda_{0}}\log\left(1/\hbar\right)$.

\begin{figure}[tbph]
\begin{centering}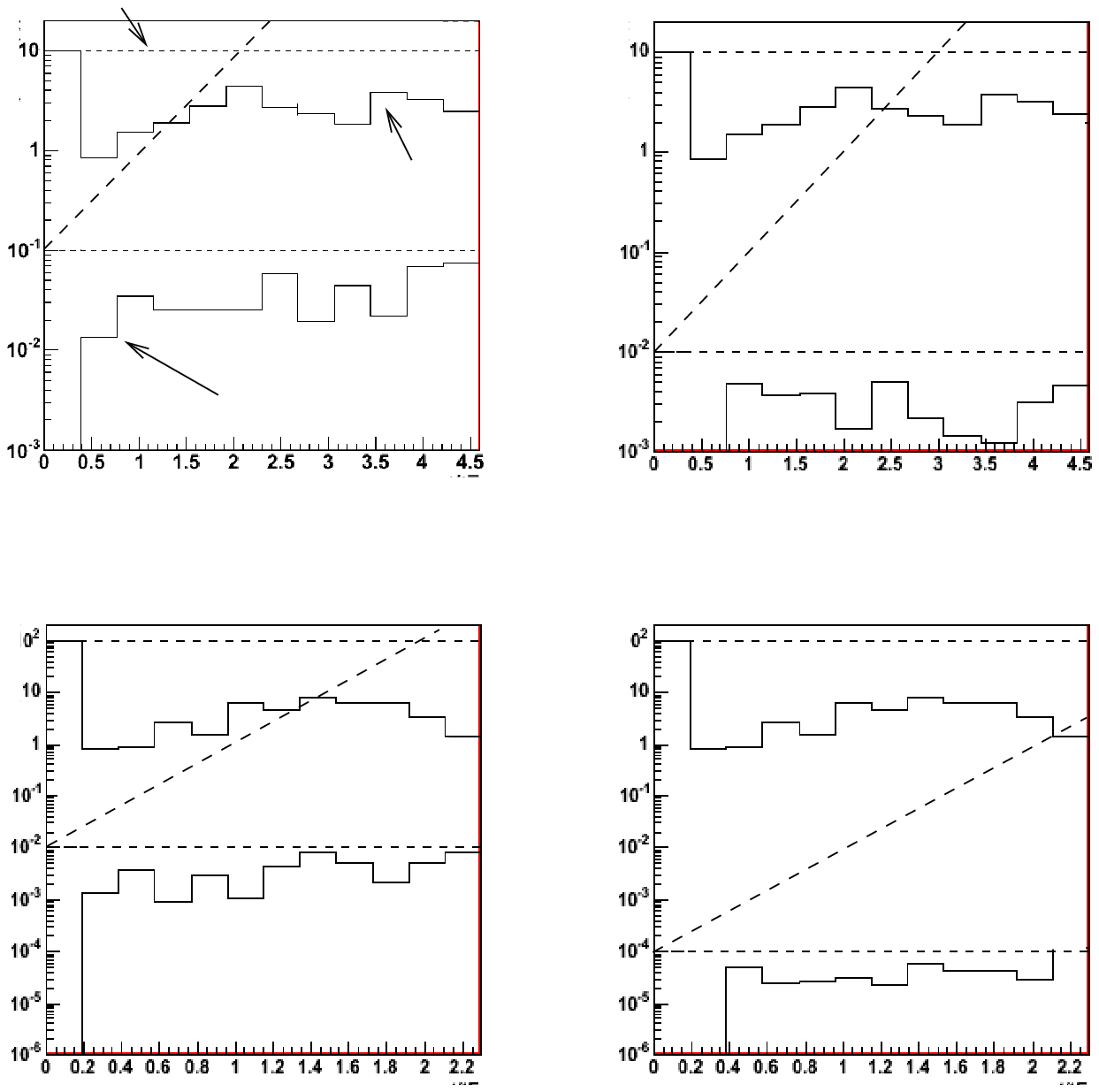\par\end{centering}

\caption{\label{cap:Numerical-results-compared}Numerical results for $\left|\mbox{Tr}\left(\hat{M}^{t}\right)\right|$
and its semiclassical approximation $T_{semi,t,J}$. We have plotted
$\mbox{Error}_{J}\left(t\right)\defi\left|\textrm{Tr}\left(\hat{M}^{t}\right)-T_{semi,t,J}\right|$
in log. scale.}
\end{figure}

On Figure \ref{cap:Numerical-results-compared}, we plot three functions
of $\left(t/t_{E}\right)$ on a logarithmic scale: $\left|\textrm{Tr}\left(\hat{M}^{t}\right)\right|$,
the observed error $\mbox{Error}_{J}\left(t\right)$ and the upper
bound of the error $\varepsilon_{J}\left(t\right)$. We also draw
the line $\textrm{Tr}_{Max}=1/\left(2\pi\hbar\right)$ (because obviously
$\left|\textrm{Tr}\left(\hat{M}^{t}\right)\right|\leq\dim\mathcal{H}_{torus}=1/\left(2\pi\hbar\right)$
), and the line of the particular value $\varepsilon_{\hbar,J}=\hbar^{J/2}$
which is expected to be an upper bound of the error from heuristic
arguments given below. The unexpected observed fact is that $\mbox{Error}_{J}\left(t\right)<\varepsilon_{\hbar,J}$,
although $\mbox{Error}_{J}\left(t\right)<\varepsilon_{J}\left(t\right)=\varepsilon_{\hbar,J}\hbar^{-t/t_{E}}$
has only been proved.

\paragraph{Remarks and observations:}

\begin{itemize}
\item $\hat{M}$ is an unitary operator, so $\left|\textrm{Tr}\left(\hat{M}^{t}\right)\right|\leq\dim\mathcal{H}_{torus}=1/\left(2\pi\hbar\right)$.
However our error estimate gives $\varepsilon_{J}\left(t\right)=\hbar^{J/2-t/t_{E}}\geq1/\left(2\pi\hbar\right)$
for $t\geq\left(\frac{J}{2}+1\right)t_{E}$. So our estimates is not
interesting for $t\geq\left(\frac{J}{2}+1\right)t_{E}$.
\item According to Eq.(\ref{eq:TMnt-TsemiJ}), the error term $\mbox{Error}_{J}\left(t\right)$
is a sum of $\mathcal{N}_{t}\simeq e^{\lambda_{0}t}$ complex numbers,
each being bounded in magnitude by $B\simeq e^{-\lambda_{0}t/2}\hbar^{J/2}$
(This appeared in the proof of Proposition \ref{pro:Tn_semi_xnt}).
With an heuristic point of view, if we think that these numbers behave
as independent random variables, then the total error is estimated
(by the central limit theorem) to be of order $\sqrt{\mathcal{N}_{t}}B\simeq\hbar^{J/2}$
as the numerical calculations shows. It would a nice result to explain
such a behaviour.
\end{itemize}

\section{\label{sec:Conclusion}Conclusion}

The main result of this paper is the use of semiclassical non stationary
normal form description of the hyperbolic dynamics first introduced
by David DeLatte \cite{delatte_92}, and which gives some invariant
semi-classical cocycles of the hyperbolic dynamics. With them, we
obtain semi-classical expressions for long time of order $t\simeq C\log\left(1/\hbar\right)$,
with any $C>0$. These cocycles are a generalisation to all order
in non linearities and in power of $\hbar$ of the Lyapounov cocycle
which controls the linear instability of the dynamics. These cocycles
allow us to have a control on the error term of the trace formula
(or other semi-classical formula) for long time. We could ask the
question if this Non-stationary Normal description presented in Section
\ref{sec:Semi-classical-De-Latte-non-stationnary} is really necessary?
the author believes that it is, because it gives Eq.(\ref{eq:quantum_non_stat_normal_form})
where $\mathcal{T}_{x}$ depends on the point $x$ and not on the
trajectory $x\left(t\right)=M^{t}x$. Without this result, we could
expect that the Normal Form construction would depend on the trajectory
$x\left(t\right)$ as a whole, and could diverge with $t$ in an uncontrollable
way. We would be then blocked.

There are many perspectives suggested by this work. Numerical calculations
presented in Section \ref{sub:Numerical-results-and} seem to show
that our bounds are not sharp, and that the actual errors are much
smaller. In particular, Gutzwiller trace formula at the leading order
$J=2$, i.e. without semi-classical normal forms corrections, could
be correct for these {}``long time'', with an error $\mathcal{O}\left(\hbar\right)$
\emph{}uniform in time. To understand this surprising fact we should
understand the way all the complex semi-classical contributions add
and compensate together. We think that the Ruelle-Pollicott thermodynamical
formalism of transfer operators in the context of prequantum dynamics,
could help us in that direction. Some preliminary results in this
direction are obtained in \cite{fred-PreQ-06}, in the case of a linear
hyperbolic map, and we hope to be able to extend these results to
non linear maps.

An other perspective is a generalization of this approach to a uniform
hyperbolic flow, like the geodesic flow on a negative curvature manifold.
We think that the approach of {}``semi-classical non stationary normal
forms'' developed in appendix \ref{sec:Proof-of-theorem} could be
generalized for such models. A similar but less obvious generalization
could be done for maximal hyperbolic sets, situations which are met
in general mixed chaotic systems, with homoclinic intersections of
stable/unstable manifolds giving horseshoes \cite{katok_hasselblatt}.

\appendix

\section{\label{sec:Proof-of-Theorem_1}Proof of Theorem \ref{pro:voisinage_trajectoire}
page \pageref{pro:voisinage_trajectoire} on semi-classical evolution
in a neighborhood of a classical trajectory for long time}

We first introduce some notations specific to this appendix.

\subsection{Some notations}

Let $D\subset\mathbb{R}^{2}$ (a measurable set). We define:\[
\hat{P}_{D}\defi\int_{x\in D}\frac{d^{2}x}{2\pi\hbar}|x\rangle\langle x|\]
where $|x\rangle$ is a coherent state defined by Eq. (\ref{e:|z>}).
One has the closure relation:\begin{equation}
\hat{P}_{D}+\hat{P}_{\mathbb{R}^{2}\backslash D}=\hat{Id}\label{eq:closure}\end{equation}

\begin{proof}
$\hat{Id}=\int_{\mathbb{R}^{2}}\frac{d^{2}x}{2\pi\hbar}|x\rangle\langle x|=\int_{D}\frac{d^{2}x}{2\pi\hbar}|x\rangle\langle x|+\int_{\mathbb{R}^{2}\backslash D}\frac{d^{2}x}{2\pi\hbar}|x\rangle\langle x|$.
\end{proof}
Let $0<\beta<1/2$ (The sharper results will correspond to $\beta$
close to $0$). Let $D\subset\mathbb{R}^{2}$ a measurable set, or
a sequence of sets $D_{\hbar}$ which depend on $\hbar$, with $\hbar\rightarrow0$.
In particular $D$ may be a point $D=\left\{ x\right\} $, $x\in\mathbb{R}^{2}$. 

\begin{defn}
\label{def:thickening}Let's define the \textbf{{}``exterior domain}''
$E\left(D\right)$, by: \begin{equation}
\boxed{E\left(D\right)=\left\{ x\in\mathbb{R}^{2}\,\mbox{such that}\,\textrm{dist}\left(x,D\right)>\hbar^{1/2-\beta}\right\} }\label{eq:def-Exclusive}\end{equation}
and its complementary, the {}``\textbf{interior}'' $I\left(D\right)$\[
\boxed{I\left(D\right)=\mathbb{R}^{2}\backslash E\left(D\right)}\]
Remarks: $I\left(D\right)$ is just the domain $D$ which has been
thickened by $\hbar^{1/2-\beta}$. In particular $D\subset I\left(D\right)$.
 In Eq.(\ref{eq:def_PD}), we have defined $\hat{P}_{x,\beta}$, which
in the present notation is $\hat{P}_{I\left\{ x\right\} }$.
\end{defn}
In this appendix, the classical dynamics is the map $M:\mathbb{R}^{2}\rightarrow\mathbb{R}^{2}$
defined in Section \ref{sec:Quantum-map}, and the quantum map $\hat{M}$
has been defined in Eq.(\ref{eq:map_M_hat}).

\subsection{Semi-classical evolution after finite time}

The results of this appendix will rely on the following lemma which
is a quite standard result on semi-classical evolution of wave packets
after finite time (with respect to $\hbar\rightarrow0$).

\begin{lem}
\label{lem:localization_wave_packet}For any $K\in\mathbb{N}$, and
$\hbar$ small enough, there exists $C>0$, such that for any $x\in\mathbb{R}^{2}$,
and any domain $D\subset\mathbb{R}^{2}$ such that $D\subset E\left(Mx\right)$,\begin{equation}
\left|\hat{P}_{D}\hat{M}|x\rangle\right|_{L^{2}}<C\hbar^{K}\label{eq:inequality_1}\end{equation}
where $|x\rangle$ is a coherent state defined in Eq.(\ref{e:|z>}).
\end{lem}

\paragraph{Remarks}

\begin{itemize}
\item In all this appendix, we will use this result, and write: $\varepsilon=C\hbar^{K}$.
At the end, $K$ will be chosen large enough. In common notations,
we can write that $\left|\hat{P}_{D}\hat{M}|x\rangle\right|_{L^{2}}=\mathcal{O}\left(\hbar^{\infty}\right)$
uniformly over $x\in\mathbb{R}^{2}$ and $D\subset E\left(Mx\right)$.
Lemma \ref{lem:localization_wave_packet} means that the evolved state
$\hat{M}|x\rangle$ is {}``localized'' at point $Mx$ at order $\mathcal{O}\left(\hbar^{\infty}\right)$.
In our case, the uniformity in $x$ is due to the fact that the dynamics
is defined on a torus (compact space).
\item Lemma \ref{lem:localization_wave_packet} follows directly from more
precise results obtained by A. Joye and G. Hagedorn in ref. \cite{joye-00}
(theorem 3.2), or M. Combescure and D. Robert \cite{combescure-97},
theorem 3.1, or A. Iantchenko \cite{ianchenko-98}, Lemma 5.
\end{itemize}
\begin{cor}
\textbf{\label{pro:basic_3}} Suppose that the domain $D\subset\mathbb{R}^{2}$
has finite measure $\mbox{S}\left(D\right)\defi\int_{x\in D}dx$.
Then for any domain $D'\subset E\left(M\left(D\right)\right)$\begin{equation}
\left\Vert \hat{P}_{D'}\hat{M}\hat{P}_{D}\right\Vert _{L^{2}}<\left(\frac{\mbox{S}\left(D\right)}{2\pi\hbar}\right)^{1/2}\varepsilon\label{eq:basic_3}\end{equation}

\end{cor}
(We don't think that the dependence on $D$ of the right hand side
is sharp).

\begin{proof}
Let $\psi\in L^{2}\left(\mathbb{R}\right)$ normalized, then\begin{eqnarray*}
\left\Vert \hat{P}_{D'}\hat{M}\hat{P}_{D}\psi\right\Vert _{L^{2}} & = & \left\Vert \hat{P}_{D'}\hat{M}\int_{x\in D}\frac{d^{2}x}{2\pi\hbar}|x\rangle\langle x|\psi\rangle\right\Vert _{L^{2}}\\
 & \leq & \int_{D}\frac{d^{2}x}{2\pi\hbar}\left|\langle x|\psi\rangle\right|\left\Vert \hat{P}_{D'}\hat{M}|x\rangle\right\Vert _{L^{2}}\\
 & \leq & \varepsilon\int_{D}\frac{d^{2}x}{2\pi\hbar}\left|\langle x|\psi\rangle\right|\leq\left(\frac{\mbox{S}\left(D\right)}{2\pi\hbar}\right)^{1/2}\varepsilon\end{eqnarray*}
In the last line we have used Eq.(\ref{eq:inequality_1}) because
$D'\subset E\left(M\left(D\right)\right)\subset E\left(Mx\right)$
if $x\in D$. We have also used Cauchy-Schwarz inequality.
\end{proof}

\subsection{Long time Semi-classical evolution}

Let $D\subset\mathbb{R}^{2}$ be a set at time $t=0$, with finite
measure. Let \[
I_{0}\left(D\right)\defi D\]

\begin{figure}[tbph]
\begin{centering}\scalebox{0.8}[0.8]{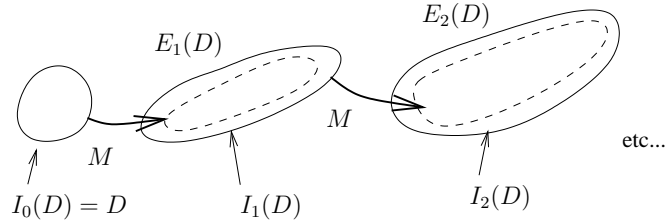}\par\end{centering}

\caption{\label{cap:Description-of-sets-E-I}Description of the sets $E_{t}\left(D\right)$
and $I_{t}\left(D\right)$, constructed inductively from $D$, by
evolution and thickening.}
\end{figure}

And for any $t\geq1$,\[
E_{t}\left(D\right)\defi E\left(MI_{t-1}\right),\qquad I_{t}\left(D\right)\defi\mathbb{R}^{2}\backslash E_{t}\left(D\right)\]
$I_{t}\left(D\right)$ is just obtained from $D$, by evolution and
thickening at each step, see figure \ref{cap:Description-of-sets-E-I}.

\begin{lem}
\label{pro:basic_4} For any time $t\geq1$:\begin{equation}
\left\Vert \hat{P}_{E_{t}\left(D\right)}\hat{M}^{t}\hat{P}_{D}\right\Vert _{L^{2}}<\varepsilon\left(\sum_{s=0}^{t-1}\left(\frac{\mbox{S}\left(I_{s}\left(D\right)\right)}{2\pi\hbar}\right)^{1/2}\right)\label{eq:basic_4}\end{equation}

\end{lem}
\begin{proof}
From Eq.(\ref{eq:closure}), we get $\hat{\mbox{Id}}=\hat{P}_{E_{t}\left(D\right)}+\hat{P}_{I_{t}\left(D\right)}$
, and $E_{t}\left(D\right)=E_{1}\left(I_{t-1}\left(D\right)\right)$.
Eq.(\ref{eq:basic_3}) gives for any $t$: \[
\left\Vert \hat{P}_{E_{1}\left(D\right)}\hat{M}\hat{P}_{D}\right\Vert _{L^{2}}<\left(\frac{\mbox{S}\left(D\right)}{2\pi\hbar}\right)^{1/2}\varepsilon\]
where $\varepsilon$ does not depend on $t\in\mathbb{R}$, and $D$.
By induction on time $t$ we get: \begin{eqnarray*}
\left\Vert \hat{P}_{E_{t}\left(D\right)}\hat{M}^{t}\hat{P}_{D}\right\Vert _{L^{2}} & = & \left\Vert \hat{P}_{E_{t}\left(D\right)}\hat{M}\left(\hat{P}_{E_{t-1}\left(D\right)}+\hat{P}_{I_{t-1}\left(D\right)}\right)\hat{M}^{t-1}\hat{P}_{D}\right\Vert _{L^{2}}\\
 & \leq & \left\Vert \hat{P}_{E_{t}\left(D\right)}\hat{M}\left(\hat{P}_{E_{t-1}\left(D\right)}\hat{M}^{t-1}\hat{P}_{D}\right)\right\Vert _{L^{2}}+\left\Vert \left(\hat{P}_{E_{t}\left(D\right)}\hat{M}\hat{P}_{I_{t-1}\left(D\right)}\right)\hat{M}^{t-1}\hat{P}_{D}\right\Vert _{L^{2}}\\
 & \leq & \varepsilon\left(\sum_{s=0}^{t-2}\left(\frac{\mbox{S}\left(I_{s}\left(D\right)\right)}{2\pi\hbar}\right)^{1/2}\right)+\varepsilon\left(\frac{\mbox{S}\left(I_{t-1}\left(D\right)\right)}{2\pi\hbar}\right)^{1/2}\end{eqnarray*}
where we have used: $\left|A+B\right|_{L^{2}}\leq\left|A\right|_{L^{2}}+\left|B\right|_{L^{2}}$,
$\left|AB\right|_{L^{2}}\leq\left|A\right|_{L^{2}}\left|B\right|_{L^{2}}$,
and $\left|\hat{P}_{D}\right|_{L^{2}}\leq1$, $\left|\hat{M}\right|_{L^{2}}=1$.
\end{proof}

\subsubsection{A particular application}

We consider now a particular application of the previous results,
which will be useful for hyperbolic dynamics. Let $t\geq1$, and suppose
that $D_{0},D_{1},\ldots,D_{t}$ is a sequence of sets. For any $s\in\left[1,t-1\right]$,
let \[
G_{s}\defi I_{1}\left(D_{s-1}\right)\backslash D_{s}\]
which corresponds to points coming from $D_{s-1}$ but not belonging
to $D_{s}$. 

\begin{figure}[tbph]
\begin{centering}\scalebox{0.8}[0.8]{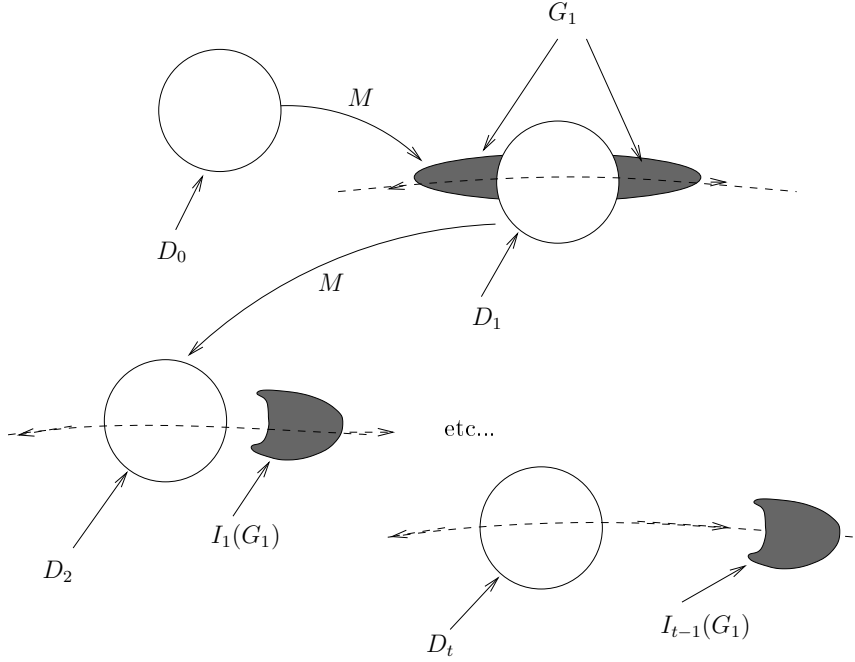}\par\end{centering}

\caption{\label{cap:schema_D_G_F}Description of the sets $I_{t-s}\left(G_{s}\right)$
used in Lemma \ref{lem:insert_PD}.}
\end{figure}

\begin{lem}
\label{lem:insert_PD}Suppose that for every $s\in\left[1,t-1\right]$,
\[
I_{t-s}\left(G_{s}\right)\cap D_{t}=\emptyset\]
which means that {}``points which leave the domain $D_{s}$ do not
come back to the domain $D_{t}$'' (despite the thickening procedure),
see Figure \ref{cap:schema_D_G_F}. Then\begin{eqnarray*}
\left\Vert \hat{P}_{D_{t}}\hat{M}^{t}\hat{P}_{D_{0}}-\hat{P}_{D_{t}}\hat{M}\hat{P}_{D_{t-1}}\ldots\hat{M}\hat{P}_{D_{1}}\hat{M}\hat{P}_{D_{0}}\right\Vert _{L^{2}}\\
\leq\frac{\varepsilon}{\sqrt{2\pi\hbar}}\sum_{s=1}^{t-1}\left(\left(\mbox{S}\left(D_{s}\right)\right)^{1/2}+\sum_{k=0}^{t-s-1}\left(\mbox{S}\left(I_{k}\left(G_{s}\right)\right)\right)^{1/2}\right)\end{eqnarray*}

\end{lem}

\begin{proof}
For any $s\in\left[1,t-1\right]$, let \[
F_{s}\defi E_{1}\left(D_{s-1}\right)\backslash D_{s},\qquad G_{s}\defi I_{1}\left(D_{s-1}\right)\backslash D_{s}\]
So that $\mathbb{R}^{2}=D_{s}\cup F_{s}\cup G_{s}$ as a disjoint
union. The hypothesis $I_{t-s}\left(G_{s}\right)\cap D_{t}=\emptyset$
means that $D_{t}\subset E_{t-s}\left(G_{s}\right)$. Then Eq.(\ref{eq:basic_4})
gives,\[
\left|\hat{P}_{D_{t}}\hat{M}^{t-s}\hat{P}_{G_{s}}\right|_{L^{2}}<\varepsilon\left(\sum_{k=0}^{t-s-1}\left(\frac{\mbox{S}\left(I_{k}\left(G_{s}\right)\right)}{2\pi\hbar}\right)^{1/2}\right)\]

Also, $F_{s}\subset E_{1}\left(D_{s-1}\right)$, so Eq.(\ref{eq:basic_3})
gives:\[
\left|\hat{P}_{F_{s}}\hat{M}\hat{P}_{D_{s-1}}\right|_{L^{2}}<\left(\frac{\mbox{S}\left(D_{s}\right)}{2\pi\hbar}\right)^{1/2}\varepsilon\]
Now we proceed by induction on $s$. We use $\hat{\mbox{Id}}=\hat{P}_{D_{1}}+\hat{P}_{F_{1}}+\hat{P}_{G_{1}}$
(from the disjoint union $\mathbb{R}^{2}=D_{s}\cup F_{s}\cup G_{s}$),
and write\[
\hat{P}_{D_{t}}\hat{M}^{t}\hat{P}_{D_{0}}=\hat{P}_{D_{t}}\hat{M}^{t-1}\left(\hat{P}_{D_{1}}+\hat{P}_{F_{1}}+\hat{P}_{G_{1}}\right)\hat{M}\hat{P}_{D_{0}}\]
so\begin{align*}
\left\Vert \hat{P}_{D_{t}}\hat{M}^{t}\hat{P}_{D_{0}}-\hat{P}_{D_{t}}\hat{M}^{t-1}\hat{P}_{D_{1}}\hat{M}\hat{P}_{D_{0}}\right\Vert _{L^{2}} & \leq\left\Vert \hat{P}_{D_{t}}\hat{M}^{t-1}\left(\hat{P}_{F_{1}}\hat{M}\hat{P}_{D_{0}}\right)\right\Vert _{L^{2}}+\left\Vert \left(\hat{P}_{D_{t}}\hat{M}^{t-1}\hat{P}_{G_{1}}\right)\hat{M}\hat{P}_{D_{0}}\right\Vert _{L^{2}}\\
 & \leq\varepsilon\left(\frac{\mbox{S}\left(D_{1}\right)}{2\pi\hbar}\right)^{1/2}+\varepsilon\left(\sum_{k=0}^{t-2}\left(\frac{\mbox{S}\left(I_{k}\left(G_{1}\right)\right)}{2\pi\hbar}\right)^{1/2}\right)\end{align*}

Similarly,\begin{eqnarray*}
\left\Vert \hat{P}_{D_{t}}\hat{M}^{t-1}\hat{P}_{D_{1}}\hat{M}\hat{P}_{D_{0}}-\hat{P}_{D_{t}}\hat{M}^{t-2}\hat{P}_{D_{2}}\hat{M}\hat{P}_{D_{1}}\hat{M}\hat{P}_{D_{0}}\right\Vert _{L^{2}} & \leq & \varepsilon\left(\frac{\mbox{S}\left(D_{2}\right)}{2\pi\hbar}\right)^{1/2}\\
 &  & +\varepsilon\left(\sum_{k=0}^{t-3}\left(\frac{\mbox{S}\left(I_{k}\left(G_{2}\right)\right)}{2\pi\hbar}\right)^{1/2}\right)\end{eqnarray*}

and therefore after few similar steps\begin{eqnarray*}
\left\Vert \hat{P}_{D_{t}}\hat{M}^{t}\hat{P}_{D_{0}}-\hat{P}_{D_{t}}\hat{M}\hat{P}_{D_{t-1}}\ldots\hat{P}_{D_{1}}\hat{M}\hat{P}_{D_{0}}\right\Vert _{L^{2}} & \leq & \varepsilon\sum_{s=1}^{t-1}\left(\frac{\mbox{S}\left(D_{s}\right)}{2\pi\hbar}\right)^{1/2}\\
 &  & +\varepsilon\left(\sum_{s=1}^{t-1}\sum_{k=0}^{t-s-1}\left(\frac{\mbox{S}\left(I_{k}\left(G_{s}\right)\right)}{2\pi\hbar}\right)^{1/2}\right)\end{eqnarray*}

\end{proof}

\subsection{Application to hyperbolic dynamics }

\subsubsection{\label{sub:Proof-1}Proof of theorem \ref{pro:voisinage_trajectoire}}

The preceding results were rather general. We will apply them to the
hyperbolic dynamics considered in this paper. In order to prove theorem
\ref{pro:voisinage_trajectoire}, we consider a trajectory $x\left(s\right)=M^{s}\left(x\left(0\right)\right)$,
$s=1\ldots t$, and disks $D_{s}$ of radius $\hbar^{1/2-\alpha}$
at each point $x\left(s\right)$. Using definitions introduced with
Lemma \ref{lem:insert_PD} page \pageref{lem:insert_PD}, we remark
that the hypothesis $I_{t-s}\left(G_{s}\right)\cap D_{t}=\emptyset$
is satisfied, because if $\beta<\alpha$, and due to hyperbolic instability
of the trajectory $x\left(s\right)$, the dynamics sends the domain
$G_{s}$ away from the domain $D_{t}$, in spite of the thickening
(see figure \ref{cap:schema_D_G_F}). We have $\mbox{S}\left(D_{s}\right)\simeq\mathcal{O}\left(\hbar^{1/2-\alpha}\right)^{2}$
but the surface $\mbox{S}\left(I_{k}\left(G_{s}\right)\right)$ grows
exponentially fast with $k$ (in the unstable direction) because of
the thickening at each time step. A rough estimate gives:\[
\mbox{S}\left(I_{k}\left(G\right)\right)=\mathcal{O}\left(\left(e^{\lambda_{max}k}\hbar^{1/2-\alpha}\right)\left(\hbar^{1/2-\alpha}\right)\right)\]
Suppose that $0<t<C\log\left(1/\hbar\right)\Leftrightarrow e^{\lambda_{max}t}<\hbar^{-C\lambda_{max}}$.
Then Lemma \ref{lem:insert_PD}, and $\varepsilon=C\hbar^{K}$, gives
\begin{eqnarray*}
\left\Vert \hat{P}_{D_{t}}\hat{M}^{t}\hat{P}_{D_{0}}-\hat{P}_{D_{t}}\hat{M}\hat{P}_{D_{t-1}}\ldots\hat{M}\hat{P}_{D_{1}}\hat{M}\hat{P}_{D_{0}}\right\Vert _{L^{2}}\\
\leq\frac{C\hbar^{K}}{\sqrt{2\pi\hbar}}\hbar^{1/2-\alpha}\sum_{s=1}^{t-1}\left(1+\sum_{k=0}^{t-s-1}e^{\lambda_{max}k/2}\right)\\
\leq\left(\frac{C}{e^{\lambda_{max}/2}-1}\right)\hbar^{K}\hbar^{-\alpha}\hbar^{-C\lambda_{max}/2}t\end{eqnarray*}

For any $K'\geq1$, the bound is $\mathcal{O}\left(\hbar^{K'}\right)$
if $K$ is chosen large enough. So we get Eq.(\ref{eq:voisinage_trajectoire}).

\subsubsection{An other consequence}

\begin{lem}
Let $0<\alpha<1/2$. For any $C>0$, any $t$, such that $\left|t\right|<C\log\left(1/\hbar\right)$,
for any $n\in\mathbb{Z}^{2}$, any $x\in\mathbb{R}^{2}$, such that
$\left|x-x_{n,t}\right|>\hbar^{1/2-\alpha}$ (which means that $x$
is at some distance of the hyperbolic fixed point $x_{n,t}$ of the
hyperbolic map $M_{n,t}$, Eq.(\ref{eq:classical_M_nt})) then\begin{equation}
\langle x|\hat{M}_{n,t}|x\rangle=\mathcal{O}\left(\hbar^{\infty}\right)\label{eq:x_Mnt_x}\end{equation}
(uniformly other $x\in\mathbb{R}^{2},n,t$), where $\hat{M}_{n,t}=\hat{T}_{-n}\hat{M}^{t}$
has been defined in Eq.(\ref{eq:Mnt}). A light generalization is
\begin{equation}
\langle x'|\hat{M}_{n,t}|x\rangle=\mathcal{O}\left(\hbar^{\infty}\right)\label{eq:x_Mnt_x_bis}\end{equation}
if $\left|x-x_{n,t}\right|>\hbar^{1/2-\alpha}$ and $\left|x-x'\right|<\hbar^{1/2-\alpha'}$,
with $0<\alpha'<\alpha$. This implies:\begin{equation}
\mbox{Tr}\left(\hat{P}_{x_{n,t},\alpha}\hat{M}_{n,t}\hat{P}_{x_{n,t},\alpha}\right)=\mbox{Tr}\left(\hat{M}_{n,t}\hat{P}_{x_{n,t},\alpha}\right)+\mathcal{O}\left(\hbar^{\infty}\right)\label{eq:Tr_PMP_et_MP}\end{equation}
where $\hat{P}_{x_{n,t},\alpha}$ is defined in Eq.(\ref{eq:def_PD}).
\end{lem}
\begin{proof}
Let $0<\gamma<\alpha'$, and let $D'_{x}$ be the disk of radius $\hbar^{1/2-\gamma}$
with center $x$. With the hypothesis $\left|x-x_{n,t}\right|>\hbar^{1/2-\alpha}$,
and $\left|x-x'\right|<\hbar^{1/2-\alpha'}$, one checks that for
$0<\beta<\gamma$, then $D'_{x'}\subset E_{t}\left(D'_{x}\right)$.
Then from Lemma \ref{pro:basic_4}, and similar estimates as in Section
\ref{sub:Proof-1}, one gets $\left|\hat{P}_{D'_{x'}}\hat{M}_{n,t}\hat{P}_{D'_{x}}\right|=\mathcal{O\left(\hbar^{\infty}\right)}$.
This implies that $\langle x'|\hat{M}_{n,t}|x\rangle=\mathcal{O}\left(\hbar^{\infty}\right)$.
To show Eq.(\ref{eq:Tr_PMP_et_MP}), write $\mbox{Tr}\left(\hat{M}_{n,t}\hat{P}_{x_{n,t},\alpha}\right)=\mbox{Tr}\left(\hat{P}_{x_{n,t},\alpha}\hat{M}_{n,t}\hat{P}_{x_{n,t},\alpha}\right)+\mbox{Tr}\left(\left(1-\hat{P}_{x_{n,t},\alpha}\right)\hat{M}_{n,t}\hat{P}_{x_{n,t},\alpha}\right)$.
But\[
\mathcal{I}\defi\mbox{Tr}\left(\left(1-\hat{P}_{x_{n,t},\alpha}\right)\hat{M}_{n,t}\hat{P}_{x_{n,t},\alpha}\right)=\int_{\left|x\right|<\hbar^{1/2-\alpha}}\int_{\left|x'\right|>\hbar^{1/2-\alpha}}\frac{dx\, dx'}{\left(2\pi\hbar\right)^{2}}\langle x|x'\rangle\langle x'|\hat{M}_{n,t}|x\rangle\]
Using the exponential decrease of $\left|\langle x|x'\rangle\right|=C\exp\left(-\left|x-x'\right|^{2}/\hbar\right)$,
and $\left|\langle x'|\hat{M}_{n,t}|x\rangle\right|\leq1$, one first
obtains: \[
\mathcal{I}=\int_{\hbar^{1/2-\alpha_{1}}<\left|x\right|<\hbar^{1/2-\alpha}}\int_{\left|x'-x\right|<\hbar^{1/2-\alpha'}}\frac{dx\, dx'}{\left(2\pi\hbar\right)^{2}}\langle x|x'\rangle\langle x'|\hat{M}_{n,t}|x\rangle+\mathcal{O}\left(\hbar^{\infty}\right)\]
with $0<\alpha_{1},\alpha'<\alpha$, and using Eq.(\ref{eq:x_Mnt_x_bis}),
one finally obtains $\mathcal{I}=\mathcal{O}\left(\hbar^{\infty}\right)$,
and deduces Eq.(\ref{eq:Tr_PMP_et_MP}).
\end{proof}

\section{\label{sec:Proof-of-Lemma_trace}Proof of Lemma \ref{lem:integrale_localisee_trace}
page \pageref{lem:integrale_localisee_trace} for the trace of an
individual orbit for long time}

\subsection{General results to control localization of wave packets }

We first give some results which will allow us to control the escape
towards infinity in $\mathbb{R}^{2}$ of the phase-space distribution
of wave packets after long time, under a hyperbolic dynamics. The
usual semiclassical Egorov Theorem which describes the propagation
of quantum observables does not work after a fixed constant times
the Ehrenfest time ($\frac{1}{2}t_{E}$ in our case, see Section \ref{sub:Characteristic-time-in}).
The aim of the Section is to develop a weaker description which will
be able to reach $t=C\log\left(1/\hbar\right)$, with any $C>0$.
Remind that the classical phase space is $x=\left(q,p\right)\in\mathbb{R}^{2}$.
The quantum Hilbert space is $\mathcal{H}_{plane}=L^{2}\left(\mathbb{R}\right)$
with Planck constant $\hbar$. 

Let $m$ be an order function which possibly decreases at infinity
(\cite{dimassi-99} p. 81, \cite{martinez-01} p. 12, \cite{zworski-03}
p. 52). 

\begin{defn}
\label{def:A-weigth-function}A \textbf{weight function} is a semi-classical
symbol $W:\mathbb{R}^{2}\rightarrow\mathbb{R}^{+,*}$ , $W\in S\left(m\right)$,
such that $\left|W\right|_{\infty}=\mbox{max}_{x}\left(W\left(x\right)\right)=1$,
and $W$ is elliptic.

Remark: $W$ Elliptic means $\left|W\right|\geq\frac{1}{C}m$, with
$C>0$, then $W^{-1}\in S\left(m^{-1}\right)$, see \cite{dimassi-99}
p.100, \cite{martinez-01} p. 13. 
\end{defn}
Example: $W\left(x\right)=m\left(x\right)=1/\left\langle x\right\rangle ^{k}$
with $\left\langle x\right\rangle =\sqrt{1+q^{2}+p^{2}}$, $k\geq1$,
or $W\left(x\right)=m\left(x\right)=\exp\left(-\left\langle x\right\rangle \right)$.

We write $\hat{W}=\mbox{Op}_{Weyl}\left(W\right)$. We check that
these operators have dense domain on $L^{2}\left(\mathbb{R}\right)$.

\begin{defn}
\label{def:psi-W-localized}If $W$ is a weight function, a sequence
of states $\psi_{\hbar}$, with $\hbar\rightarrow0$, is \textbf{W-localized}
if $\left\Vert \psi_{\hbar}\right\Vert =1$ and if there exists $C>0$
(independent of $\hbar$) such that\[
\left\Vert \hat{W}^{-1}\psi_{\hbar}\right\Vert _{L^{2}}\leq C\]

\end{defn}
The intuitive idea of this definition, is that a W-localized quantum
state has a distribution in phase space which is bounded by the function
$W\left(x\right)$. This definition is very similar with weight functions
used to treat tunnelling effect.

Remind that the dynamics considered in this paper is the map $M:\mathbb{R}^{2}\rightarrow\mathbb{R}^{2}$,
defined in Eq.(\ref{eq:map_M}) as a product $M=M_{1}M_{0}$, with
$M_{0}\in\mbox{SL}(2,\mathbb{R})$ linear hyperbolic and $M_{1}$
bounded. We also defined $\hat{M}$ in Eq.(\ref{eq:map_M_hat}).

\begin{prop}
\label{pro:W-Wt}Suppose that $W_{0}\in S\left(m_{0}\right)$, $W_{1}\in S\left(m_{1}\right)$
are weight functions such that\[
W_{1}\left(x\right)\geq W_{0}\left(M^{-1}\left(x\right)\right),\qquad\forall x\in\mathbb{R}^{2}\]
and $m_{1}\left(x\right)\geq m_{0}\left(M_{0}^{-1}\left(x\right)\right),\forall x$.
Suppose that $\psi_{0}$ is $W_{0}$-localized. Then\[
\psi_{1}=\hat{M}\psi_{0}\]
is $W_{1}$-localized.
\end{prop}
\begin{proof}
Write\[
\left\Vert \hat{W}_{1}^{-1}\psi_{1}\right\Vert =\left\Vert \hat{W}_{1}^{-1}\hat{M}\psi_{0}\right\Vert =\left\Vert \hat{W}_{1}^{-1}\hat{M}\hat{W}_{0}\hat{W}_{0}^{-1}\psi_{0}\right\Vert \]

From Egorov theorem (\cite{dimassi-99} p.125., \cite{martinez-01}
page 138, \cite{zworski-03} p.139) \[
\hat{M}\hat{W}_{0}\hat{M}^{-1}=\hat{W}'+\hbar\hat{R}\]
where $\hat{W}'=Op\left(W'\right)$ and $W'$ has Weyl symbol $W'\left(x\right)=W_{0}\left(M^{-1}\left(x\right)\right)$,
and belongs to $S\left(m'\right)$, with order function $m'=m\circ M_{0}^{-1}$.
The remainder $\hat{R}$ has Weyl symbol $R\in S\left(m'\right)$.

Then \begin{align*}
\left\Vert \hat{W}_{1}^{-1}\hat{M}_{1}\hat{W}_{0}\hat{W}_{0}^{-1}\psi_{0}\right\Vert  & =\left\Vert \hat{W}_{1}^{-1}\left(\hat{W}'+\hbar\hat{R}\right)\hat{M}\hat{W}_{0}^{-1}\psi_{0}\right\Vert \\
 & \leq\left\Vert \hat{W}_{1}^{-1}\left(\hat{W}'+\hbar\hat{R}\right)\right\Vert \left\Vert \hat{M}\right\Vert \left\Vert \hat{W}_{0}^{-1}\psi_{0}\right\Vert \end{align*}
By hypothesis, $W_{1}\geq W'\Leftrightarrow\left|W_{1}^{-1}W'\right|_{\infty}\leq1$,
and $W_{1}^{-1}\left(W'+\hbar R\right)\in S\left(m_{1}^{-1}m'\right)=S\left(1\right)$.
By Calderon-Vaillancourt Theorem (\cite{dimassi-99} p.85, \cite{martinez-01}
page 43, \cite{zworski-03} p.69), $\left\Vert \hat{W}_{1}^{-1}\left(\hat{W}'+\hbar\hat{R}\right)\right\Vert \leq1+\mathcal{O}\left(\hbar\right)$.
Also $\left\Vert \hat{M}\right\Vert =1$, and $\left\Vert \hat{W}_{0}^{-1}\psi_{0}\right\Vert \leq C$.
We get:\[
\left\Vert \hat{W}_{1}^{-1}\psi_{1}\right\Vert \leq C.\left(1+\mathcal{O}\left(\hbar\right)\right)\]

\end{proof}
\begin{prop}
\label{pro:PSi_0_Psit_with_W}Suppose that $W_{0}$ , $W_{1}$ are
weight functions, and that $\psi_{0}$ is $W_{0}$-localized, and
$\psi_{1}$ is $W_{1}$-localized. Then there exists $C>0$ and $\hbar_{0}>0$
such that for any $\hbar<\hbar_{0}$:\[
\left|\langle\psi_{0}|\psi_{1}\rangle\right|\leq C.\left|W_{0}W_{1}\right|_{\infty}\]

\end{prop}
\begin{proof}
We suppose $\left\Vert \hat{W}_{0}^{-1}\psi_{0}\right\Vert \leq C_{0}$
and $\left\Vert \hat{W}_{1}^{-1}\psi_{1}\right\Vert \leq C_{1}$.
Then\begin{align*}
\left|\langle\psi_{0}|\psi_{1}\rangle\right| & =\left|\langle\psi_{0}|\hat{W}_{0}^{-1}\hat{W}_{0}\hat{W}_{1}\hat{W}_{1}^{-1}|\psi_{1}\rangle\right|\\
 & \leq\left\Vert \hat{W}_{0}^{-1}\psi_{0}\right\Vert \left\Vert \hat{W}_{1}^{-1}\psi_{1}\right\Vert \left\Vert \hat{W}_{0}\hat{W}_{1}\right\Vert \end{align*}
From Calderon-Vaillancourt Theorem, $\left\Vert \hat{W}_{0}\hat{W}_{1}\right\Vert \leq\left|W_{0}W_{1}\right|_{\infty}+\mathcal{O}\left(\hbar\right)$.
We get:\[
\left|\langle\psi_{0}|\psi_{1}\rangle\right|\leq C_{0}C_{1}\left|W_{0}W_{1}\right|_{\infty}+\mathcal{O}\left(\hbar\right)\]
But $\left|W_{0}W_{1}\right|_{\infty}\leq1$, so \[
\left|\langle\psi_{0}|\psi_{1}\rangle\right|\leq C.\left|W_{0}W_{1}\right|_{\infty}\]
with $C>0$.
\end{proof}

\subsection{Application}

The previous paragraph will allow us to obtain the following Lemma:

\begin{lem}
\label{lem:decroissance_x0_infini}There exists $R>0$, such that
for any $C>0$, $0<\alpha<1/2$, there exists $\hbar_{0}$, $C_{1}$,
such that for any $t\leq C/\hbar$, $\hbar<\hbar_{0}$, $\left|x_{0}\right|>R$,
one has \[
\left|\langle x_{0}|\hat{M}^{t}|x_{0}\rangle\right|\leq C_{1}e^{-\frac{\left(\left|x_{0}\right|-R\right)}{\hbar^{1/2-\alpha}}}\]
where $|x_{0}\rangle$ is a coherent state at position $x_{0}=\left(q_{0},p_{0}\right)\in\mathbb{R}^{2}$.
\end{lem}
This last results means that the auto-correlation function $\left|\langle x_{0}|\hat{M}^{t}|x_{0}\rangle\right|$
decreases exponentially fast as $x_{0}$ goes to infinity. This is
due to the hyperbolicity of the classical map $M$ on $\mathbb{R}^{2}$.
The proof is given below. We deduce:

\begin{cor}
\label{lem:integrale_disque_R}There exists $R>0$, $C>0$, such that
for any $1\leq t\leq C/\hbar$, $\hbar<\hbar_{0}$, one has\[
\int_{\left|x_{0}\right|>R}\langle x_{0}|\hat{M}^{t}|x_{0}\rangle\frac{dx_{0}}{2\pi\hbar}=\mathcal{O}\left(\hbar^{\infty}\right)\]
\begin{equation}
\int_{\left|x-x_{n,t}\right|>R}\langle x|\hat{M}_{n,t}|x\rangle\frac{dx}{2\pi\hbar}=\mathcal{O}\left(\hbar^{\infty}\right)\label{eq:Integrale_R}\end{equation}

\end{cor}
The second equation is a easy generalisation of the first one, where
we considered the map Eq.(\ref{eq:Mnt}). This last result improves
the result of convergence Eq.(\ref{eq:Integrale}), because it shows
that the integral is semiclassically negligible outside a disk of
\emph{fixed} radius $R$.

\subsubsection{Proof of Lemma \ref{lem:decroissance_x0_infini}}

In all the paragraph, let us denote by $x=\left(q,p\right)\in\mathbb{R}^{2}$
the unstable/stable coordinates in which the matrix $M_{0}$ , Eq.(\ref{eq:M0}),
is diagonal. $M_{0}:\left(q,p\right)\rightarrow\left(e^{\lambda_{0}}q,e^{-\lambda_{0}}p\right)$.

\paragraph{A first result on the classical map $M$:}

Remind that the perturbation $M_{1}$ is bounded: $\left|M_{1}\left(x\right)-x\right|\leq C$,
for any $x\in\mathbb{R}^{2}$. Without further assumption on $M_{1}$
(we don't assume here that $M=M_{1}M_{0}$ is hyperbolic), we have
the following simple result Lemma on the map $M=M_{1}M_{0}$:

\begin{lem}
\label{lem:escape_by_M}For any $\varepsilon$, such that $0<\varepsilon<\lambda_{0}$,
let $\lambda=\lambda_{0}-\varepsilon>0$, $R=\frac{Ce^{-\lambda_{0}}}{\left(1-e^{-\varepsilon}\right)}>0$.
For any point $x=\left(q,p\right)$, let $x'=\left(q',p'\right)=M\left(x\right)$.
If $q>R$, then $q'>e^{\lambda}q$. 
\end{lem}
This result guaranties that under the map $M$, points escape exponentially
fast towards infinity if $q$ is large enough. 

\begin{proof}
We have $x'=M_{1}M_{0}\left(x\right)$, so $\left|q'-e^{\lambda_{0}}q\right|\leq C$,
so $q'\geq e^{\lambda_{0}}q-C$. And $e^{\lambda_{0}}q-C>e^{\lambda}q\Leftrightarrow q>R$.
\end{proof}

\paragraph{Choice of Weight functions $W_{t}$:}

In order to apply later on Proposition \ref{pro:W-Wt}, let $x_{0}=\left(q_{0},p_{0}\right)\in\mathbb{R}^{2}$,
with $q_{0}>R$. Let $0<\alpha<1/2$ (by rescaling, we use semiclassical
calculus at the scale $\hbar^{1/2-\alpha}$, \cite{dimassi-99} p.82,
\cite{zworski-03} p.52). For any $t\in\mathbb{N}$, let us choose
a weight function $W_{t}\left(q\right)$ (independent of $p$ and
increasing in $q$), defined by $W_{t}\left(q\right)=1$, for $q\geq q_{0}e^{\lambda t}$,
$W_{t}\left(q\right)=\exp\left(\left(qe^{-\lambda t}-q_{0}\right)/\hbar^{1/2-\alpha}\right)$
for $e^{\lambda t}R<q<q_{0}e^{\lambda t}$, and $W_{t}\left(q\right)=W_{-\infty}\defi e^{-\left(q_{0}-R\right)/\hbar^{1/2-\alpha}}$
for $q<e^{\lambda t}R$. We smooth $W_{t}\left(q\right)$ in the vicinity
of $R$ and $q_{0}$ and let it be still increasing but $C^{\infty}$.
For any $t$, the function $W_{t}$ is a weight function according
to definition \ref{def:A-weigth-function}, but at the semiclassical
scale $\tilde{x}=x/\left(\hbar^{1/2-\alpha}\right)$ . From Lemma
\ref{lem:escape_by_M}, we check that it satisfies: 

\begin{lem}
\label{lem:Wt-Wt}For any $x=\left(q,p\right)\in\mathbb{R}^{2}$,
any $t\geq1$, then
\end{lem}
\[
W_{t}\left(x\right)\geq W_{t-1}\left(M^{-1}\left(x\right)\right)\]

~

\begin{lem}
Let $\psi_{0}=|x_{0}\rangle$ be a coherent state at position $x_{0}\in\mathbb{R}^{2}$,
with $q_{0}>R$. Then $\psi_{0}$ is $W_{0}$-localized (according
to definition \ref{def:psi-W-localized}), and for any $0\leq t\leq C/\hbar$,
$\psi_{t}\defi\hat{M}^{t}\psi_{0}$ is $W_{t}$-localized.
\end{lem}
\begin{proof}
$\psi_{0}$ is $W_{0}$-localized because as explained in Section
\ref{sub:Standard-coherent-states}, a Gaussian wave packet $\psi_{0}\left(q\right)$
is exponentially localized in $q$, around $q_{0}$. Then with Lemma
\ref{lem:Wt-Wt} and Lemma \ref{pro:W-Wt}, we deduce iteratively
that $\psi_{t}$ is $W_{t}$-localized for any $t$. We have to be
careful, and observe that in the proof of Lemma \ref{pro:W-Wt}, there
appears $\left\Vert \hat{W}_{t}^{-1}\psi_{t}\right\Vert \leq C_{t-1}.\left(1+\mathcal{O}\left(\hbar\right)\right)$.
Under iterations, this gives $C_{0}\left(1+\mathcal{O}\left(\hbar\right)\right)^{t}$,
which is still $\mathcal{O}\left(1\right)$ if $t\leq C/\hbar$. 
\end{proof}

\paragraph{Last step of the proof:}

Now let us consider another weight function $W'\left(q\right)$ (again
independent of $p$) defined by $W'\left(q\right)=1$ for $q<q_{0}$,
$W'\left(q\right)=\exp\left(-\left(q-q_{0}\right)/\hbar^{1/2-\alpha}\right)$
if $q>q_{0}$. We modify $W'$ in the vicinity of $q_{0}$, so that
it is $C^{\infty}$. As above, it is clear that the coherent state
$\psi_{0}=|x_{0}\rangle$ is $W'$-localized. Then we apply Proposition
\ref{pro:PSi_0_Psit_with_W} to deduce that $\left|\langle\psi_{0}|\psi_{t}\rangle\right|\leq C.\left|W'W_{t}\right|_{\infty}$.
Observe that $\left|W'W_{t}\right|_{\infty}=W_{-\infty}=e^{-\left(q_{0}-R\right)/\hbar^{1/2-\alpha}}$
so this gives\begin{equation}
\left|\langle x_{0}|\hat{M}^{t}|x_{0}\rangle\right|\leq Ce^{-\left(q_{0}-R\right)/\hbar^{1/2-\alpha}}\label{eq:Decroiss}\end{equation}
The same analysis can be done for $q_{0}<R$, and similarly for $\left|p_{0}\right|>R$
(in that last case, this is the stable direction, we have to work
in the past, writing $\langle x_{0}|\hat{M}^{t}|x_{0}\rangle=\langle\psi_{-t}|\psi_{0}\rangle$).
Finally we deduce Lemma \ref{lem:decroissance_x0_infini}.

\subsection{A more refined estimate}

Corollary \ref{lem:integrale_disque_R} is not precise enough to give
Lemma \ref{lem:integrale_localisee_trace} we are looking for. Indeed,
instead of a disk of finite radius $R$, we have to reduce the integral
Eq.(\ref{eq:Integrale}) to a disk of smaller radius $\hbar^{1/2-\alpha}$.
Let $0<\alpha<1/2$, and $t<C\log\left(1/\hbar\right)$, with $C>0$.
Decompose Eq.(\ref{eq:Integrale}) in three parts: \begin{align*}
\mbox{T}\left(\hat{M}_{n,t}\right)= & \int_{\left|x-x_{n,t}\right|<\hbar^{1/2-\alpha}}\langle x|\hat{M}_{n,t}|x\rangle\frac{dx}{2\pi\hbar}\\
 & +\int_{\hbar^{1/2-\alpha}<\left|x-x_{n,t}\right|<R}\langle x|\hat{M}_{n,t}|x\rangle\frac{dx}{2\pi\hbar}+\int_{R<\left|x-x_{n,t}\right|}\langle x|\hat{M}_{n,t}|x\rangle\frac{dx}{2\pi\hbar}\end{align*}

From Eq.(\ref{eq:Integrale_R}) and Eq.(\ref{eq:x_Mnt_x}), the last
two integrals are $\mathcal{O}\left(\hbar^{\infty}\right)$. One obtains
\[
T\left(\hat{M}_{n,t}\right)=\int_{x\in D_{x_{n,t}}}\frac{dx}{2\pi\hbar}\langle x|\hat{M}_{n,t}|x\rangle+\mathcal{O}\left(\hbar^{\infty}\right)=\mbox{Tr}\left(\hat{M}_{n,t}\hat{P}_{x_{n,t}}\right)+\mathcal{O}\left(\hbar^{\infty}\right)\]
where $\hat{P}_{x_{n,t}}$ is defined in Eq.(\ref{eq:def_PD}). Eq.(\ref{eq:Tr_PMP_et_MP})
gives $\mbox{Tr}\left(\hat{M}_{n,t}\hat{P}_{x_{n,t}}\right)=\mbox{Tr}\left(\hat{P}_{x_{n,t}}\hat{M}_{n,t}\hat{P}_{x_{n,t}}\right)+\mathcal{O}\left(\hbar^{\infty}\right)$.
Finally Lemma \ref{lem:integrale_localisee_trace} is proved.

\subsubsection{Estimates of Trace norm in terms of Operator norm}

We consider operators $\hat{P}_{x,\alpha}$ defined in Eq.(\ref{eq:def_PD}),
with $x=0$.

\begin{lem}
\label{lem:norme_1_et_norme_2} Suppose that $\hat{A}$ is bounded:
$\left\Vert \hat{A}\right\Vert _{L^{2}}=\mathcal{O}\left(1\right)$,
and $\left\Vert \hat{P}_{\alpha}\hat{A}\right\Vert _{L^{2}}=\mathcal{O}\left(\hbar^{K}\right)$
with $K\geq0$. Let $\alpha'>\alpha$. Then in Trace norm:\[
\left\Vert \hat{P}_{\alpha'}\hat{A}\right\Vert _{1}=\mathcal{O}\left(\hbar^{K-\alpha'}\right),\qquad\left\Vert \hat{P}_{\alpha'}\hat{A}\hat{P}_{\alpha'}\right\Vert _{1}=\mathcal{O}\left(\hbar^{K-\alpha'}\right)\]

Similarly if $\left\Vert \hat{B}\right\Vert _{L^{2}}=\mathcal{O}\left(1\right)$,
and $\left\Vert \hat{P}_{\alpha}\hat{B}\hat{P}_{\alpha}\right\Vert _{L^{2}}=\mathcal{O}\left(\hbar^{K}\right)$
then\begin{equation}
\left\Vert \hat{P}_{\alpha'}\hat{B}\hat{P}_{\alpha'}\right\Vert _{1}=\mathcal{O}\left(\hbar^{K-\alpha'}\right)\label{eq:estimate_norme_trace}\end{equation}

\end{lem}
\begin{proof}
We first give basic estimates:\begin{equation}
\left\Vert \hat{P}_{\alpha}\right\Vert _{1}=\int_{\left|x\right|<\hbar^{1/2-\alpha}}\frac{1}{2\pi\hbar}=\mathcal{O}\left(\hbar^{-2\alpha}\right)\label{eq:trace_P_alpha}\end{equation}
If $\alpha'>\alpha$,\[
\left\Vert \hat{P}_{\alpha}\left(1-\hat{P}_{\alpha'}\right)\right\Vert _{L^{2}}=\mathcal{O}\left(\hbar^{\infty}\right)\]
and\[
\left\Vert \hat{P}_{\alpha}\right\Vert _{L^{2}}<1\]
One has $\left\Vert \hat{P}_{\alpha'}\hat{P}_{\alpha}\hat{A}\right\Vert _{1}\leq\left\Vert \hat{P}_{\alpha'}\right\Vert _{1}\left\Vert \hat{P}_{\alpha}\hat{A}_{1}\right\Vert _{L^{2}}=\mathcal{O}\left(\hbar^{K-\alpha'}\right)$
(see \cite{gohberg-00}, prop 5.4 page 62). Write $1=1-\hat{P}_{\alpha}+\hat{P}_{\alpha}$,
and \[
\left\Vert \hat{P}_{\alpha'}\hat{A}\right\Vert _{1}\leq\left\Vert \hat{P}_{\alpha'}\hat{P}_{\alpha}\hat{A}\right\Vert _{1}+\left\Vert \hat{P}_{\alpha'}\left(1-\hat{P}_{\alpha}\right)\hat{A}\right\Vert _{1}\]
where \[
\left\Vert \hat{P}_{\alpha'}\left(1-\hat{P}_{\alpha}\right)\hat{A}\right\Vert _{1}\leq\left\Vert \hat{P}_{\alpha'}\left(1-\hat{P}_{\alpha}\right)\right\Vert _{1}\left\Vert \hat{A}\right\Vert _{L^{2}}\leq\left\Vert \hat{P}_{\alpha'}\left(1-\hat{P}_{\alpha}\right)\right\Vert _{L^{2}}\left\Vert \hat{A}\right\Vert _{L^{2}}=\mathcal{O}\left(\hbar^{\infty}\right)\]
Therefore $\left\Vert \hat{P}_{\alpha'}\hat{A}\right\Vert _{1}=\mathcal{O}\left(\hbar^{K-\alpha'}\right)$
and $\left\Vert \hat{P}_{\alpha'}\hat{A}\hat{P}_{\alpha'}\right\Vert _{1}\leq\left\Vert \hat{P}_{\alpha'}\hat{A}\right\Vert _{1}\left\Vert \hat{P}_{\alpha'}\right\Vert _{L^{2}}=\mathcal{O}\left(\hbar^{K-\alpha'}\right)$.
For the second estimate, let $\hat{A}=\hat{B}\hat{P}_{\alpha}$. This
gives $\left\Vert \hat{P}_{\alpha'}\hat{B}\hat{P}_{\alpha}\right\Vert _{1}=\left\Vert \hat{P}_{\alpha'}\hat{A}\right\Vert _{1}=\mathcal{O}\left(\hbar^{K-\alpha'}\right)$.
As above, one can show that $\left\Vert \hat{P}_{\alpha'}\hat{B}\hat{P}_{\alpha'}\right\Vert _{1}\leq\left\Vert \hat{P}_{\alpha'}\hat{B}\hat{P}_{\alpha}\right\Vert _{1}+\mathcal{O}\left(\hbar^{\infty}\right)$.
\end{proof}

\section{Useful results for finite time evolution}

In this appendix we collect well known results on semiclassical finite
time evolution. We state them in the context of the present paper.
They are used in other parts of the paper.

\subsection{Truncation of Taylor series of the Hamiltonian}

\selectlanguage{french}
\vspace{0.cm}\begin{center}\fbox{\parbox{16cm}{

\selectlanguage{english}
\begin{thm}
\label{thm:Taylor_approx_of_H}Let $H_{1}\left(x,t\right)$ and $H_{2}\left(x,t\right)$
be two symbols on $x=\left(q,p\right)\in\mathbb{R}^{2}$, such that
$0$ is a fixed point (i.e. $\left(DH_{1}\right)_{x=0}=0$), and suppose
that they have identical Taylor series in $\left(x,\hbar\right)$
at the origin $\left(0,0,0\right)$, up to order $J$:\[
H_{1}\left(x,t\right)=H_{2}\left(x,t\right)+\mathcal{O}\left(\left(x,\hbar\right)^{J+1}\right)\]
(i.e. Taylor terms $\hbar^{l}q^{a}p^{b}$ are identical if $l+a+b\leq J$).
Let $\hat{U}_{1}=\exp\left(-i\hat{H}_{1}/\hbar\right)$ and $\hat{U}_{2}=\exp\left(-i\hat{H}_{2}/\hbar\right)$,
with $\hat{H}_{j}=Op_{Weyl}\left(H_{j}\right)$, $j=1,2$. Let $0<\alpha<1/2$,
and $\hat{P}_{\alpha}$ be defined by Eq.(\ref{eq:def_PD}). Then\begin{equation}
\left\Vert \left(\hat{U}_{1}-\hat{U}_{2}\right)\hat{P}_{\alpha}\right\Vert _{L^{2}}\leq C\hbar^{A}\label{eq:error_Taylor_symbol}\end{equation}
with $A=\left(\frac{1}{2}-\alpha'\right)\left(J+1\right)-1$, for
any $\alpha'>\alpha$.
\end{thm}
\selectlanguage{french}
}}\end{center}\vspace{0.cm}

\selectlanguage{english}
This result is interesting in the semiclassical limit if $A>0\Leftrightarrow J>\frac{1+2\alpha}{1-2\alpha}$,
for example if $J=2$ and $0<\alpha<1/6$. This theorem will be used
page \pageref{eq:H_normal_form} for the proof of Theorem \ref{pro:semi-classical-normal-form}.

\subsection{Proof of Theorem \ref{thm:Taylor_approx_of_H}}

We will use the Duhamel formula \cite{joye_99}(which is not semiclassical):

\begin{prop}
Suppose that $\hat{H}\left(t\right)$ is a self-adjoint operator in
a Hilbert space $\mathcal{H}$, and $\psi\left(t\right)\in\mathcal{H}$
is solution of \[
i\hbar\frac{d\psi}{dt}=\hat{H}\left(t\right)\psi+\xi\left(t\right)\]
where $\xi\left(t\right)\in\mathcal{H}$, $\left\Vert \xi\left(t\right)\right\Vert \leq\mu\left(t\right)$,
and $\left\Vert \psi\left(0\right)\right\Vert =1$. Suppose that $\varphi\left(t\right)$
is solution of \[
i\hbar\frac{d\varphi}{dt}=\hat{H}\left(t\right)\varphi\]
with $\varphi\left(0\right)=\psi\left(0\right)$. Then\begin{equation}
\left\Vert \psi\left(t\right)-\varphi\left(0\right)\right\Vert \leq\frac{1}{\hbar}\int_{0}^{t}\mu\left(s\right)ds\label{eq:Duhamel}\end{equation}

\end{prop}
We can deduce a {}``semiclassical version'' of the Duhamel formula:

\begin{cor}
\label{cor:Duhamel}If $H_{1},H_{2}\in S\left(1\right)$ are real
bounded semiclassical symbols (cf \cite{dimassi-99}, p.81), and if
$K=H_{1}-H_{2}\in S^{-k}\left(1\right)$ (i.e. $K\in\hbar^{k}S\left(1\right)$)
, and $\hat{U}_{1}=\exp\left(-i\hat{H}_{1}/\hbar\right)$, $\hat{U}_{2}=\exp\left(-i\hat{H}_{2}/\hbar\right)$,
with $\hat{H}_{j}=Op_{Weyl}\left(H_{j}\right)$, $j=1,2$, then\[
\left\Vert \hat{U}_{1}-\hat{U}_{2}\right\Vert _{L^{2}}\leq\mathcal{O}\left(\hbar^{k-1}\right)\]

\end{cor}
\begin{proof}
$K\in S^{-k}\left(1\right)$, thus $\left\Vert \hat{K}\right\Vert _{L^{2}}=\mathcal{O}\left(\hbar^{k}\right)$,
from Calderon-Vaillancourt Theorem. Let $\psi\in\mathcal{H}$, be
normalized: $\left\Vert \psi\right\Vert =1$, and $\psi_{1}\left(t\right)$,
$\psi_{2}\left(t\right)$, solutions of $i\hbar\frac{d\psi_{j}}{dt}=\hat{H}_{j}\psi_{j}$,
$j=1,2$, with $\psi_{1}\left(0\right)=\psi_{2}\left(0\right)=\psi$
. This gives $i\hbar d\psi_{2}/dt=\left(\hat{H}_{1}-\hat{K}\right)\psi_{2}=\hat{H}_{1}\psi_{2}+\xi\left(t\right)$,
with $\left\Vert \xi\left(t\right)\right\Vert \leq\left\Vert \hat{K}\right\Vert _{L^{2}}=\mathcal{O}\left(\hbar^{k}\right)$.
From Eq.(\ref{eq:Duhamel}), we deduce $\left\Vert \hat{U}_{2}\psi-\hat{U}_{1}\psi\right\Vert _{L^{2}}\leq\frac{1}{\hbar}\mathcal{O}\left(\hbar^{k}\right)$.
\end{proof}

\paragraph{Micro-Localization Lemma:}

\begin{lem}
\label{pro:Lemme_de_microloc}Let $0<\alpha<\alpha'<1/2$, and $H\left(x,t\right)$
a real symbol, with fixed point $0$: $DH_{/x=0}=0$. Let $\hat{H}\left(t\right)=Op_{Weyl}\left(H\left(t\right)\right)$,
and $\hat{H}'\left(t\right)=\hat{H}\left(t\right)\hat{P}_{\alpha'}$
be the truncated Hamiltonian, where $\hat{P}_{\alpha'}$is defined
in Eq.(\ref{eq:def_PD}). Let $\hat{U}\left(t\right)$ and $\hat{U}'\left(t\right)$
be the evolution operators defined by $i\hbar\frac{d\hat{U}\left(t\right)}{dt}=\hat{H}\left(t\right)\hat{U}\left(t\right)$,
$i\hbar\frac{d\hat{U}'\left(t\right)}{dt}=\hat{H}'\left(t\right)\hat{U}'\left(t\right)$,
and $\hat{U}\left(0\right)=\hat{U}'\left(0\right)=\hat{Id}$. Then
for any finite time $t$,\[
\left\Vert \left(\hat{U}\left(t\right)-\hat{U}'\left(t\right)\right)\hat{P}_{\alpha}\right\Vert _{L^{2}}=\mathcal{O}\left(\hbar^{\infty}\right)\]

\end{lem}

\paragraph{Remarks:}

Remind that $\hat{P}_{\alpha}$ truncates on a disk of radius $\hbar^{1/2-\alpha}$,
and notice that $\hbar^{1/2}\ll\hbar^{1/2-\alpha}\ll\hbar^{1/2-\alpha'}$
in the semiclassical limit. Proposition \ref{pro:Lemme_de_microloc}
means that the evolution of a quantum states in the vicinity of a
fixed point (or more generally of a trajectory) after finite time,
depends only of $H\left(x\right)$ in the vicinity of this fixed point.
The proof of Proposition \ref{pro:Lemme_de_microloc} relies on Egorov
Theorem, and can be found for example in \cite{ianchenko-98}, Lemma
3, Lemma 4.

\paragraph{Proof of Theorem \ref{thm:Taylor_approx_of_H}:}

\begin{proof}
We use notations of Lemma \ref{pro:Lemme_de_microloc} and Corollary
\ref{cor:Duhamel}, and combine them. Let $H'_{j}=H_{j}\chi'$, $j=1,2$,
where $\chi'\left(x\right)$ truncates at radius $r'=\hbar^{1/2-\alpha'}$
(i.e. $\chi'\left(x\right)=1$, if $\left|x\right|<r'$, $\chi'\left(x\right)=0$,
if $\left|x\right|>2r'$). Similarly, let $\chi\left(x\right)$ which
truncates at radius $r=\hbar^{1/2-\alpha}$. Let $\hat{U}_{j}=\exp\left(-i\hat{H}_{j}/\hbar\right)$,
$\hat{U}'_{j}=\exp\left(-i\hat{H}'_{j}/\hbar\right)$. By hypothesis,
we have $K=\left(H'_{2}-H'_{1}\right)\in S_{\alpha'}^{-k}\left(1\right)$,
with $\hbar^{k}=\hbar^{\left(1/2-\alpha'\right)J}$ i.e. $k=\left(1/2-\alpha'\right)J$.
From Corollary \ref{cor:Duhamel}, $\left\Vert \hat{U}'_{1}-\hat{U}'_{2}\right\Vert _{L^{2}}=\mathcal{O}\left(\hbar^{k-1}\right)$.
Let $\hat{\chi}=Op_{Weyl}\left(\chi\right)$. One has\begin{align*}
\left\Vert \left(\hat{U}_{1}-\hat{U}_{2}\right)\hat{\chi}\right\Vert _{L^{2}} & =\left\Vert \left(\hat{U}_{1}-\hat{U}'_{1}+\hat{U}'_{1}-\hat{U}'_{2}+\hat{U}'_{2}-\hat{U}_{2}\right)\hat{\chi}\right\Vert \\
 & \leq\left\Vert \left(\hat{U}_{1}-\hat{U}'_{1}\right)\hat{\chi}\right\Vert +\left\Vert \left(\hat{U}'_{1}-\hat{U}'_{2}\right)\hat{\chi}\right\Vert +\left\Vert \left(\hat{U}'_{2}-\hat{U}_{2}\right)\hat{\chi}\right\Vert \\
 & \leq\mathcal{O}\left(\hbar^{\infty}\right)+\mathcal{O}\left(\hbar^{k-1}\right)+\mathcal{O}\left(\hbar^{\infty}\right)\end{align*}

\end{proof}

\section{Proof of Theorem \ref{pro:semi-classical-normal-form} on Semiclassical
Non Stationary Normal Forms\label{sec:Proof-of-theorem}}

In this appendix we develop the Theory of Semi-classical Non-Stationary
Normal form for hyperbolic map on the torus, which extends the results
of David DeLatte \cite{delatte_92} from the classical to the semi-classical
case.

\subsection{Description of the dynamics:}

Let us write $x=\left(q,p\right)\in\mathbb{R}^{2}$. Remind that the
classical map $M=M_{1}M_{0}:\mathbb{R}^{2}\rightarrow\mathbb{R}^{2}$
was defined in Eq.(\ref{eq:map_M}) as the result of a first flow
$M_{0}$ generated by the quadratic Hamiltonian $H_{0}$ for time
interval $t\in\left[0,1\right]$, followed by the {}``perturbation''
$M_{1}$ generated by the Hamiltonian $H_{1}$ on time interval $t\in\left[1,2\right]$.
$H_{1}\left(x\right)$ is periodic with respect to $\mathbb{Z}^{2}\subset\mathbb{R}^{2}$.
So the dynamics is defined by a \emph{time dependant} Hamiltonian
function written $H\left(x,t\right)$ with period $2$ in $t$.

For any $t,t'\in\mathbb{R}$, we denote $M_{t,t'}:\mathbb{R}^{2}\rightarrow\mathbb{R}^{2}$
the flow generated by the Hamiltonian $H\left(x,t\right)$, in the
time interval $\left[t,t'\right]$. In particular $M=M_{0,2}$.

Using the Weyl quantization procedure, we have defined the corresponding
quantum Hamiltonian $\hat{H}\left(t\right)=Op_{Weyl}\left(H\left(x,t\right)\right)$,
in Section \ref{sub:Quantum-mechanics}. Conversely $H\left(x,t\right)$
is called the total semiclassical symbol of $\hat{H}\left(t\right)$.
The unitary evolution operator $\hat{M}_{t,t'}$ is defined by $\frac{d\hat{M}_{t,t'}}{dt'}=-\frac{i}{\hbar}\hat{H}\left(t'\right)\hat{M}_{t,t'}$,
and $\hat{M}_{t,t}=\hat{Id}$.

\paragraph{Stable and unstable tangent vectors for intermediate time $t$:}

In Section \ref{sub:Unstable-and-stable}, we explained that for every
point $x\in\mathbb{T}^{2}$ , there is a basis of tangent vectors
$\left(u_{x},s_{x}\right)\in T_{x}\mathbb{T}^{2}$, continuous with
respect to $x\in\mathbb{T}^{2}$, and tangent to the unstable/stable
foliation. We can define this basis for intermediate time $t\in\left[0,2\right]$
(and then for any time $t\in\mathbb{R}$ by requiring periodicity),
giving vectors $u_{x,t},s_{x,t}$ as follows: the direction of $u_{x,t}$
(respect. $s_{x,t}$) is the direction $\left(DM_{0,t}\right)_{x'}u_{x'}$
with $x=M_{0,t}\left(x'\right)$ (respect. for $s_{x'}$), and we
fix the choice of $u_{x,t}$ (respect. $s_{x,t}$) by requiring that

\begin{equation}
u_{x,t}\wedge s_{x,t}=1\label{eq:ux_sx_t}\end{equation}
i.e. they form a symplectic basis, and we also impose that $\left\Vert u_{x,t}\right\Vert =\left\Vert s_{x,t}\right\Vert $. 

Let us define the lattices $\Gamma_{0}=\mathbb{Z}^{2}$, $\Gamma_{t}=M_{0,t}\left(\Gamma_{0}\right)$
for $t\in\left[0,1\right]$, and $\Gamma_{t}=\Gamma_{0}$ for $t\in\left[1,2\right]$.
Let us define $\mathbb{T}_{t}\defi\mathbb{R}^{2}/\Gamma_{t}$. If
we consider $t\in S^{1}=\mathbb{R}/\left(2\mathbb{Z}\right)$, then
the phase space and time containing points $\left(x,t\right)$ forms
a non trivial bundle of tori $\mathcal{B}\rightarrow S^{1}$ over
$S^{1}$, whose fiber over $t\in S^{1}$ is the torus $\mathbb{T}_{t}^{2}=\mathbb{R}^{2}/\Gamma_{t}$.
$u_{x,t}$ and $s_{x,t}$ are continuous in $x\in\mathbb{R}^{2}$
and periodic with respect to $\Gamma_{t}$, and also continuous and
periodic in $t\in\mathbb{R}$. In other words they are continuous
functions of $\left(x,t\right)\in\mathcal{B}$. In this section periodicity
in $x$ and $t$ will always refer to the lattice $\Gamma_{t}$ and
to $S^{1}=\mathbb{R}/\left(2\mathbb{Z}\right)$ respectively.

Define the expansion rate $\lambda\left(x,t,t'\right)$ by

\[
\left(DM_{t,t'}\right)_{x}\left(u_{x,t}\right)=e^{\lambda\left(x,t,t'\right)}u_{M_{t,t'}\left(x\right),t'}\]

and the local expansion rate $\eta\left(x,t\right)$ by\begin{equation}
\eta\left(x,t\right)=\left(\frac{d\lambda}{dt'}\right)_{t'=t}.\label{eq:taux_expansion}\end{equation}

$\eta\left(x,t\right)$ is a piece-wise continuous function and periodic.
In particular\[
\lambda_{x}=\int_{0}^{2}\eta\left(x\left(t\right),t\right)dt=\lambda\left(x,0,2\right)\]

gives the expansion rate already defined in Eq.(\ref{eq:DM}).

\subsection{Preliminaries on Time-dependant canonical transformations}

In order to simplify the description of the dynamics, and obtain the
non-stationary normal form, we will perform time dependant canonical
transformations at the semi-classical level (\cite{martinez-01},
\cite{zworski-03}, \cite{dimassi-99}). In this paragraph, we prepare
some results and fix the notations.

\subsubsection{Moyal formula}

We remind that if $A,B$ are two symbols and $\hat{A}=Op_{Weyl}\left(A\right)$
and $\hat{B}=Op_{Weyl}\left(B\right)$, then the Moyal product $\sharp$
between symbols correspond to product of operators and is defined
by (\cite{martinez-01} p. 41)\begin{equation}
A\sharp B=\sigma_{Weyl}\left(\hat{A}\hat{B}\right)\label{eq:start_product}\end{equation}
Where $\sigma_{Weyl}$ gives the symbol of an operator. The Moyal
formula gives the symbol of a product of operators:

\begin{equation}
A\sharp B=\sum_{n\geq0}\left(i\hbar\right)^{n}\frac{1}{2^{n}n!}\left(A\mathcal{D}^{n}B\right)=AB+i\hbar\frac{1}{2}\left\{ A,B\right\} +\ldots\label{eq:Moyal_product}\end{equation}

where $\mathcal{D}=\frac{\overleftarrow{\partial}}{\partial p}\vec{\frac{\partial}{\partial q}}-\frac{\overleftarrow{\partial}}{\partial q}\vec{\frac{\partial}{\partial p}}$,
and $\left\{ A,B\right\} =\left(A\mathcal{D}B\right)=\frac{\partial A}{\partial p}\frac{\partial B}{\partial q}-\frac{\partial A}{\partial q}\frac{\partial B}{\partial p}$
is the Poisson Bracket. As a consequence the symbol of the commutator
is given by\begin{align}
\mathfrak{X}_{A}\left(B\right)\defi\sigma_{Weyl}\left(\frac{1}{i\hbar}\left[\hat{A},\hat{B}\right]\right) & =\sum_{n\geq0}\left(i\hbar\right)^{2n}\frac{2}{2^{2n+1}\left(2n+1\right)!}\left(A\mathcal{D}^{2n+1}B\right)\label{eq:Moyal}\\
 & =\left\{ A,B\right\} +\frac{i\hbar^{2}}{6}\left(A\mathcal{D}^{3}B\right)+\ldots\nonumber \end{align}
The first term, which is the classical Poisson Bracket will also be
written: \[
\mathfrak{X}_{class,A}\left(B\right)\defi\left\{ A,B\right\} \]

These formula exist in a more general framework of star-product, and
all what follows can be applied to any star-product, in particular
to any choice of quantization procedure.

\subsubsection{Time dependant semiclassical canonical transformation}

Let $G_{t}\left(x\right)$ be a semiclassical symbol which depends
on $t$, $\hat{G}_{t}=\mbox{Op}_{Weyl}\left(G_{t}\left(x\right)\right)$
and $\hat{U}_{t}=\exp\left(-i\hat{G}_{t}/\hbar\right)$ the unitary
transformation in $L^{2}\left(\mathbb{R}\right)$ generated by $\hat{G}_{t}$
for any fixed value of $t$. 

\selectlanguage{french}
\vspace{0.cm}\begin{center}\fbox{\parbox{16cm}{

\selectlanguage{english}
\begin{prop}
\label{pro:Time_dependant_canon_transf}For any $\left(t,t'\right)$,
the transformed evolution operator:\begin{equation}
\hat{M}'_{t,t'}=\hat{U}_{t'}^{-1}\hat{M}_{t,t'}\hat{U}_{t}\label{eq:relation_commutt}\end{equation}
is generated by a transformed Hamiltonian symbol $H'$, (i.e. $d\hat{M}'_{t,t'}/dt'=-\frac{i}{\hbar}\hat{H}'\left(t'\right)\hat{M}'_{t,t'}$,
and $\hat{M}'_{t,t}=\hat{Id}$, with $\hat{H}'=\mbox{Op}_{Weyl}\left(H'\right)$),
given by

\begin{equation}
H'=\exp\left(-\mathfrak{X}_{G_{t}}\right)H+R\label{eq:H_prime}\end{equation}
with:\[
R=\sum_{k\geq0}\frac{\left(-1\right)^{k}}{\left(k+1\right)!}\mathfrak{X}_{G_{t}}^{k}\left(\frac{\partial G_{t}}{\partial t}\right)=\left(\frac{\partial G_{t}}{\partial t}\right)-\frac{1}{2}\mathfrak{X}_{G}\left(\frac{\partial G_{t}}{\partial t}\right)+\ldots\]

\end{prop}
\selectlanguage{french}
}}\end{center}\vspace{0.cm}

\selectlanguage{english}
Notice that the term $R$ is due to the time dependence of the canonical
transformation, and that $\exp\left(-\mathfrak{X}_{class,G}\right)H=H\circ U_{t}^{-1}$
is the function $H$ after the classical canonical transformation
generated by $G_{t}$.

\begin{proof}
For the proof, we consider the enlarged phase space $\left(t,\tau,q,p\right)\in\mathbb{R}^{4}$
with the symplectic two form $\omega=dq\wedge dp+dt\wedge d\tau$,
where $t$ is considered here as a dynamical variable, and $\tau$
is its conjugated variable. This is a well known and useful trick
both in classical and quantum mechanics which allows to map a time
dependant dynamics onto a time independent dynamics, so that we can
use well known tools. With $\hat{\tau}\defi-i\hbar\partial/\partial t$,
and $\hat{\mathbf{H}}\defi\hat{\tau}+\hat{H}$, the time-dependant
Schrödinger equation $i\hbar\frac{\partial\psi}{\partial t}=\hat{H}\psi$
can be written \begin{equation}
\hat{\mathbf{H}}\psi=\left(\hat{\tau}+\hat{H}\right)\psi=0\label{eq:H_psi=0}\end{equation}
i.e. as the time-independent Schrödinger equation for $\psi\left(q,t\right)$.
We will use bold fonts when dealing with the enlarged phase space
or enlarged Hilbert space $L^{2}\left(\mathbb{R}^{2}\right)$. Let
$\hat{\mathbf{G}}=Op_{Weyl}\left(\mathbf{G}\right)$, where $\mathbf{G}\left(q,p,t,\tau\right)=G_{t}\left(q,p\right)$
is independent of $\tau$, let $\hat{\mathbf{U}}=\exp\left(-i\hat{\mathbf{G}}/\hbar\right)$,
and consider the transformed Hamiltonian\begin{equation}
\hat{\mathbf{H}}'\defi\hat{\mathbf{U}}^{-1}\mathbf{\hat{H}}\hat{\mathbf{U}}\label{eq:Hbold_prime}\end{equation}
whose total symbol is:\[
\mathbf{H}'\defi\exp\left(-\mathfrak{X}_{\mathbf{G}}\right)\mathbf{H}=\mathbf{H}-\mathfrak{X}_{\mathbf{G}}\left(\mathbf{H}\right)+\frac{1}{2!}\mathfrak{X}_{\mathbf{G}}^{2}\left(\mathbf{H}\right)+\ldots\]
where $\mathfrak{X}_{\mathbf{G}}\left(\mathbf{H}\right)=\frac{2}{\hbar}\left(\sum_{k\geq0}\left(-1\right)^{k}\frac{1}{\left(2k+1\right)!}\mathbf{G}\left(\frac{\hbar\mathbf{D}}{2}\right)^{2k+1}\mathbf{H}\right)$,
with $\mathbf{D}=\left(\frac{\overleftarrow{\partial}}{\partial\tau}\frac{\vec{\partial}}{\partial t}-\frac{\overleftarrow{\partial}}{\partial t}\frac{\vec{\partial}}{\partial\tau}\right)+\left(\frac{\overleftarrow{\partial}}{\partial p}\frac{\vec{\partial}}{\partial q}-\frac{\overleftarrow{\partial}}{\partial p}\frac{\vec{\partial}}{\partial p}\right)=\mathcal{D}_{t}+\mathcal{D}_{q}$.
Because $\mathbf{G}$ does not depend on $\tau$, we have $\mathbf{G}\,\mathbf{D}\,\mathbf{H}=\left(-\frac{\partial\mathbf{G}}{\partial t}\right)+\mathbf{G}\mathcal{D}_{q}\mathbf{H}=\left(-\frac{\partial G_{t}}{\partial t}\right)+G_{t}\mathcal{D}_{q}H$
and for $k\geq2$, $\mathbf{G}\,\mathbf{D}^{k}\,\mathbf{H}=\mathbf{G}\left(\mathcal{D}_{t}+\mathcal{D}_{q}\right)^{k}\mathbf{H}=G_{t}\mathcal{D}_{q}^{k}H$.
Then $\mathfrak{X}_{\mathbf{G}}\left(\mathbf{H}\right)=\left(-\frac{\partial G_{t}}{\partial t}\right)+\mathfrak{X}_{G_{t}}\left(H\right)$,
and $\mathfrak{X}_{\mathbf{G}}^{k}\left(\mathbf{H}\right)=-\mathfrak{X}_{G_{t}}^{k-1}\left(\frac{\partial G_{t}}{\partial t}\right)+\mathfrak{X}_{G_{t}}^{k}\left(H\right)$,
for $k\geq1$. Therefore we get \[
\mathbf{H}'=\exp\left(-\mathfrak{X}_{\mathbf{G}}\right)\mathbf{H}=\tau+H'\]
with\[
H'=\exp\left(-\mathfrak{X}_{G_{t}}\right)H+R\]
with $R=-\sum_{k\geq1}\frac{\left(-1\right)^{k}}{k!}\mathfrak{X}_{G_{t}}^{k-1}\left(\frac{\partial G_{t}}{\partial t}\right)$. 

Now, for any $u\in\mathbb{R}$, let $\hat{\mathbf{M}}_{u}=\exp\left(-iu\hat{\mathbf{H}}/\hbar\right)$
and $\hat{\mathbf{M}}'_{u}=\exp\left(-iu\hat{\mathbf{H}}'/\hbar\right)$.
From Eq.(\ref{eq:Hbold_prime}), we have the conjugation relation
$\hat{\mathbf{M}}'_{u}=\hat{\mathbf{U}}^{-1}\hat{\mathbf{M}}_{u}\hat{\mathbf{U}}$.
But $\left(\hat{\mathbf{U}}\psi\right)\left(q,t\right)=\left(\hat{U}_{t}\psi\right)\left(q,t\right)$
and $\left(\hat{\mathbf{M}}_{u}\psi\right)\left(q,t\right)=\left(\hat{M}_{t-u,t}\psi\right)\left(q,t-u\right)$.
So the conjugation relation can be written: $\hat{M}'_{t-u,t}=\hat{U}_{t}^{-1}\hat{M}_{t-u,t}\hat{U}_{t-u}$,
or $\hat{M}'_{t,t'}=\hat{U}_{t'}^{-1}\hat{M}_{t,t'}\hat{U}_{t}$,
for any $t,t'$.
\end{proof}

\subsubsection{Translations}

If $x=\left(q,p\right)\in\mathbb{R}^{2}$, and $x'=\left(q',p'\right)\in\mathbb{R}^{2}$,
we write $x\wedge x'=qp'-pq'$. Let $A\left(t\right)=\left(a\left(t\right),b\left(t\right)\right)\in\mathbb{R}^{2}$,
$t\in\mathbb{R}$, be any path, and as a special case of Proposition
\ref{pro:Time_dependant_canon_transf}, suppose that $G_{t}$ is a
linear function of $x=\left(q,p\right)$:\begin{equation}
G_{t}\left(q,p\right)=a\left(t\right)p-b\left(t\right)q=A\left(t\right)\wedge x\label{eq:G_translation}\end{equation}

We get:

\begin{equation}
H'\left(x,t\right)=\frac{dA}{dt}\wedge\left(x-\frac{1}{2}A\right)+H\left(x-A\left(t\right),t\right)\label{eq:Hp_translation}\end{equation}

\begin{proof}
We have $\left(\frac{\partial G_{t}}{\partial t}\right)=\frac{da}{dt}p-\frac{db}{dt}q=\frac{dA}{dt}\wedge x$,
and since $G_{t}$ is linear, $\mathfrak{X}_{G}=\mathfrak{X}_{class,G}=\left\{ G_{t},.\right\} $.
Then $\mathfrak{X}_{G_{t}}\left(\frac{\partial G_{t}}{\partial t}\right)=\left\{ G_{t},\frac{\partial G_{t}}{\partial t}\right\} =\frac{da}{dt}b-a\frac{db}{dt}=\frac{dA}{dt}\wedge A$,
so $\mathbf{R}=\frac{dA}{dt}\wedge\left(x-\frac{1}{2}A\right)$. Also
we have $\exp\left(-\mathfrak{X}_{G_{t}}\right)H=\exp\left(-\mathfrak{X}_{class,G_{t}}\right)H=H\left(q-a\left(t\right),p-b\left(t\right)\right)$,
i.e. the translated function, as transformed by the classical canonical
transformation. 
\end{proof}

\subsubsection{Symplectic linear transformations}

Suppose that $G_{t}\left(q,p\right)$ is quadratic in $q,p$ variables,
with time-dependant coefficients. Then except from the quadratic terms
who may change, the transformed Hamiltonian $H'$ given in Proposition
\ref{pro:Time_dependant_canon_transf} is just the classical transform
of $H$: \begin{equation}
H'=\exp\left(-\mathfrak{X}_{class,G_{t}}\right)H+\textrm{quadratic terms}\label{eq:Hprime_G_quadratic}\end{equation}

\begin{proof}
Again $\mathfrak{X}_{G_{t}}=\mathfrak{X}_{class,G_{t}}=\left\{ G_{t},.\right\} $.
Also, for any quadratic homogeneous functions $A,B$, then $\left\{ A,B\right\} $
is also quadratic. We deduce that $\mathfrak{X}_{class.G_{t}}^{k}\left(\frac{\partial G_{t}}{\partial t}\right)$
and $R$ are quadratic homogeneous function of $\left(q,p\right)$.
Then Eq.(\ref{eq:H_prime}) gives Eq.(\ref{eq:Hprime_G_quadratic}). 
\end{proof}

\subsubsection{Time-dependant transformations of a special kind}

When dealing with next order terms, we will only need symbols $G_{t}$
of the simple factorized form:\[
G_{t}\left(x\right)=g\left(t\right)\tilde{G}\left(x\right)\]

Then\begin{equation}
H'=\left(\frac{\partial G_{t}}{\partial t}\right)+\exp\left(-\mathfrak{X}_{G_{t}}\right)H\label{eq:Hprime_pour_Gt}\end{equation}
since $\mathfrak{X}_{G_{t}}\left(\frac{\partial G_{t}}{\partial t}\right)=g\left(t\right)g'\left(t\right)\mathfrak{X}_{\tilde{G}}\left(\tilde{G}\right)=0$.

\subsection{Pre-Normalization}

\begin{figure}[tbph]
\begin{centering}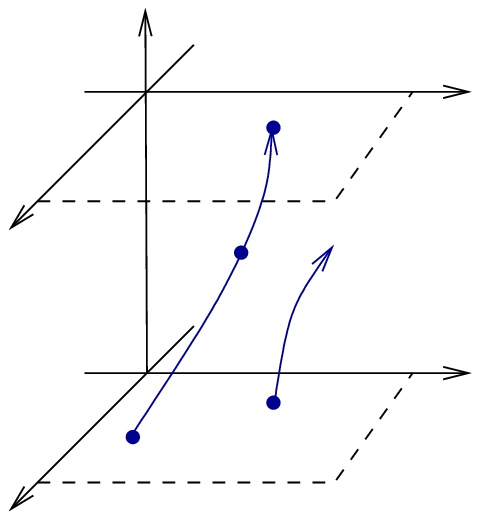\par\end{centering}

\caption{\label{cap:Trajectory--xt}Trajectory $x\left(t\right)$.}
\end{figure}

Consider a given point $x_{0}=\left(q_{0},p_{0}\right)\in\mathbb{R}^{2}$
at time $t=0$, and $x\left(t\right)=\left(q\left(t\right),p\left(t\right)\right)=M_{0,t}x_{0}$
the trajectory passing through it. See figure \ref{cap:Trajectory--xt}.
We will sometimes use the condensed notation $\mathbf{x}=\left(x\left(t\right),t\right)\in\mathbb{R}^{3}$.

During Pre-normalization, we will perform successively two canonical
transformations (a translation and a linear symplectic transformation)
in order to simplify the Taylor expansion of $H\left(q,p,t\right)$
along the trajectory $x\left(t\right)$, up to degree two in $\left(q,p\right)$.
Higher degrees will be treated in a second stage called the normalization
process. We will use the notation $x'=\left(q',p'\right)$ for the
new coordinates and $H'_{\mathbf{x}}\left(q',p'\right)$ for the transformed
Hamiltonian, which depends on time through $\mathbf{x}=\left(x\left(t\right),t\right)$.

Notice that if $F\left(\mathbf{x}\right)=F\left(x,t\right)$ is a
periodic function with respect to $t$, then it fulfills the obvious
relation:\begin{equation}
F\left(x\left(t\right),t\right)_{t=2}=F\left(\left(Mx\right)\left(t\right),t\right)_{t=0}\label{eq:periodicity}\end{equation}

\subsubsection{Translation}

We decide to map the trajectory $x\left(t\right)=\left(q\left(t\right),p\left(t\right)\right)$
onto the origin $x'=\left(q',p'\right)=0$, using a time-dependant
translation, generated by Eq.(\ref{eq:G_translation}), with $A\left(t\right)=-x\left(t\right)$.
From Eq.(\ref{eq:Hp_translation}), the value of the transformed Hamiltonian
$H'$ at the origin is then\[
H_{\mathbf{x}}'\left(x'=0\right)=-\mathcal{A}_{\mathbf{x}}\]
with \begin{equation}
\mathcal{A}_{\mathbf{x}}=\frac{1}{2}\frac{dx}{dt}\wedge x-H\left(x\right)=\frac{1}{2}\left(p\frac{dq}{dt}-q\frac{dp}{dt}\right)-H\left(x\right).\label{eq:action_Ax}\end{equation}
We recognize the time derivative of the classical action along the
trajectory $x\left(t\right)$ (see \cite{fred-PreQ-06} for a geometric
interpretation). From Eq.(\ref{eq:Hp_translation}) and using classical
Hamilton equations for $x\left(t\right)$, one obtains that the derivative
of $H_{\mathbf{x}}'$ at the origin is zero: $\left(\partial_{x'}H'\right)\left(x'=0\right)=0$,
as expected because the origin is now a fixed point. Higher derivatives
are equal to derivatives of $H$ at $x\left(t\right)$:\[
\left(\partial_{x'}^{k}H'\right)_{x'=0}=\left(\partial_{x}^{k}H\right)_{x=x\left(t\right)},\qquad\forall k\geq2\]

\subsubsection{Linear Symplectic transformation}

Similarly to Eq.(\pageref{eq:Qx}), let $Q_{\mathbf{x}}\in SL\left(2,\mathbb{R}\right)$
be the symplectic matrix which transforms the canonical basis of $\mathbb{R}^{2}$
to the basis $\left(u_{\mathbf{x}},s_{\mathbf{x}}\right)$ at point
$x\left(t\right)$ and time $t$ defined above. We use this canonical
transformation in order to simplify the quadratic terms of the Hamiltonian.
This transformation is generated by a quadratic function $G$, and
from Eq.(\ref{eq:Hprime_G_quadratic}), the transformed Hamiltonian
$H'$ differs from $H$ only by different quadratic terms. From Eq.(\ref{eq:DM})
and Eq.(\ref{eq:taux_expansion}), the new quadratic terms are \[
\left(H_{\mathbf{x}}'\left(q',p'\right)\right)_{degree\,2}=\eta_{\mathbf{x}}q'p'\]

\subsubsection{Result of the pre-normalization process}

\begin{figure}[tbph]
\begin{centering}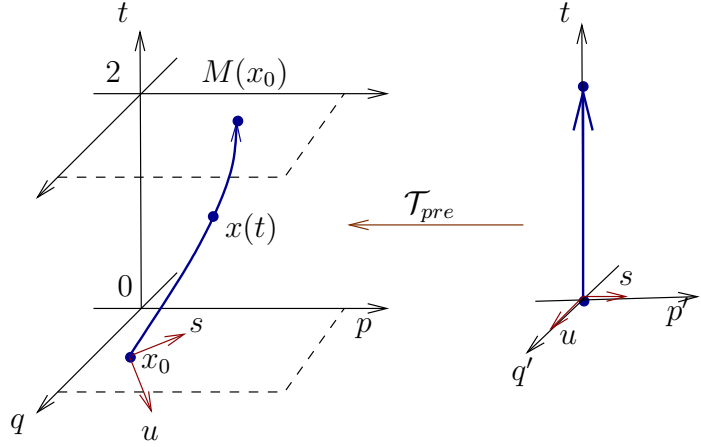\par\end{centering}

\caption{\label{cap:Trajectory--xt_bis}The Pre-normalization transformation
$\mathcal{T}_{pre,\mathbf{x}}^{-1}$ translates the trajectory to
the origin, and transforms the unstable/stable directions onto the
principal axis $\left(q',p'\right)$.}
\end{figure}

Up to now, we have chosen an initial point $x_{0}\in\mathbb{R}^{2}$,
and have performed successively a translation and a linear symplectic
transform:\[
\mathcal{T}{}_{pre,\mathbf{x}}\defi T_{x\left(t\right)}Q_{\mathbf{x}}\]
so that the trajectory $x\left(t\right)$ is mapped onto the origin,
and the new Hamiltonian $H'_{\mathbf{x}}\left(q',p'\right)$ has a
Taylor expansion in variables $x'=\left(q',p'\right)$ at the origin
of the form:\[
\left(H'_{\mathbf{x}}\right)_{deg\,\leq2}\left(q',p'\right)=-\mathcal{A}_{\mathbf{x}}+\eta_{\mathbf{x}}q'p'\]

\[
\left(H'_{\mathbf{x}}\right)_{deg\,\geq3}\left(q',p'\right)=\left(H\circ\mathcal{T}_{pre,\mathbf{x}}\right)\left(q',p'\right)\]
Terms of degree lower or equal to two are now in normal form, which
means that the stable and unstable manifold are tangential to the
principal axis, see figure \ref{cap:Trajectory--xt_bis}.

Eq.(\ref{eq:relation_commutt}) with $t=0$, $t'=2$, gives

\[
\hat{M}=\hat{\mathcal{T}}_{pre,Mx_{0}}\hat{M}'_{x_{0}}\hat{\mathcal{T}}_{pre,x_{0}}^{-1}\]
where $\mathcal{T}_{pre,x_{0}}=\mathcal{T}_{pre,\left(x_{0},t=0\right)}$.
This relation gives Theorem \ref{eq:quantum_non_stat_normal_form}
page \pageref{eq:quantum_non_stat_normal_form} at the linear order
$J=2$ (we will explain below how to transform the time dependent
term $\eta_{\mathbf{x}}q'p'$ into the time independent $\lambda_{0,1,x_{0}}q'p'$,
when treating resonant terms).

Notice that the Hamiltonian $H'_{\mathbf{x}}\left(q',p'\right)$ in
pre-normal form we have obtained, is continuous with respect to $\mathbf{x}=\left(x,t\right)$,
and periodic in $t$, as explained in Eq.(\ref{eq:periodicity}).
The action $\mathcal{A}_{\mathbf{x}}$ is not periodic in $x$ with
respect to the lattice $\Gamma_{t}$, but the terms of higher degrees
are.

\subsection{Normalization}

We go on with the normalization process, starting now from the Hamiltonian
$H_{\mathbf{x}}\left(q',p'\right)$ given by the pre-normalization
above, and which has a Taylor series in $x'=\left(q',p'\right)$ at
the origin of the form:\begin{equation}
H_{\mathbf{x}}\left(q',p'\right)=-\mathcal{A}_{\mathbf{x}}+\eta_{\mathbf{x}}q'p'+\sum_{l\geq0}\sum_{\alpha+\beta\geq0}v_{\alpha,\beta,l,\mathbf{x}}\hbar^{l}q'^{\alpha}p'^{\beta}\label{eq:Hx}\end{equation}
In this expansion, every term is a periodic and continuous function
of $\mathbf{x}=\left(x,t\right)\in\mathbb{R}^{3}$, (except for $\mathcal{A}_{\mathbf{x}}$
which is not periodic in $x$), and $v_{\alpha,\beta,l,\mathbf{x}}=0$,
if $l=0$ and $\alpha+\beta\leq2$. We will transform Eq.(\ref{eq:Hx})
into a normal form, by successive iterations. At the first stage,
we have no {}``quantum terms'', i.e. $v_{\alpha,\beta,l,\mathbf{x}}=0$
if $l\geq1$. But during the normalization process such terms will
possibly appear. We will check that due to the special form of the
Moyal formula Eq.(\ref{eq:Moyal}), only terms with even value of
$l$ will appear.

\paragraph{Notations:}

To simplify the notations, we will write $\eta\left(t\right)\defi\eta_{\left(x\left(t\right),t\right)}$,
$v_{\alpha,\beta,l}\left(t\right)\defi v_{\alpha,\beta,l,\left(x\left(t\right),t\right)}$
etc.. in order to show the time dependence along the trajectory $x\left(t\right)$,
and keep in mind that they also depend continuously and periodically
on $x$ and $t$. Let

\begin{equation}
n\defi\alpha+\beta+2l\in\mathbb{N}\label{eq:def_degree}\end{equation}
be the degree of the monomial $\hbar^{l}q'^{\alpha}p'^{\beta}$. We
will use the notation $\mathcal{O}\left(n\right)\defi\mathcal{O}\left(\left(\hbar,q',p'\right)^{n}\right)$
which means higher degree terms with respect to this graduation.

\subsubsection{Homological equation}

Suppose that we want to suppress the monomial term $v_{\alpha,\beta,l}\left(t\right)\hbar^{l}q'^{\alpha}p'^{\beta}$
with degree $n=\alpha+\beta+2l$, from the Hamiltonian Eq.(\ref{eq:Hx}).
For that purpose, we consider a time dependant canonical transformation
generated by the symbol $G_{\mathbf{x}}\left(t,q',p'\right)=g_{\mathbf{x}}\left(t\right)\hbar^{l}q'^{\alpha}p'^{\beta}\chi\left(q',p'\right)$,
where $g_{\mathbf{x}}\left(t\right)$ is a still unknown function,
but supposed to be continuous and periodic with $\mathbf{x}=\left(q,p,t\right)$.
The function $\chi\left(q',p'\right)$ is a $C_{0}^{\infty}$ cut-off
function%
\footnote{The cut-off function $\chi$ is necessary otherwise the canonical
transformation generated by $G$ is not defined other all $\left(q',p'\right)\in\mathbb{R}^{2}$.
Think for example of $G=C\, p'q'^{2}$.%
}with compact support such that $\chi\left(q',p'\right)=1$ for $\left(q',p'\right)\in D_{1}=\left\{ \left(q',p'\right)\slash\left|q'\right|^{2}+\left|p'\right|^{2}<1\right\} $.
From Eq.(\ref{eq:Hprime_pour_Gt}), the new Hamiltonian will be\[
H'=\left(\frac{\partial G}{\partial t}\right)+\exp\left(-\mathfrak{X}_{G}\right)H\]

Consider $\left(q',p'\right)\in D_{1}$. We have \begin{eqnarray*}
\mathfrak{X}_{G}\left(H\right)\left(q',p'\right) & = & \mathfrak{X}_{G}\left(\eta\left(t\right)q'p'\right)+\mathcal{O}\left(n+1\right)\\
 & = & \eta\left(t\right)\left\{ \mathbf{G},q'p'\right\} +\mathcal{O}\left(n+1\right)\\
 & = & \eta\left(t\right)\left(\beta-\alpha\right)g\left(t\right)\hbar^{l}q'^{\alpha}p'^{\beta}+\mathcal{O}\left(n+1\right)\end{eqnarray*}

So $H'$ contains the monomial\[
K\left(q',p'\right)=\left(v_{\alpha,\beta,l}\left(t\right)+\frac{dg}{dt}+\eta\left(t\right)\left(\alpha-\beta\right)g\left(t\right)\right)\hbar^{l}q'^{\alpha}p'^{\beta}\]
and differs from $H$ by this term and terms of higher orders only.
We wish to suppress this term, by solving $K=0$. This gives\begin{equation}
g_{\mathbf{x}}\left(t\right)=\left(C\left(x_{0}\right)-\int_{0}^{t}v_{\mathbf{x}}\left(s\right)e^{\left(\alpha-\beta\right)\int_{0}^{s}\eta_{\mathbf{x}}ds'}ds\right)e^{-\left(\alpha-\beta\right)\int_{0}^{t}\eta_{\mathbf{x}}ds}\label{eq:sol_homologique}\end{equation}
where $C\left(x_{0}\right)$ is a constant which depends only on $x_{0}=M_{-t}\left(x\right)\in\mathbb{R}^{2}$
(i.e. the point of the trajectory at $t=0$). The continuity in $\mathbf{x}$
is obvious, but we have to impose the periodicity relation Eq.(\ref{eq:periodicity}),
which writes:\[
g_{x_{0}\left(2\right)}\left(2\right)=g_{\left(Mx_{0}\right)}\left(0\right)\]

This gives a relation for the function $C\left(x_{0}\right)$, called
the \textbf{homological equation} (this is usual in normal form calculations
(see \cite{arnold-ed2} chap 5, \cite{katok_hasselblatt} p.100):\begin{equation}
\boxed{C\left(Mx_{0}\right)=\left(C\left(x_{0}\right)-P\left(x_{0}\right)\right)e^{-\left(\alpha-\beta\right)\lambda_{x_{0}}}}\label{eq:homological_equation}\end{equation}
with\[
\lambda_{x_{0}}=\int_{0}^{2}\eta_{\mathbf{x}\left(t\right)}dt,\]
which is the expanding rate along a piece of trajectory, and with\[
P\left(x_{0}\right)=\int_{0}^{2}v_{\mathbf{x}}\left(s\right)e^{\left(\alpha-\beta\right)\int_{0}^{s}\eta_{\mathbf{x}}ds'}ds.\]

We have to check now that the above homological equation has a solution
$C\left(x_{0}\right)$ continuous with respect to $x_{0}\in\mathbb{R}^{2}$.
We have to consider two cases.

\paragraph{The {}``non resonant case'' $\alpha\neq\beta$:}

If $\alpha>\beta$, we can express $C\left(x_{0}\right)$ from the
past image $M^{-t}x_{0}$ and iterate:

\begin{eqnarray*}
C\left(x_{0}\right) & = & \left(C\left(M^{-1}x_{0}\right)-P\left(M^{-1}x_{0}\right)\right)e^{-\left(\alpha-\beta\right)\lambda_{M^{-1}x_{0}}}\\
 & = & \left(C\left(M^{-2}x_{0}\right)-P\left(M^{-2}x_{0}\right)\right)e^{-\left(\alpha-\beta\right)\left(\lambda_{M^{-2}x_{0}}+\lambda_{M^{-1}x_{0}}\right)}-P\left(M^{-1}x_{0}\right)e^{-\left(\alpha-\beta\right)\lambda_{M^{-1}x_{0}}}\end{eqnarray*}

etc...

Similarly, if $\alpha<\beta$, we use images of $x_{0}$ in the future:

\begin{eqnarray*}
C\left(x_{0}\right) & = & P\left(x_{0}\right)+C\left(Mx_{0}\right)e^{-\left(\beta-\alpha\right)\lambda_{x_{0}}}\\
 & = & P\left(x_{0}\right)+e^{-\left(\beta-\alpha\right)\lambda_{x_{0}}}\left(P\left(Mx_{0}\right)+C\left(M^{2}x_{0}\right)e^{-\left(\beta-\alpha\right)\lambda_{Mx_{0}}}\right)\end{eqnarray*}

etc... The following Lemma shows that these iterations converge nicely.

\begin{lem}
\label{lem:(David-Delatte)}(David DeLatte \cite{delatte_92}). For
$\alpha>\beta$, the solution of the homological equation Eq.(\ref{eq:homological_equation})
is\[
C\left(x_{0}\right)=-\sum_{t=1}^{\infty}P\left(M^{-t}x_{0}\right)e^{-\left(\alpha-\beta\right)\Lambda_{-t}\left(x_{0}\right)}\]
which is a continuous function of $x_{0}$, and with\[
\Lambda_{-t}\left(x_{0}\right)=\sum_{-t\leq k\leq-1}\lambda_{M^{k}x_{0}},\qquad\textrm{for}\,\,\, t\geq1\]
is the expanding rate along a piece of trajectory of length $t$.

For $\alpha<\beta$, the solution

\[
C\left(x_{0}\right)=\sum_{t=0}^{+\infty}P\left(M^{t}x_{0}\right)e^{-\left(\beta-\alpha\right)\Lambda_{t}\left(x_{0}\right)}\]
is continuous in $x_{0}$, with\begin{eqnarray*}
\Lambda_{t}\left(x_{0}\right) & = & \sum_{0\leq k\leq t-1}\lambda_{M^{k}x_{0}},\qquad\textrm{for}\,\,\,\, t\geq1\\
 & = & 0,\qquad\qquad\textrm{for}\,\,\,\, t=0\end{eqnarray*}

\end{lem}
\begin{proof}
From uniform hyperbolicity assumption, there exist $\lambda_{min}>0$,
$t_{min}>0$ such that $\forall x_{0}\in\mathbb{T}^{2},t>t_{min}\quad\Lambda_{t}\left(x_{0}\right)>\left|t\right|\lambda_{min}$.
$P\left(x\right)$ is a continuous function of $x\in\mathbb{T}^{2}$,
so bounded by $\left|P\left(x\right)\right|<P_{max}$. Now if $C^{\left(N\right)}\left(x\right)=-\sum_{t=1}^{N}\left(\ldots\right)$
is the partial sum of $C\left(x\right)$ as given above, with $N\in\mathbb{N}$,
then $\forall x\in\mathbb{T}^{2},\left|C^{\left(N+1\right)}\left(x\right)-C^{\left(N\right)}\left(x\right)\right|<e^{-\lambda_{min}N}P_{max}$
. This implies that $C^{\left(N\right)}\left(x\right)$ converges
uniformly for $N\rightarrow\infty$, and that $C\left(x_{0}\right)$
is continuous in $x_{0}$.
\end{proof}

\paragraph{The {}``resonant case'' $\alpha=\beta$:}

In that case we cannot solve the Homological equation, so we keep
the monomial term of $H$. We can however make a small simplification
of the Hamiltonian. Let us suppose that $H$ has a resonant term $\eta_{l,j,\mathbf{x}}\hbar^{l}\left(q'p'\right)^{j}$,
where $\eta_{l,j,\mathbf{x}}$ is time dependant because of $\mathbf{x}=\left(x\left(t\right),t\right)$.
Let us define: \[
\lambda_{l,j,\left(x_{0}\right)}\defi\int_{0}^{2}\eta_{l,j,\mathbf{x}}dt\]

We decompose the resonant term into $\eta_{l,j,\mathbf{x}}=\tilde{\eta}_{l,j,\mathbf{x}}+\frac{1}{2}\lambda_{l,j,\left(x_{0}\right)}$.
From Eq.(\ref{eq:sol_homologique}), the first time-dependent term
$\tilde{\eta}_{l,j,\mathbf{x}}\hbar^{l}\left(q'p'\right)^{j}$ can
be eliminated after a time dependant canonical transformation given
by $g_{\mathbf{x}}\left(t\right)=-\int_{0}^{t}\tilde{\eta}_{l,j,\mathbf{x}}dt$.
There remains a time-independent term $\frac{1}{2}\lambda_{l,j,\left(x_{0}\right)}\hbar^{l}\left(q'p'\right)^{j}$.
In particular the resonant term $\left(l,j\right)=\left(0,0\right)$
is the action term Eq.(\ref{eq:action_Ax}), and gives\begin{equation}
\lambda_{0,0,\left(x\right)}=-\int_{0}^{2}\mathcal{A}_{\mathbf{x}}dt=-\int_{0}^{2}\left(\frac{1}{2}\left(p\frac{dq}{dt}-q\frac{dp}{dt}\right)-H\left(x\right)\right)dt\label{eq:action_l00}\end{equation}

\subsubsection{Result of the normalization process, and final proof of Theorem \ref{pro:semi-classical-normal-form}}

By successive iterations of the normalization process described above,
we can discard every non resonant term of the Hamiltonian Eq.(\pageref{eq:Hx}),
up to a given degree. We have performed a sequence of time-dependant
canonical transformations in a given order indexed by $n=1,2,\ldots n_{max}$,
and $\mathbf{x}$:\[
\hat{\mathcal{T}}_{\mathbf{x}}=T_{x\left(t\right)}Q_{\mathbf{x}}\hat{U}_{G_{1,\mathbf{x}}}\hat{U}_{G_{2,\mathbf{x}}}\ldots\hat{U}_{G_{n_{max},\mathbf{x}}}\]

In this product, each term $\hat{U}_{G_{n,\mathbf{x}}}$ is generated
by a bounded symbol $G_{n,\mathbf{x}}\left(q',p',t\right)$ with compact
support, equal to $G_{n,\mathbf{x}}\left(t,q',p'\right)=g_{l,a,b,\mathbf{x}}\left(t\right)\hbar^{l}q'^{\alpha}p'^{\beta}$
inside the disk $D_{1}$ of radius 1. The resulting time-dependant
Hamiltonian $H_{\mathbf{x}}\left(q',p'\right)$ is a total symbol
which grows quadratically at infinity and has a Taylor expansion in
$x'=\left(q',p'\right)$ at the origin of the form:\[
H_{\mathbf{x}}\left(q',p'\right)=\frac{1}{2}\sum_{2\left(l+j\right)\leq J}\lambda_{l,j,\left(x\right)}\hbar^{l}\left(q'p'\right)^{j}+\mathcal{O}\left(J+1\right)\]

i.e. the Taylor terms of degree less or equal to $J$ are in normal
form, with time-independent coefficients $\lambda_{l,j,\left(x\right)}$.
We can apply Eq.(\ref{eq:relation_commutt}), with time $t=0$, $t'=2$,
and deduce that\[
\hat{M}=\hat{\mathcal{T}}_{M\left(x_{0}\right)}\hat{M}_{J,x_{0}}\hat{\mathcal{T}}_{x_{0}}^{-1}\]
where $\hat{M}_{J,x_{0}}$ is generated by $\hat{H}_{x_{0}\left(t\right)}=Op_{Weyl}\left(H_{x_{0}\left(t\right)}\right)$
on time two, i.e. $\hat{M}_{J,x_{0}}=\hat{M}_{J,x_{0},t=2}$, solution
of $d\hat{M}_{J,x_{0},t}/dt=-\left(i/\hbar\right)\hat{H}_{x_{0}\left(t\right)}\hat{M}_{J,x_{0},t}$,
and $\hat{M}_{J,x_{0},t=0}=\hat{Id}$. Also $\hat{\mathcal{T}}_{x}$
means $\hat{\mathcal{T}}_{\mathbf{x}=\left(x,t=0\right)}$, and involves
$g_{l,a,b,\left(x\right)}\defi g_{l,a,b,\mathbf{x}}\left(t=0\right)$.

In order to simplify the notations, we rescale time by a factor $1/2$,
so that $t\in\left[0,1\right]$, and correspondingly we multiply $H_{x_{0}\left(t\right)}$
by a factor two. We obtain finally:\begin{equation}
H_{x}\left(q',p'\right)=K_{x}\left(q',p'\right)+\mathcal{O}\left(J+1\right)\label{eq:H_normal_form}\end{equation}
\[
K_{x}\left(q',p'\right)=\sum_{l+j\leq J/2}\lambda_{l,j,\left(x\right)}\hbar^{l}\left(q'p'\right)^{j}\]

The functions $\lambda_{l,j,\left(x\right)}$, and $g_{l,a,b,\left(x\right)}$
are continuous with respect to $x\in\mathbb{R}^{2}$. They are also
periodic in $x$ with respect to the lattice $\mathbb{Z}^{2}$, except
for the action $\lambda_{0,0,\left(x\right)}$ . To deduce Eq.(\pageref{eq:quantum_non_stat_normal_form})
of Theorem \ref{pro:semi-classical-normal-form}, we still have to
compare $\hat{M}$ with the unitary map generated by the normal form
part $K_{x}$ only, dropping out the terms of degree higher than $J$,
and estimate the error. This is exactly what gives Theorem \ref{thm:Taylor_approx_of_H},
page \pageref{thm:Taylor_approx_of_H}. Thus, we have finally proved
Theorem \ref{pro:semi-classical-normal-form}.

\section{Trace of a semi-classical hyperbolic normal form\label{sub:Appendix:-Calculus_Normal_Forms}}

In this appendix, we give explicit semi-classical expressions for
the trace and matrix elements of a hyperbolic normal form. They are
already obtained in \cite{iantchenko-02b}, but we adapt them to the
present context. Consider the Hilbert space $\mathcal{H}_{plane}=L^{2}\left(\mathbb{R}\right)$,
and the quadratic normal form Hamiltonian:\[
\hat{K}_{0}=Op_{Weyl}\left(qp\right)=\frac{1}{2}\left(\hat{p}\hat{q}+\hat{q}\hat{p}\right)=\hbar\hat{I}\]
with \[
\hat{I}\equiv-i\frac{1}{2}\left(\frac{\partial}{\partial q}q+q\frac{\partial}{\partial q}\right)\]
Let

\begin{equation}
\hat{K}\defi\sum_{l,j\geq0}\tilde{\mu}_{l,j}\hbar^{l}\hat{K}_{0}^{j}\label{eq:H_H0}\end{equation}
where $\tilde{\mu}_{l,j}\in\mathbb{R}$, but $\tilde{\mu}_{0,1}>0$.
In this paper, the operator $\hat{K}$ appears in Eq.(\ref{eq:K_post_n})
and Eq.(\ref{eq:Ktilde_series_post_n}), as a result of the (post)
normalization procedure along a classical trajectory. Remind that
$\tilde{\mu}_{l,j}=\mu_{l,j}$, if $l\leq1$. The operator $\hat{K}$
generates a unitary propagator operator after time $t$:\begin{equation}
\hat{N}\left(t\right)\defi\exp\left(-it\hat{K}/\hbar\right)\label{eq:M_exp_H}\end{equation}
\foreignlanguage{french}{The truncation operators $\hat{P}_{\alpha}$
has been defined in Eq.(\pageref{eq:def_PD}).}

\selectlanguage{french}
\vspace{0.cm}\begin{center}\fbox{\parbox{16cm}{

\selectlanguage{english}
\begin{prop}
\label{pro:appendix_trace_normale_form}For any $0<\alpha<1/2$, one
has a semi-classical expression:\begin{equation}
\textrm{Tr}\left(\hat{P}_{\alpha}\hat{N}\left(t\right)\hat{P}_{\alpha}\right)=T_{semi}+\mathcal{O}\left(\hbar^{\infty}\right)\label{eq:Append_Trace_forme_normale}\end{equation}
\[
T_{semi}\defi\exp\left(-it\frac{\mu_{0,0}}{\hbar}\right)\exp\left(-it\mu_{1,0}\right)\frac{1}{2\sinh\left(\frac{\mu_{0,1}t}{2}\right)}\mathcal{E}\]
with a semi-classical series

\[
\mathcal{E}=1+\sum_{s\geq1}\hbar^{s}E_{s}=1+\hbar E_{1}+\hbar^{2}E_{2}+\ldots\]
where $E_{s}$ depends on $t$ and $\tilde{\mu}_{l,j}$ , with $\left(l+j\right)\leq s+1$,
and are given explicitly in the proof. We have the control: \begin{equation}
\left|E_{s}\right|\leq t^{s}E_{max,s}\label{eq:bound_Es}\end{equation}
where $E_{max,s}$ does not depend on $t$. 
\end{prop}
\selectlanguage{french}
}}\end{center}\vspace{0.cm}

\selectlanguage{english}
We prove this Proposition below. The idea and techniques of the proof
are taken from \cite{iantchenko-02b} and related work cited in this
paper. In this Proposition, the time $t$ is fixed with respect to
$\hbar$, but the result remains valid for $\left|t\right|=\mathcal{O}\left(\log\left(1/\hbar\right)\right)$,
because the semiclassical expansion remains unchanged.

\subsection{A first useful formal asymptotic formula}

Define $\mu\left(X\right)\defi\sum_{l,j\geq0}\tilde{\mu}_{l,j}\hbar^{l}X^{j}$.
From Eq.(\ref{eq:H_H0}) and Eq.(\ref{eq:M_exp_H}), one can write
$\hat{N}_{t}=\exp\left(-\frac{i}{\hbar}t\mu\left(\hbar\hat{I}\right)\right)$.
Define: \[
r\left(X\right)\defi\mu\left(X\right)-\left(\mu_{0,0}+\hbar\mu_{1,0}+\mu_{0,1}X\right)=\sum_{l+j\geq2}\tilde{\mu}_{l,j}\hbar^{l}X^{j}\]

This decomposition is natural because for a given $I\in\mathbb{R}$,
the expression: \begin{equation}
S=-\frac{it}{\hbar}r\left(\hbar I\right)=-it\sum_{l+j\geq2}\tilde{\mu}_{l,j}\hbar^{l+j-1}I^{j}=\mathcal{O}\left(\hbar\right)\label{eq:S_tilde}\end{equation}

is an asymptotic series in $\hbar$:\begin{equation}
S=\sum_{s\geq1}\hbar^{s}S_{s}\label{eq:equ_S}\end{equation}

with

\[
S_{s}=\left(-it\sum_{j=0}^{s+1}\tilde{\mu}_{\left(l=s-j+1\right),j}I^{j}\right),\qquad s\geq1\]
The first terms are explicitely:

\[
S_{1}=-it\left(\tilde{\mu}_{2,0}+\tilde{\mu}_{1,1}I+\tilde{\mu}_{0,2}I^{2}\right)\]
\[
S_{2}=-it\left(\tilde{\mu}_{3,0}+\tilde{\mu}_{2,1}I+\tilde{\mu}_{1,2}I^{2}+\tilde{\mu}_{0,3}I^{3}\right),\]
Then

\begin{equation}
\exp\left(-\frac{i}{\hbar}t\mu\left(\hbar I\right)\right)=\exp\left(-\frac{i}{\hbar}t\mu_{0,0}\right)\exp\left(-it\mu_{1,0}I\right)\exp\left(-t\mu_{0,1}\right)\mathcal{E}\left(I\right)\label{eq:devel_S}\end{equation}
with the series in $\hbar$ and $I$:\begin{equation}
\mathcal{E}\left(I\right)\defi\exp\left(S\right)=1+\sum_{s\geq1}\hbar^{s}E_{s}\left(I\right)\label{eq:def_E}\end{equation}
where $E_{s}\left(I\right)$ is a polynomial of degree $2s$ in $I$:
\[
E_{s}\left(I\right)=\sum_{j=0}^{2s}E_{s,j}I^{j}\]
The first two terms $E_{s}\left(I\right)$ are:\begin{align}
E_{1}\left(I\right) & =S_{1}=-it\left(\tilde{\mu}_{2,0}+\tilde{\mu}_{1,1}I+\tilde{\mu}_{0,2}I^{2}\right)\label{eq:E1_E2}\\
E_{2}\left(I\right) & =S_{2}+\frac{1}{2}S_{1}^{2}=-\left(it\tilde{\mu}_{3,0}+\frac{1}{2}t^{2}\tilde{\mu}_{2,0}^{2}\right)-I\left(it\tilde{\mu}_{2,1}+t^{2}\tilde{\mu}_{2,0}\tilde{\mu}_{1,1}\right)\\
 & \quad-I^{2}\left(it\tilde{\mu}_{1,2}+\frac{1}{2}t^{2}\tilde{\mu}_{1,1}^{2}+t^{2}\tilde{\mu}_{2,0}\tilde{\mu}_{0,2}\right)-I^{3}\left(it\tilde{\mu}_{0,3}+t^{2}\tilde{\mu}_{1,1}\tilde{\mu}_{0,2}\right)\\
 & \quad-I^{4}\left(\frac{1}{2}t^{2}\tilde{\mu}_{0,2}^{2}\right).\\
 & \mbox{etc..}\end{align}

However, in this paper, due to the special choice of Weyl quantization,
we have $\tilde{\mu}_{l,j}=0$ if $l$ is odd. This simplifies:\[
E_{1}\left(I\right)=-it\left(\tilde{\mu}_{2,0}+\tilde{\mu}_{0,2}I^{2}\right)\]
\[
E_{2}\left(I\right)=-\left(\frac{1}{2}t^{2}\tilde{\mu}_{2,0}^{2}\right)-I\left(it\tilde{\mu}_{2,1}\right)-I^{2}\left(t^{2}\tilde{\mu}_{2,0}\tilde{\mu}_{0,2}\right)-I^{3}\left(it\tilde{\mu}_{0,3}\right)-I^{4}\left(\frac{1}{2}t^{2}\tilde{\mu}_{0,2}^{2}\right).\]

We get:

\[
\hat{N}\left(t\right)=\exp\left(-it\frac{\tilde{\mu}_{0,0}}{\hbar}\right)\exp\left(-it\tilde{\mu}_{1,0}\right)\exp\left(-\frac{it}{\hbar}r\left(\hbar\hat{I}\right)\right)\hat{N}_{0}\left(t\right)\]
with\begin{equation}
\hat{N}_{0}\left(t\right)\defi\exp\left(-it\frac{\tilde{\mu}_{0,1}}{\hbar}\hat{K}_{0}\right)=\exp\left(-i\mu\hat{I}\right)\label{eq:N0_t}\end{equation}
and\[
\mu\defi\tilde{\mu}_{0,1}t\]

We have for any $k\in\mathbb{N}$, \[
\left(i\frac{\partial}{\partial\mu}\right)^{k}\hat{N}_{0}\left(t\right)=\left(\hat{I}\right)^{k}\hat{N}_{0}\left(t\right)\]
so for any formal series $\mathcal{E}\left(I\right)$ in $I$, this
gives:\[
\mathcal{E}\left(i\frac{\partial}{\partial\mu}\right)\hat{N}_{0}\left(t\right)=\mathcal{E}\left(\hat{I}\right)\hat{N}_{0}\left(t\right)\]
It means that we may substitute $\hat{I}\Leftrightarrow\left(i\frac{\partial}{\partial\mu}\right)$
in the formal series. We deduce the formal result \begin{equation}
\boxed{\hat{N}\left(t\right)=\exp\left(-it\frac{\tilde{\mu}_{0,0}}{\hbar}\right)\exp\left(-it\tilde{\mu}_{1,0}\right)\mathcal{E}\left(i\frac{\partial}{\partial\mu}\right)\hat{N}_{0}\left(t\right)}\label{eq:U_asympt}\end{equation}

\paragraph{Remark:}

In the right hand side of Eq.(\ref{eq:U_asympt}), only $\hat{N}_{0}\left(t\right)$
is an operator.

\subsection{Semi-classical expression of the Trace}

We first compute the Trace of Eq.(\ref{eq:N0_t}) in the linear case.
From Eq.(\ref{eq:Integrale}), the functional $\mbox{T}\left(\hat{N}_{0}\right)$
is well defined. It's value is \begin{equation}
T_{\mu}\defi\mbox{T}\left(\hat{N}_{0}\left(t\right)\right)=\frac{1}{2\sinh\left(\frac{\mu}{2}\right)},\qquad\mu=\lambda_{0,1}t\label{eq:Trace_lin}\end{equation}
Remark: it can also be written $T_{\mu}=1/\sqrt{\left|\mbox{det}\left(N_{0}\left(t\right)-I\right)\right|}$
where $N_{0}\left(t\right)\in SL\left(2,\mathbb{R}\right)$ is the
classical map.

\begin{proof}
From its Schwartz kernel $\langle x'|\hat{N}_{0}\left(t\right)|x\rangle=e^{\mu/2}\delta\left(x'-e^{\mu}x\right)$,
and a suitable regularization, one has $\mbox{T}\left(\hat{N}_{0}\left(t\right)\right)=e^{\mu/2}\int dx\delta\left(x-e^{\mu}x\right)=e^{\mu/2}\frac{1}{\left|1-e^{\mu}\right|}=\frac{1}{2\sinh\left(\frac{\mu}{2}\right)}$
.
\end{proof}
In Lemma \ref{lem:integrale_localisee_trace} page \pageref{lem:integrale_localisee_trace},
we showed that the value $T_{\mu}$ comes micro-locally from the origin:\[
\textrm{Tr}\left(\hat{P}_{\alpha}\hat{N}_{0}\left(t\right)\hat{P}_{\alpha}\right)=T_{\mu}+\mathcal{O}\left(\hbar^{\infty}\right)\]

From Eq.(\ref{eq:def_E}), $\mathcal{E}\left(i\frac{\partial}{\partial\mu}\right)\, T_{\mu}=1+\sum_{s,j}\hbar^{s}E_{s,j}\left(i\frac{\partial}{\partial\mu}\right)^{j}T_{\mu}=T_{\mu}\mathcal{E}$
with\[
\mathcal{E}=1+\sum_{s}\hbar^{s}E_{s},\qquad E_{s}=\sum_{j=0}^{2s}E_{s,j}I^{\left(j\right)}\]
and\[
I^{\left(j\right)}\defi\frac{\left(i\right)^{j}}{T_{\mu}}\frac{d^{j}T_{\mu}}{d\mu^{j}}\]
In other words, we have substituted $I^{j}$ by $I^{\left(j\right)}$
in the series Eq.(\ref{eq:def_E}). Finally, we deduce from Eq.(\ref{eq:U_asympt})
that:

\begin{align*}
\textrm{Tr}\left(\hat{P}_{\alpha}\hat{N}\left(t\right)\hat{P}_{\alpha}\right) & =\exp\left(-it\frac{\tilde{\mu}_{0,0}}{\hbar}\right)\exp\left(-it\tilde{\mu}_{1,0}\right)\,\mathcal{E}\left(i\frac{\partial}{\partial\mu}\right)\, T_{\mu}+\mathcal{O}\left(\hbar^{\infty}\right)\\
 & =\exp\left(-it\frac{\tilde{\mu}_{0,0}}{\hbar}\right)\exp\left(-it\tilde{\mu}_{1,0}\right)\,\mathcal{E}\, T_{\mu}+\mathcal{O}\left(\hbar^{\infty}\right)\end{align*}

We have obtained Eq.(\ref{eq:Append_Trace_forme_normale}). We can
calculate the first terms $I^{\left(j\right)}$. With $s\defi\sinh\left(\frac{\mu}{2}\right)$
and $c\defi\cosh\left(\frac{\mu}{2}\right)$, we obtain:

\begin{equation}
I^{\left(1\right)}=-i\frac{c}{2s},\quad I^{\left(2\right)}=-\frac{1}{4s^{2}}\left(2c^{2}-s^{2}\right)\label{eq:I1_I2}\end{equation}
\[
I^{\left(3\right)}=-\frac{i}{8s^{3}}\left(-6c^{3}+5cs^{2}\right),\qquad I^{\left(4\right)}=\frac{1}{16s^{4}}\left(24c^{4}-28s^{2}c^{2}+5s^{4}\right).\]

\paragraph{Expressions for large values of $t$:}

If $t\gg1$, then $\mu=\lambda_{0,1}t\gg1$, and \[
T_{\mu}=e^{-\mu/2}\left(1-e^{-\mu}\right)^{-1}=\sum_{k\geq0}\exp\left(-\mu\left(k+\frac{1}{2}\right)\right)\]
thus\begin{eqnarray*}
I^{(j)} & =\left(i\right)^{j}\frac{1}{T_{\mu}} & \partial_{\mu}^{j}T_{\mu}=\left(-i\right)^{j}\frac{1}{T_{\mu}}\sum_{k\geq0}\left(k+\frac{1}{2}\right)^{j}\exp\left(-\mu\left(k+\frac{1}{2}\right)\right)\\
 &  & =\left(\frac{-i}{2}\right)^{j}+\sum_{k\geq1}\left(\left(k+\frac{1}{2}\right)^{j}-\left(k-\frac{1}{2}\right)^{j}\right)e^{-k\mu}\\
 &  & =\left(\frac{-i}{2}\right)^{j}+\mathcal{O}\left(e^{-\tilde{\mu}_{0,1}t}\right)\end{eqnarray*}
and we deduce in particular that $I^{\left(j\right)}$ is uniformly
bounded for $t\in[1,+\infty[$. From Eq.(\ref{eq:def_E}), we deduce
that \[
\left|E_{s}\right|\leq t^{s}E_{max,s}\]
where $E_{max,s}$ does not depend on $t$.\bibliographystyle{plain}
\bibliography{/home/faure/articles/articles}

\end{document}

%% file: evolution_paquet_onde.pstex_t
\begin{picture}(0,0)%
\includegraphics{evolution_paquet_onde}%
\end{picture}%
\setlength{\unitlength}{3947sp}%
\begingroup\makeatletter\ifx\SetFigFont\undefined%
\gdef\SetFigFont#1#2#3#4#5{%
  \reset@font\fontsize{#1}{#2pt}%
  \fontfamily{#3}\fontseries{#4}\fontshape{#5}%
  \selectfont}%
\fi\endgroup%
\begin{picture}(8219,6640)(-239,-5714)
\put(  1,614){\makebox(0,0)[lb]{\smash{{\SetFigFont{12}{14.4}{\familydefault}{\mddefault}{\updefault}{\color[rgb]{0,0,0}$p$}%
}}}}
\put(1876,-1186){\makebox(0,0)[lb]{\smash{{\SetFigFont{12}{14.4}{\familydefault}{\mddefault}{\updefault}{\color[rgb]{0,0,0}$q$}%
}}}}
\put(1576,-1261){\makebox(0,0)[lb]{\smash{{\SetFigFont{12}{14.4}{\familydefault}{\mddefault}{\updefault}{\color[rgb]{0,0,0}$0.4$}%
}}}}
\put(-74,389){\makebox(0,0)[lb]{\smash{{\SetFigFont{12}{14.4}{\familydefault}{\mddefault}{\updefault}{\color[rgb]{0,0,0}$0.4$}%
}}}}
\put(976,-1261){\makebox(0,0)[lb]{\smash{{\SetFigFont{12}{14.4}{\familydefault}{\mddefault}{\updefault}{\color[rgb]{0,0,0}$0$}%
}}}}
\put(  1,-286){\makebox(0,0)[lb]{\smash{{\SetFigFont{12}{14.4}{\familydefault}{\mddefault}{\updefault}{\color[rgb]{0,0,0}$0$}%
}}}}
\put(-224,-961){\makebox(0,0)[lb]{\smash{{\SetFigFont{12}{14.4}{\familydefault}{\mddefault}{\updefault}{\color[rgb]{0,0,0}$-0.4$}%
}}}}
\put( 76,-1261){\makebox(0,0)[lb]{\smash{{\SetFigFont{12}{14.4}{\familydefault}{\mddefault}{\updefault}{\color[rgb]{0,0,0}$-0.4$}%
}}}}
\put(226,764){\makebox(0,0)[lb]{\smash{{\SetFigFont{12}{14.4}{\familydefault}{\mddefault}{\updefault}{\color[rgb]{0,0,0}$t=0,$}%
}}}}
\put(826,764){\makebox(0,0)[lb]{\smash{{\SetFigFont{12}{14.4}{\familydefault}{\mddefault}{\updefault}{\color[rgb]{0,0,0} $t/t_E=0.$}%
}}}}
\put(2176,764){\makebox(0,0)[lb]{\smash{{\SetFigFont{12}{14.4}{\familydefault}{\mddefault}{\updefault}{\color[rgb]{0,0,0}$t=1,$}%
}}}}
\put(2776,764){\makebox(0,0)[lb]{\smash{{\SetFigFont{12}{14.4}{\familydefault}{\mddefault}{\updefault}{\color[rgb]{0,0,0} $t/t_E=0.14$}%
}}}}
\put(4126,764){\makebox(0,0)[lb]{\smash{{\SetFigFont{12}{14.4}{\familydefault}{\mddefault}{\updefault}{\color[rgb]{0,0,0}$t=2,$}%
}}}}
\put(4726,764){\makebox(0,0)[lb]{\smash{{\SetFigFont{12}{14.4}{\familydefault}{\mddefault}{\updefault}{\color[rgb]{0,0,0} $t/t_E=0.28$}%
}}}}
\put(6076,764){\makebox(0,0)[lb]{\smash{{\SetFigFont{12}{14.4}{\familydefault}{\mddefault}{\updefault}{\color[rgb]{0,0,0}$t=3,$}%
}}}}
\put(6676,764){\makebox(0,0)[lb]{\smash{{\SetFigFont{12}{14.4}{\familydefault}{\mddefault}{\updefault}{\color[rgb]{0,0,0} $t/t_E=0.42$}%
}}}}
\put(226,-1486){\makebox(0,0)[lb]{\smash{{\SetFigFont{12}{14.4}{\familydefault}{\mddefault}{\updefault}{\color[rgb]{0,0,0}$t=4,$}%
}}}}
\put(826,-1486){\makebox(0,0)[lb]{\smash{{\SetFigFont{12}{14.4}{\familydefault}{\mddefault}{\updefault}{\color[rgb]{0,0,0} $t/t_E=0.56$}%
}}}}
\put(2176,-1486){\makebox(0,0)[lb]{\smash{{\SetFigFont{12}{14.4}{\familydefault}{\mddefault}{\updefault}{\color[rgb]{0,0,0}$t=5,$}%
}}}}
\put(2776,-1486){\makebox(0,0)[lb]{\smash{{\SetFigFont{12}{14.4}{\familydefault}{\mddefault}{\updefault}{\color[rgb]{0,0,0} $t/t_E=0.70$}%
}}}}
\put(4126,-1486){\makebox(0,0)[lb]{\smash{{\SetFigFont{12}{14.4}{\familydefault}{\mddefault}{\updefault}{\color[rgb]{0,0,0}$t=6,$}%
}}}}
\put(6076,-1486){\makebox(0,0)[lb]{\smash{{\SetFigFont{12}{14.4}{\familydefault}{\mddefault}{\updefault}{\color[rgb]{0,0,0}$t=7,$}%
}}}}
\put(4726,-1486){\makebox(0,0)[lb]{\smash{{\SetFigFont{12}{14.4}{\familydefault}{\mddefault}{\updefault}{\color[rgb]{0,0,0} $t/t_E=0.84$}%
}}}}
\put(6676,-1486){\makebox(0,0)[lb]{\smash{{\SetFigFont{12}{14.4}{\familydefault}{\mddefault}{\updefault}{\color[rgb]{0,0,0} $t/t_E=0.98$}%
}}}}
\put(226,-3661){\makebox(0,0)[lb]{\smash{{\SetFigFont{12}{14.4}{\familydefault}{\mddefault}{\updefault}{\color[rgb]{0,0,0}$t=8,$}%
}}}}
\put(2176,-3661){\makebox(0,0)[lb]{\smash{{\SetFigFont{12}{14.4}{\familydefault}{\mddefault}{\updefault}{\color[rgb]{0,0,0}$t=9,$}%
}}}}
\put(4126,-3661){\makebox(0,0)[lb]{\smash{{\SetFigFont{12}{14.4}{\familydefault}{\mddefault}{\updefault}{\color[rgb]{0,0,0}$t=10,$}%
}}}}
\put(826,-3661){\makebox(0,0)[lb]{\smash{{\SetFigFont{12}{14.4}{\familydefault}{\mddefault}{\updefault}{\color[rgb]{0,0,0} $t/t_E=1.1$}%
}}}}
\put(2776,-3661){\makebox(0,0)[lb]{\smash{{\SetFigFont{12}{14.4}{\familydefault}{\mddefault}{\updefault}{\color[rgb]{0,0,0} $t/t_E=1.3$}%
}}}}
\put(4801,-3661){\makebox(0,0)[lb]{\smash{{\SetFigFont{12}{14.4}{\familydefault}{\mddefault}{\updefault}{\color[rgb]{0,0,0} $t/t_E=1.4$}%
}}}}
\put(6076,-3661){\makebox(0,0)[lb]{\smash{{\SetFigFont{12}{14.4}{\familydefault}{\mddefault}{\updefault}{\color[rgb]{0,0,0}$t=15,$}%
}}}}
\put(6751,-3661){\makebox(0,0)[lb]{\smash{{\SetFigFont{12}{14.4}{\familydefault}{\mddefault}{\updefault}{\color[rgb]{0,0,0} $t/t_E=2.2$}%
}}}}
\end{picture}%

%% file: time_range.pstex_t
\begin{picture}(0,0)%
\includegraphics{time_range}%
\end{picture}%
\setlength{\unitlength}{3947sp}%
\begingroup\makeatletter\ifx\SetFigFont\undefined%
\gdef\SetFigFont#1#2#3#4#5{%
  \reset@font\fontsize{#1}{#2pt}%
  \fontfamily{#3}\fontseries{#4}\fontshape{#5}%
  \selectfont}%
\fi\endgroup%
\begin{picture}(7527,2244)(211,-2455)
\put(2026,-1186){\makebox(0,0)[lb]{\smash{{\SetFigFont{12}{14.4}{\rmdefault}{\mddefault}{\updefault}{\color[rgb]{0,0,0}$t_E/6$}%
}}}}
\put(3076,-1186){\makebox(0,0)[lb]{\smash{{\SetFigFont{12}{14.4}{\rmdefault}{\mddefault}{\updefault}{\color[rgb]{0,0,0}$t_E/2$}%
}}}}
\put(4051,-1186){\makebox(0,0)[lb]{\smash{{\SetFigFont{12}{14.4}{\rmdefault}{\mddefault}{\updefault}{\color[rgb]{0,0,0}$3t_E/2$}%
}}}}
\put(4651,-1186){\makebox(0,0)[lb]{\smash{{\SetFigFont{12}{14.4}{\rmdefault}{\mddefault}{\updefault}{\color[rgb]{0,0,0}$2t_E$}%
}}}}
\put(5851,-1186){\makebox(0,0)[lb]{\smash{{\SetFigFont{12}{14.4}{\rmdefault}{\mddefault}{\updefault}{\color[rgb]{0,0,0}$C.t_E$}%
}}}}
\put(6751,-1186){\makebox(0,0)[lb]{\smash{{\SetFigFont{12}{14.4}{\rmdefault}{\mddefault}{\updefault}{\color[rgb]{0,0,0}$t_H$}%
}}}}
\put(7651,-1036){\makebox(0,0)[lb]{\smash{{\SetFigFont{12}{14.4}{\rmdefault}{\mddefault}{\updefault}{\color[rgb]{0,0,0}$t$}%
}}}}
\end{picture}%

%% file: flot_hyperbolique.pstex_t
\begin{picture}(0,0)%
\includegraphics{flot_hyperbolique}%
\end{picture}%
\setlength{\unitlength}{3947sp}%
\begingroup\makeatletter\ifx\SetFigFont\undefined%
\gdef\SetFigFont#1#2#3#4#5{%
  \reset@font\fontsize{#1}{#2pt}%
  \fontfamily{#3}\fontseries{#4}\fontshape{#5}%
  \selectfont}%
\fi\endgroup%
\begin{picture}(4224,2733)(3589,-2473)
\put(6976,-1486){\makebox(0,0)[lb]{\smash{{\SetFigFont{12}{14.4}{\rmdefault}{\mddefault}{\updefault}{\color[rgb]{0,0,0}$q$}%
}}}}
\put(5176, 89){\makebox(0,0)[lb]{\smash{{\SetFigFont{12}{14.4}{\rmdefault}{\mddefault}{\updefault}{\color[rgb]{0,0,0}$p$}%
}}}}
\put(4126, 14){\makebox(0,0)[lb]{\smash{{\SetFigFont{12}{14.4}{\rmdefault}{\mddefault}{\updefault}{\color[rgb]{0,0,0}$e^{-\lambda_0}<1$}%
}}}}
\put(6676, 14){\makebox(0,0)[lb]{\smash{{\SetFigFont{12}{14.4}{\rmdefault}{\mddefault}{\updefault}{\color[rgb]{0,0,0}$e^{\lambda_0}>1$}%
}}}}
\end{picture}%

%% file: cat_map.pstex_t
\begin{picture}(0,0)%
\includegraphics{cat_map}%
\end{picture}%
\setlength{\unitlength}{3947sp}%
\begingroup\makeatletter\ifx\SetFigFont\undefined%
\gdef\SetFigFont#1#2#3#4#5{%
  \reset@font\fontsize{#1}{#2pt}%
  \fontfamily{#3}\fontseries{#4}\fontshape{#5}%
  \selectfont}%
\fi\endgroup%
\begin{picture}(3474,1224)(1789,-3823)
\put(3076,-3736){\makebox(0,0)[lb]{\smash{{\SetFigFont{12}{14.4}{\rmdefault}{\mddefault}{\updefault}{\color[rgb]{0,0,0}$M_0$ on $\mathbb{T}^2$}%
}}}}
\end{picture}%

%% file: variete.pstex_t
\begin{picture}(0,0)%
\includegraphics{variete}%
\end{picture}%
\setlength{\unitlength}{3947sp}%
\begingroup\makeatletter\ifx\SetFigFont\undefined%
\gdef\SetFigFont#1#2#3#4#5{%
  \reset@font\fontsize{#1}{#2pt}%
  \fontfamily{#3}\fontseries{#4}\fontshape{#5}%
  \selectfont}%
\fi\endgroup%
\begin{picture}(2877,2785)(886,-2684)
\put(901,-286){\makebox(0,0)[lb]{\smash{{\SetFigFont{12}{14.4}{\rmdefault}{\mddefault}{\updefault}{\color[rgb]{0,0,0}$p$}%
}}}}
\put(3376,-2611){\makebox(0,0)[lb]{\smash{{\SetFigFont{12}{14.4}{\rmdefault}{\mddefault}{\updefault}{\color[rgb]{0,0,0}$q$}%
}}}}
\put(1651,-586){\makebox(0,0)[lb]{\smash{{\SetFigFont{12}{14.4}{\rmdefault}{\mddefault}{\updefault}{\color[rgb]{0,0,0}$s_x$}%
}}}}
\put(2701,-811){\makebox(0,0)[lb]{\smash{{\SetFigFont{12}{14.4}{\rmdefault}{\mddefault}{\updefault}{\color[rgb]{0,0,0}$u_x$}%
}}}}
\put(2026,-1411){\makebox(0,0)[lb]{\smash{{\SetFigFont{12}{14.4}{\rmdefault}{\mddefault}{\updefault}{\color[rgb]{0,0,0}$x$}%
}}}}
\end{picture}%

%% file: points_periodiques.pstex_t
\begin{picture}(0,0)%
\includegraphics{points_periodiques}%
\end{picture}%
\setlength{\unitlength}{3947sp}%
\begingroup\makeatletter\ifx\SetFigFont\undefined%
\gdef\SetFigFont#1#2#3#4#5{%
  \reset@font\fontsize{#1}{#2pt}%
  \fontfamily{#3}\fontseries{#4}\fontshape{#5}%
  \selectfont}%
\fi\endgroup%
\begin{picture}(2952,2874)(586,-2548)
\put(601, 14){\makebox(0,0)[lb]{\smash{{\SetFigFont{12}{14.4}{\familydefault}{\mddefault}{\updefault}{\color[rgb]{0,0,0}$p$}%
}}}}
\put(3151,-2461){\makebox(0,0)[lb]{\smash{{\SetFigFont{12}{14.4}{\familydefault}{\mddefault}{\updefault}{\color[rgb]{0,0,0}$q$}%
}}}}
\end{picture}%

%% file: normalization_final.pstex_t
\begin{picture}(0,0)%
\includegraphics{normalization_final}%
\end{picture}%
\setlength{\unitlength}{3947sp}%
\begingroup\makeatletter\ifx\SetFigFont\undefined%
\gdef\SetFigFont#1#2#3#4#5{%
  \reset@font\fontsize{#1}{#2pt}%
  \fontfamily{#3}\fontseries{#4}\fontshape{#5}%
  \selectfont}%
\fi\endgroup%
\begin{picture}(5424,1983)(1039,-2323)
\put(1126,-736){\makebox(0,0)[lb]{\smash{{\SetFigFont{12}{14.4}{\rmdefault}{\mddefault}{\updefault}{\color[rgb]{0,0,0}$p$}%
}}}}
\put(1426,-1261){\makebox(0,0)[lb]{\smash{{\SetFigFont{12}{14.4}{\rmdefault}{\mddefault}{\updefault}{\color[rgb]{0,0,0}$s_x$}%
}}}}
\put(1726,-1936){\makebox(0,0)[lb]{\smash{{\SetFigFont{12}{14.4}{\rmdefault}{\mddefault}{\updefault}{\color[rgb]{0,0,0}$x$}%
}}}}
\put(2476,-1636){\makebox(0,0)[lb]{\smash{{\SetFigFont{12}{14.4}{\rmdefault}{\mddefault}{\updefault}{\color[rgb]{0,0,0}$u_x$}%
}}}}
\put(2851,-2236){\makebox(0,0)[lb]{\smash{{\SetFigFont{12}{14.4}{\rmdefault}{\mddefault}{\updefault}{\color[rgb]{0,0,0}$q$}%
}}}}
\put(2701,-811){\makebox(0,0)[lb]{\smash{{\SetFigFont{12}{14.4}{\rmdefault}{\mddefault}{\updefault}{\color[rgb]{0,0,0}$M(x)$}%
}}}}
\put(1951,-1261){\makebox(0,0)[lb]{\smash{{\SetFigFont{12}{14.4}{\rmdefault}{\mddefault}{\updefault}{\color[rgb]{0,0,0}$M$}%
}}}}
\put(2251,-511){\makebox(0,0)[lb]{\smash{{\SetFigFont{12}{14.4}{\rmdefault}{\mddefault}{\updefault}{\color[rgb]{0,0,0}$s_{Mx}$}%
}}}}
\put(3301,-811){\makebox(0,0)[lb]{\smash{{\SetFigFont{12}{14.4}{\rmdefault}{\mddefault}{\updefault}{\color[rgb]{0,0,0}$u_{Mx}$}%
}}}}
\put(6301,-1636){\makebox(0,0)[lb]{\smash{{\SetFigFont{12}{14.4}{\rmdefault}{\mddefault}{\updefault}{\color[rgb]{0,0,0}$q'$}%
}}}}
\put(5176,-586){\makebox(0,0)[lb]{\smash{{\SetFigFont{12}{14.4}{\rmdefault}{\mddefault}{\updefault}{\color[rgb]{0,0,0}$p'$}%
}}}}
\put(3901,-2011){\makebox(0,0)[lb]{\smash{{\SetFigFont{12}{14.4}{\rmdefault}{\mddefault}{\updefault}{\color[rgb]{0,0,0}$\mathcal{T}_x$}%
}}}}
\put(3976,-961){\makebox(0,0)[lb]{\smash{{\SetFigFont{12}{14.4}{\rmdefault}{\mddefault}{\updefault}{\color[rgb]{0,0,0}$\mathcal{T}_{Mx}$}%
}}}}
\put(5701,-1036){\makebox(0,0)[lb]{\smash{{\SetFigFont{12}{14.4}{\rmdefault}{\mddefault}{\updefault}{\color[rgb]{0,0,0}$N_x$}%
}}}}
\put(5251,-1561){\makebox(0,0)[lb]{\smash{{\SetFigFont{12}{14.4}{\rmdefault}{\mddefault}{\updefault}{\color[rgb]{0,0,0}$0$}%
}}}}
\end{picture}%

%% file: trace_numerique.pstex_t
\begin{picture}(0,0)%
\includegraphics{trace_numerique}%
\end{picture}%
\setlength{\unitlength}{3947sp}%
\begingroup\makeatletter\ifx\SetFigFont\undefined%
\gdef\SetFigFont#1#2#3#4#5{%
  \reset@font\fontsize{#1}{#2pt}%
  \fontfamily{#3}\fontseries{#4}\fontshape{#5}%
  \selectfont}%
\fi\endgroup%
\begin{picture}(6494,6498)(2970,-6280)
\put(5551,-2761){\makebox(0,0)[lb]{\smash{{\SetFigFont{12}{14.4}{\rmdefault}{\mddefault}{\updefault}{\color[rgb]{0,0,0}$t/t_E$}%
}}}}
\put(8851,-6061){\makebox(0,0)[lb]{\smash{{\SetFigFont{12}{14.4}{\rmdefault}{\mddefault}{\updefault}{\color[rgb]{0,0,0}$t/t_E$}%
}}}}
\put(5551,-6061){\makebox(0,0)[lb]{\smash{{\SetFigFont{12}{14.4}{\rmdefault}{\mddefault}{\updefault}{\color[rgb]{0,0,0}$t/t_E$}%
}}}}
\put(7726,-3436){\makebox(0,0)[lb]{\smash{{\SetFigFont{12}{14.4}{\rmdefault}{\mddefault}{\updefault}{\color[rgb]{0,0,0}$1/h$}%
}}}}
\put(8101,-5611){\makebox(0,0)[lb]{\smash{{\SetFigFont{12}{14.4}{\rmdefault}{\mddefault}{\updefault}{\color[rgb]{0,0,0}Error$(t)$}%
}}}}
\put(8176,-4636){\makebox(0,0)[lb]{\smash{{\SetFigFont{12}{14.4}{\rmdefault}{\mddefault}{\updefault}{\color[rgb]{0,0,0}$\varepsilon_J(t)$}%
}}}}
\put(8856,-2804){\makebox(0,0)[lb]{\smash{{\SetFigFont{12}{14.4}{\rmdefault}{\mddefault}{\updefault}{\color[rgb]{0,0,0}$t/t_E$}%
}}}}
\put(7126,-2911){\makebox(0,0)[lb]{\smash{{\SetFigFont{12}{14.4}{\familydefault}{\mddefault}{\updefault}{\color[rgb]{0,0,0}$h=0.1,   J=4$}%
}}}}
\put(3751,-6211){\makebox(0,0)[lb]{\smash{{\SetFigFont{12}{14.4}{\familydefault}{\mddefault}{\updefault}{\color[rgb]{0,0,0}$h=0.01,   J=2$}%
}}}}
\put(6976,-6211){\makebox(0,0)[lb]{\smash{{\SetFigFont{12}{14.4}{\familydefault}{\mddefault}{\updefault}{\color[rgb]{0,0,0}$h=0.01,   J=4$}%
}}}}
\put(3751,-2911){\makebox(0,0)[lb]{\smash{{\SetFigFont{12}{14.4}{\familydefault}{\mddefault}{\updefault}{\color[rgb]{0,0,0}$h=0.1,   J=2$}%
}}}}
\put(3676,-136){\makebox(0,0)[lb]{\smash{{\SetFigFont{12}{14.4}{\rmdefault}{\mddefault}{\updefault}{\color[rgb]{0,0,0}$1/h$}%
}}}}
\put(4576,-136){\makebox(0,0)[lb]{\smash{{\SetFigFont{12}{14.4}{\rmdefault}{\mddefault}{\updefault}{\color[rgb]{0,0,0}$\varepsilon_J(t)$}%
}}}}
\put(5401,-1036){\makebox(0,0)[lb]{\smash{{\SetFigFont{12}{14.4}{\rmdefault}{\mddefault}{\updefault}{\color[rgb]{0,0,0}Tr$(\hat{M}^t)$}%
}}}}
\put(4351,-2386){\makebox(0,0)[lb]{\smash{{\SetFigFont{12}{14.4}{\rmdefault}{\mddefault}{\updefault}{\color[rgb]{0,0,0}Error$(t)$}%
}}}}
\put(6976,-361){\makebox(0,0)[lb]{\smash{{\SetFigFont{12}{14.4}{\rmdefault}{\mddefault}{\updefault}{\color[rgb]{0,0,0}$1/h$}%
}}}}
\put(6901,-661){\makebox(0,0)[lb]{\smash{{\SetFigFont{12}{14.4}{\rmdefault}{\mddefault}{\updefault}{\color[rgb]{0,0,0}Tr$(\hat{M}^t)$}%
}}}}
\put(7576,-1261){\makebox(0,0)[lb]{\smash{{\SetFigFont{12}{14.4}{\rmdefault}{\mddefault}{\updefault}{\color[rgb]{0,0,0}$\varepsilon_J(t)$}%
}}}}
\put(7126,-2386){\makebox(0,0)[lb]{\smash{{\SetFigFont{12}{14.4}{\rmdefault}{\mddefault}{\updefault}{\color[rgb]{0,0,0}Error$(t)$}%
}}}}
\put(3676,-3886){\makebox(0,0)[lb]{\smash{{\SetFigFont{12}{14.4}{\rmdefault}{\mddefault}{\updefault}{\color[rgb]{0,0,0}Tr$(\hat{M}^t)$}%
}}}}
\put(3676,-5236){\makebox(0,0)[lb]{\smash{{\SetFigFont{12}{14.4}{\rmdefault}{\mddefault}{\updefault}{\color[rgb]{0,0,0}Error$(t)$}%
}}}}
\put(5401,-3736){\makebox(0,0)[lb]{\smash{{\SetFigFont{12}{14.4}{\rmdefault}{\mddefault}{\updefault}{\color[rgb]{0,0,0}$\varepsilon_J(t)$}%
}}}}
\put(3751,-3436){\makebox(0,0)[lb]{\smash{{\SetFigFont{12}{14.4}{\rmdefault}{\mddefault}{\updefault}{\color[rgb]{0,0,0}$1/h$}%
}}}}
\put(6901,-4336){\makebox(0,0)[lb]{\smash{{\SetFigFont{12}{14.4}{\rmdefault}{\mddefault}{\updefault}{\color[rgb]{0,0,0}Tr$(\hat{M}^t)$}%
}}}}
\put(4576,-1411){\makebox(0,0)[lb]{\smash{{\SetFigFont{12}{14.4}{\rmdefault}{\mddefault}{\updefault}{\color[rgb]{0,0,0}$\varepsilon_{\hbar,J}$}%
}}}}
\put(8251,-1936){\makebox(0,0)[lb]{\smash{{\SetFigFont{12}{14.4}{\rmdefault}{\mddefault}{\updefault}{\color[rgb]{0,0,0}$\varepsilon_{\hbar,J}$}%
}}}}
\put(4351,-4636){\makebox(0,0)[lb]{\smash{{\SetFigFont{12}{14.4}{\rmdefault}{\mddefault}{\updefault}{\color[rgb]{0,0,0}$\varepsilon_{\hbar,J}$}%
}}}}
\put(8026,-5236){\makebox(0,0)[lb]{\smash{{\SetFigFont{12}{14.4}{\rmdefault}{\mddefault}{\updefault}{\color[rgb]{0,0,0}$\varepsilon_{\hbar,J}$}%
}}}}
\end{picture}%

%% file: set_Ei_Ii.pstex_t
\begin{picture}(0,0)%
\includegraphics{set_Ei_Ii}%
\end{picture}%
\setlength{\unitlength}{3947sp}%
\begingroup\makeatletter\ifx\SetFigFont\undefined%
\gdef\SetFigFont#1#2#3#4#5{%
  \reset@font\fontsize{#1}{#2pt}%
  \fontfamily{#3}\fontseries{#4}\fontshape{#5}%
  \selectfont}%
\fi\endgroup%
\begin{picture}(5214,1728)(661,-2005)
\put(1276,-1561){\makebox(0,0)[lb]{\smash{{\SetFigFont{12}{14.4}{\rmdefault}{\mddefault}{\updefault}{\color[rgb]{0,0,0}$M$}%
}}}}
\put(3151,-1261){\makebox(0,0)[lb]{\smash{{\SetFigFont{12}{14.4}{\rmdefault}{\mddefault}{\updefault}{\color[rgb]{0,0,0}$M$}%
}}}}
\put(676,-1936){\makebox(0,0)[lb]{\smash{{\SetFigFont{12}{14.4}{\rmdefault}{\mddefault}{\updefault}{\color[rgb]{0,0,0}$I_0(D)=D$}%
}}}}
\put(2476,-1936){\makebox(0,0)[lb]{\smash{{\SetFigFont{12}{14.4}{\rmdefault}{\mddefault}{\updefault}{\color[rgb]{0,0,0}$I_1(D)$}%
}}}}
\put(4276,-1861){\makebox(0,0)[lb]{\smash{{\SetFigFont{12}{14.4}{\rmdefault}{\mddefault}{\updefault}{\color[rgb]{0,0,0}$I_2(D)$}%
}}}}
\put(3901,-436){\makebox(0,0)[lb]{\smash{{\SetFigFont{12}{14.4}{\rmdefault}{\mddefault}{\updefault}{\color[rgb]{0,0,0}$E_2(D)$}%
}}}}
\put(1801,-661){\makebox(0,0)[lb]{\smash{{\SetFigFont{12}{14.4}{\rmdefault}{\mddefault}{\updefault}{\color[rgb]{0,0,0}$E_1(D)$}%
}}}}
\end{picture}%

%% file: schema_D_G_F.pstex_t
\begin{picture}(0,0)%
\includegraphics{schema_D_G_F}%
\end{picture}%
\setlength{\unitlength}{3947sp}%
\begingroup\makeatletter\ifx\SetFigFont\undefined%
\gdef\SetFigFont#1#2#3#4#5{%
  \reset@font\fontsize{#1}{#2pt}%
  \fontfamily{#3}\fontseries{#4}\fontshape{#5}%
  \selectfont}%
\fi\endgroup%
\begin{picture}(6680,5193)(-2602,-5170)
\put(691,-5101){\makebox(0,0)[lb]{\smash{{\SetFigFont{12}{14.4}{\rmdefault}{\mddefault}{\updefault}{\color[rgb]{0,0,0}$D_t$}%
}}}}
\put(-2319,-4526){\makebox(0,0)[lb]{\smash{{\SetFigFont{12}{14.4}{\rmdefault}{\mddefault}{\updefault}{\color[rgb]{0,0,0}$D_2$}%
}}}}
\put(-999,-4246){\makebox(0,0)[lb]{\smash{{\SetFigFont{12}{14.4}{\rmdefault}{\mddefault}{\updefault}{\color[rgb]{0,0,0}$I_1(G_1)$}%
}}}}
\put(2531,-4956){\makebox(0,0)[lb]{\smash{{\SetFigFont{12}{14.4}{\rmdefault}{\mddefault}{\updefault}{\color[rgb]{0,0,0}$I_{t-1}(G_1)$}%
}}}}
\put(-1424,-2011){\makebox(0,0)[lb]{\smash{{\SetFigFont{12}{14.4}{\rmdefault}{\mddefault}{\updefault}{\color[rgb]{0,0,0}$D_0$}%
}}}}
\put(1651,-136){\makebox(0,0)[lb]{\smash{{\SetFigFont{12}{14.4}{\rmdefault}{\mddefault}{\updefault}{\color[rgb]{0,0,0}$G_1$}%
}}}}
\put( 76,-811){\makebox(0,0)[lb]{\smash{{\SetFigFont{12}{14.4}{\rmdefault}{\mddefault}{\updefault}{\color[rgb]{0,0,0}$M$}%
}}}}
\put(1051,-2536){\makebox(0,0)[lb]{\smash{{\SetFigFont{12}{14.4}{\rmdefault}{\mddefault}{\updefault}{\color[rgb]{0,0,0}$D_1$}%
}}}}
\put(-159,-2266){\makebox(0,0)[lb]{\smash{{\SetFigFont{12}{14.4}{\rmdefault}{\mddefault}{\updefault}{\color[rgb]{0,0,0}$M$}%
}}}}
\put(832,-3406){\makebox(0,0)[lb]{\smash{{\SetFigFont{12}{14.4}{\rmdefault}{\mddefault}{\updefault}{\color[rgb]{0,0,0}etc...}%
}}}}
\end{picture}%

%% file: espace_qp_t.pstex_t
\begin{picture}(0,0)%
\includegraphics{espace_qp_t}%
\end{picture}%
\setlength{\unitlength}{3947sp}%
\begingroup\makeatletter\ifx\SetFigFont\undefined%
\gdef\SetFigFont#1#2#3#4#5{%
  \reset@font\fontsize{#1}{#2pt}%
  \fontfamily{#3}\fontseries{#4}\fontshape{#5}%
  \selectfont}%
\fi\endgroup%
\begin{picture}(2427,2620)(1261,-2459)
\put(1951,-286){\makebox(0,0)[lb]{\smash{{\SetFigFont{12}{14.4}{\rmdefault}{\mddefault}{\updefault}{\color[rgb]{0,0,0}$2$}%
}}}}
\put(1951,-1636){\makebox(0,0)[lb]{\smash{{\SetFigFont{12}{14.4}{\rmdefault}{\mddefault}{\updefault}{\color[rgb]{0,0,0}$0$}%
}}}}
\put(1276,-2386){\makebox(0,0)[lb]{\smash{{\SetFigFont{12}{14.4}{\rmdefault}{\mddefault}{\updefault}{\color[rgb]{0,0,0}$q$}%
}}}}
\put(1951, 14){\makebox(0,0)[lb]{\smash{{\SetFigFont{12}{14.4}{\rmdefault}{\mddefault}{\updefault}{\color[rgb]{0,0,0}$t$}%
}}}}
\put(2776,-1936){\makebox(0,0)[lb]{\smash{{\SetFigFont{12}{14.4}{\rmdefault}{\mddefault}{\updefault}{\color[rgb]{0,0,0}$M(x_0)$}%
}}}}
\put(3526,-1636){\makebox(0,0)[lb]{\smash{{\SetFigFont{12}{14.4}{\rmdefault}{\mddefault}{\updefault}{\color[rgb]{0,0,0}$p$}%
}}}}
\put(2176,-1111){\makebox(0,0)[lb]{\smash{{\SetFigFont{12}{14.4}{\rmdefault}{\mddefault}{\updefault}{\color[rgb]{0,0,0}$x(t)$}%
}}}}
\put(2776,-586){\makebox(0,0)[lb]{\smash{{\SetFigFont{12}{14.4}{\rmdefault}{\mddefault}{\updefault}{\color[rgb]{0,0,0}$M(x_0)$}%
}}}}
\put(2026,-2161){\makebox(0,0)[lb]{\smash{{\SetFigFont{12}{14.4}{\rmdefault}{\mddefault}{\updefault}{\color[rgb]{0,0,0}$x_0$}%
}}}}
\end{picture}%

%% file: espace_qp_t_bis.pstex_t
\begin{picture}(0,0)%
\includegraphics{espace_qp_t_bis}%
\end{picture}%
\setlength{\unitlength}{3947sp}%
\begingroup\makeatletter\ifx\SetFigFont\undefined%
\gdef\SetFigFont#1#2#3#4#5{%
  \reset@font\fontsize{#1}{#2pt}%
  \fontfamily{#3}\fontseries{#4}\fontshape{#5}%
  \selectfont}%
\fi\endgroup%
\begin{picture}(4377,2836)(1261,-2600)
\put(1951, 89){\makebox(0,0)[lb]{\smash{{\SetFigFont{12}{14.4}{\rmdefault}{\mddefault}{\updefault}{\color[rgb]{0,0,0}$t$}%
}}}}
\put(2476,-286){\makebox(0,0)[lb]{\smash{{\SetFigFont{12}{14.4}{\rmdefault}{\mddefault}{\updefault}{\color[rgb]{0,0,0}$M(x_0)$}%
}}}}
\put(2626,-1261){\makebox(0,0)[lb]{\smash{{\SetFigFont{12}{14.4}{\rmdefault}{\mddefault}{\updefault}{\color[rgb]{0,0,0}$x(t)$}%
}}}}
\put(2401,-1861){\makebox(0,0)[lb]{\smash{{\SetFigFont{12}{14.4}{\rmdefault}{\mddefault}{\updefault}{\color[rgb]{0,0,0}$s$}%
}}}}
\put(2101,-2536){\makebox(0,0)[lb]{\smash{{\SetFigFont{12}{14.4}{\rmdefault}{\mddefault}{\updefault}{\color[rgb]{0,0,0}$u$}%
}}}}
\put(1876,-286){\makebox(0,0)[lb]{\smash{{\SetFigFont{12}{14.4}{\rmdefault}{\mddefault}{\updefault}{\color[rgb]{0,0,0}$2$}%
}}}}
\put(1951,-1636){\makebox(0,0)[lb]{\smash{{\SetFigFont{12}{14.4}{\rmdefault}{\mddefault}{\updefault}{\color[rgb]{0,0,0}$0$}%
}}}}
\put(1276,-2461){\makebox(0,0)[lb]{\smash{{\SetFigFont{12}{14.4}{\rmdefault}{\mddefault}{\updefault}{\color[rgb]{0,0,0}$q$}%
}}}}
\put(3451,-1861){\makebox(0,0)[lb]{\smash{{\SetFigFont{12}{14.4}{\rmdefault}{\mddefault}{\updefault}{\color[rgb]{0,0,0}$p$}%
}}}}
\put(5401,-1786){\makebox(0,0)[lb]{\smash{{\SetFigFont{12}{14.4}{\rmdefault}{\mddefault}{\updefault}{\color[rgb]{0,0,0}$p'$}%
}}}}
\put(4426,-2161){\makebox(0,0)[lb]{\smash{{\SetFigFont{12}{14.4}{\rmdefault}{\mddefault}{\updefault}{\color[rgb]{0,0,0}$q'$}%
}}}}
\put(4726,-1936){\makebox(0,0)[lb]{\smash{{\SetFigFont{12}{14.4}{\rmdefault}{\mddefault}{\updefault}{\color[rgb]{0,0,0}$u$}%
}}}}
\put(5101,-1561){\makebox(0,0)[lb]{\smash{{\SetFigFont{12}{14.4}{\rmdefault}{\mddefault}{\updefault}{\color[rgb]{0,0,0}$s$}%
}}}}
\put(3751,-1111){\makebox(0,0)[lb]{\smash{{\SetFigFont{12}{14.4}{\rmdefault}{\mddefault}{\updefault}{\color[rgb]{0,0,0}$\mathcal{T}_{pre}$}%
}}}}
\put(2101,-2086){\makebox(0,0)[lb]{\smash{{\SetFigFont{12}{14.4}{\rmdefault}{\mddefault}{\updefault}{\color[rgb]{0,0,0}$x_0$}%
}}}}
\put(4726, 89){\makebox(0,0)[lb]{\smash{{\SetFigFont{12}{14.4}{\rmdefault}{\mddefault}{\updefault}{\color[rgb]{0,0,0}$t$}%
}}}}
\end{picture}%